\documentclass[twocolumn,times]{aastex63}
\usepackage{CJK}
\usepackage{amsmath}
\usepackage{mathtools}
\usepackage[caption=false]{subfig}

\shorttitle{Calibrating the DCF method}
\shortauthors{Liu et al.}

\begin{document}
\begin{CJK*}{UTF8}{gbsn}
\title{Calibrating the Davis-Chandrasekhar-Fermi method with numerical simulations: uncertainties in estimating the magnetic field strength from statistics of field orientations}

\correspondingauthor{Keping Qiu}
\email{kpqiu@nju.edu.cn}

\author[0000-0002-4774-2998]{Junhao Liu (刘峻豪)}
\affil{School of Astronomy and Space Science, Nanjing University, 163 Xianlin Avenue, Nanjing, Jiangsu 210023, People's Republic of China}
\affil{Key Laboratory of Modern Astronomy and Astrophysics (Nanjing University), Ministry of Education, Nanjing, Jiangsu 210023, People's Republic of China}
\affil{Center for Astrophysics $\vert$ Harvard \& Smithsonian, 60 Garden Street, Cambridge, MA 02138, USA}
\affil{East Asian Observatory, 660 N. A`oh\={o}k\={u} Place, University Park, Hilo, HI 96720, USA}
\email{liujh@smail.nju.edu.cn, liujunhao42@outlook.com}

\author[0000-0003-2384-6589]{Qizhou Zhang}
\affil{Center for Astrophysics $\vert$ Harvard \& Smithsonian, 60 Garden Street, Cambridge, MA 02138, USA}

\author{Beno\^{i}t Commer\c{c}on}
\affil{Univ Lyon, Ens de Lyon, Univ Lyon1, CNRS, Centre de Recherche Astrophysique de Lyon UMR5574, F-69007, Lyon, France}

\author{Valeska Valdivia}
\affil{Laboratoire AIM, Paris-Saclay, CEA/IRFU/SAp - CNRS - Universit\'{e} Paris Diderot, 91191 Gif-sur-Yvette Cedex, France}

\author{Ana{\"e}lle Maury}
\affil{Laboratoire AIM, Paris-Saclay, CEA/IRFU/SAp - CNRS - Universit\'{e} Paris Diderot, 91191 Gif-sur-Yvette Cedex, France}
\affil{Center for Astrophysics $\vert$ Harvard \& Smithsonian, 60 Garden Street, Cambridge, MA 02138, USA}

\author[0000-0002-5093-5088]{Keping Qiu}
\affil{School of Astronomy and Space Science, Nanjing University, 163 Xianlin Avenue, Nanjing, Jiangsu 210023, People's Republic of China}
\affil{Key Laboratory of Modern Astronomy and Astrophysics (Nanjing University), Ministry of Education, Nanjing, Jiangsu 210023, People's Republic of China}


\begin{abstract}
The Davis-Chandrasekhar-Fermi (DCF) method is widely used to indirectly estimate the magnetic field strength from the plane-of-sky field orientation. In this work, we present a set of 3D MHD simulations and synthetic polarization images using radiative transfer of clustered massive star-forming regions. We apply the DCF method on the synthetic polarization maps to investigate its reliability in high-density molecular clumps and dense cores where self-gravity is significant. We investigate the validity of the assumptions of the DCF method step by step and compare the model and estimated field strength to derive the correction factors for the estimated uniform and total (rms) magnetic field strength at clump and core scales. The correction factors in different situations are catalogued. We find the DCF method works well in strong field cases. However, the magnetic field strength in weak field cases could be significantly overestimated by the DCF method when the turbulent magnetic energy is smaller than the turbulent kinetic energy. We investigate the accuracy of the angular dispersion function (ADF, a modified DCF method) method on the effects that may affect the measured angular dispersion and find that the ADF method correctly accounts for the ordered field structure, the beam-smoothing, and the interferometric filtering, but may not be applicable to account for the signal integration along the line of sight in most cases. Our results suggest that the DCF methods should be avoided to be applied below $\sim$0.1 pc scales if the effect of line-of-sight signal integration is not properly addressed.
\end{abstract}


\section{Introduction}
Magnetic fields play an important role in the star formation process \citep{2012ARA&A..50...29C}. It may provide support against gravitational collapse \citep{1987ARA&A..25...23S}, prevent or regulate the fragmentation of dense clumps\footnote{Following the nomenclature in \citet{2009ApJ...696..268Z}, we refer to a molecular clump as an entity of $\sim$ 1 pc and a dense core as an entity of $\sim$ 0.1 pc} and cores \citep{2011ApJ...742L...9C, 2014ApJ...792..116Z}, and affect the isotropy of turbulence \citep{1995ApJ...438..763G}. However, the relative importance of the magnetic field compared to the turbulence and the gravity in star formation is still uncertain \citep{2012ARA&A..50...29C}. Correctly estimating the strength of magnetic field is crucial to compare the magnetic energy with the turbulent energy and the gravitational energy, and to distinguish between star formation theories with a strong field \citep{2006ApJ...646.1043M} or with a weak field \citep{2004RvMP...76..125M}. 

There are several ways to observationally determine the magnetic field strength in star-forming regions. Zeeman splitting observations of spectral lines can provide information on the direction and strength of the line-of-sight magnetic field component and is the only technique to directly measure the magnetic field strength \citep{2019FrASS...6...66C}. However, Zeeman splitting observations require particular line tracers and are very time-consuming if the line is weak. The recently developed velocity gradient technique proposes to trace the plane-of-sky magnetic field orientations from spectral line observations with the assumption that the velocity gradient is perpendicular to the magnetic field \citep{2017ApJ...835...41G}. On the other hand, polarized thermal dust emissions at mm/submm wavelengths have been proven to be a promising way to trace the orientation of the plane-of-sky magnetic field if the shortest axis of irregular dust grains is aligned with magnetic field lines \citep{1949PhRv...75.1605D, 1988ApL&C..26..263H, 2007JQSRT.106..225L, 2007MNRAS.378..910L, 2015ARA&A..53..501A}. Once the plane-of-sky magnetic field orientation is obtained for the observed region either from dust polarization or velocity gradient of molecular lines, the Davis-Chandrasekhar-Fermi (DCF) method \citep{1951PhRv...81..890D, 1953ApJ...118..113C} or the polarization-intensity gradient method \citep{2012ApJ...747...79K} can be applied to indirectly estimate the magnetic field strength. 

Specifically, the DCF method and its modified forms have been the most widely used method to determine the magnetic field strength in star-forming regions. The method is based on the assumptions that there is a mean and prominent magnetic field component, the perturbation on the mean field component is due to turbulence, and the magnetic field perturbation can be traced by angular dispersions on polarizaton maps. It also assumes that the turbulence is isotropic and there is an equipartition between the turbulent components of the kinetic energy and the magnetic energy. With these assumptions, the magnetic field strength can be derived from the DCF method once the angular dispersion on polarization maps, the turbulent velocity dispersion, and the gas density are estimated. 

The DCF methods are subject to uncertainties from various effects that may affect the measured angular dispersion and lead to underestimation or overestimation of the magnetic field strength. These effects include the signal integration and averaging along the line of sight \citep{1990ApJ...362..545Z, 1991ApJ...373..509M}, the contribution from the turbulent magnetic field to the angular dispersion at scales smaller than the turbulent correlation scale \citep{2009ApJ...696..567H, 2009ApJ...706.1504H}, and the contribution from the ordered magnetic field to the angular dispersion \citep{2006Sci...313..812G, 2008ApJ...679..537F, 2009ApJ...696..567H, 2015ApJ...799...74P, 2017ApJ...846..122P}. There are also additional observational uncertainties from the beam-smoothing effect \citep{2001ApJ...561..800H, 2008ApJ...679..537F, 2009ApJ...706.1504H} and the interferometric filtering effect \citep{2016ApJ...820...38H}. Moreover, none-satisfaction of the assumptions of the DCF method, i.e., the turbulence is not isotropic, the angular dispersion does not correctly trace the perturbation of the magnetic field, or the energy is not in an equipartion, could also lead to inaccurate results.

There were several numerical studies to incorporate one or several of the above mentioned effects \citep{2001ApJ...546..980O, 2001ApJ...559.1005P, 2001ApJ...561..800H, 2008ApJ...679..537F, 2020arXiv201015141S}. However, none of these simulations have explored physical conditions comparable to high-density and small-scale star-forming regions where self-gravity is dominant. On the other hand, the angular dispersion function method \citep[ADF.][]{2008ApJ...679..537F, 2009ApJ...696..567H, 2009ApJ...706.1504H, 2016ApJ...820...38H} analytically takes into account all the above mentioned effects that affect the measured angular dispersion, but has not been tested in comparison with numerical simulations. Thus, it is of great importance to test the raw DCF method and the ADF method with synthetic observations of small-scale self-gravitating MHD simulations. 

In this work we perform high-resolution MHD simulations and radiative transfer simulations of massive star-forming clusters. We study the accuracy of the assumptions of the DCF method step by step and apply the raw DCF method and the ADF method to synthetic observations of the simulations. We describe our simulations in Section \ref{section:simulations}. We present the results of the simulation and the application of the DCF method in Section \ref{section:results}. In Section \ref{section:discussion} we discuss the uncertainties of the DCF method and compare our work with previous simulation works. Section \ref{section:summary} is a summary of this work. 

\section{Simulations} \label{section:simulations}
We adapt several simulations from \citet{2018AA...615A..94F} and perform a set of new 3D ideal radiative magneto-hydrodynamic (MHD) numerical simulations to simulate the gravitational collapse of massive star-forming regions with the RAMSES code \citep{2002A&A...385..337T, 2006A&A...457..371F, 2011ApJ...742L...9C, 2014A&A...563A..11C}. RAMSES uses a second-order finite volume scheme to integrate the MHD equations, combined with a Constrained Transport scheme for the induction equation \citep{2006A&A...457..371F}, and an implicit solver for the radiative transfer \citep{2011ApJ...742L...9C, 2014A&A...563A..11C}.  The initial conditions of our simulations are very similar to those in \citet{2011A&A...528A..72H}, \citet{2011ApJ...742L...9C}, and \citet{2016A&A...593L..14F}. We consider dense spherical structures with a initial mass of 300 $M_{\odot}$ and a initial radius of $r_0$. The initial Plummer-like density profile is given by $\rho(r) = \rho_c/(1+(r/r_c)^2)$, where $\rho_c$ is the central density and $r_c$ is the extent of the central plateau. The density contrast between the centre and the border of the structure is 10. The initial magnetic field is along the x axis and its strength is proportional to the column density. The initial magnetic critical parameter $\mu$, which is defined as the mass-to-flux ratio over its critical value \citep{1976ApJ...210..326M}, ranges from 1.2 to 200 in our models. We include initial (super)sonic turbulence by seeding an initial turbulent velocity field with random-phase Kolmogorov-like power spectrum (see \citet{2011A&A...528A..72H}). We vary the initial turbulence level by changing the amplitude of the velocity dispersion, but we use the same velocity perturbation power spectrum at initialization. The magnetic fields and density are not initialized with perturbations, but the onset of turbulence and gravitational collapse modifies their distribution. We allow the turbulence to decay (the initial free fall time is always shorter than the turbulent crossing time). To follow the gravitational collapse, we employ adaptive mesh refinement based on the local Jeans length (see \citet{1997ApJ...489L.179T}). We use a Jeans criterion which ensures mass greater than the Jeans mass is always resolved by at least 8 cells. We refine the mesh until we reach our maximum resolution, which corresponds to the level 14 in RAMSES. Depending on the initial thermal support (either set with initial temperature $T$=10 K or 20 K), the maximum resolution achieved is either 12.5 AU or 25 AU (as indicated in Table \ref{tab:mhdmodel}). When the maximum level of refinement is reached, sink particles (i.e., protostars) are introduced to avoid further refinement of  the grid to follow the collapse. We use the implementation presented in \citet{2014MNRAS.445.4015B}, with slight modifications on the creation and accretion schemes. The sink particles accrete the gas that sits in their accretion volumes (sphere of radius 4 cells at the maximum level of refinement) and that is Jeans unstable. We consider that half of the mass accreted into the sink particles actually goes into stellar material. The luminosity, radius, and effective temperature of the protostars are then computed using mass-radius and mass-luminosity empirical relations of main sequence stars \citep[e.g.,][]{2005bookW}. This protostellar luminosity is then injected within the accretion volume in the computational domain as a source term \citep[e.g.,][]{2009Sci...323..754K}. We do not account for accretion luminosity.  The initial physical parameters of different simulation models are listed in Table \ref{tab:mhdmodel}. The physical parameters would evolve temporally and locally and deviate from the initial condition during the simulation. We analyse the simulations at two different time snapshots: when the first sink forms and when the star formation efficiency (SFE) reaches 15\%.

\begin{deluxetable*}{ccccccccc}[t!]
\tablecaption{Initial parameters of MHD models \label{tab:mhdmodel}}
\tablecolumns{9}
\tablewidth{0pt}
\tablehead{
\colhead{Model} &
\colhead{$T$\tablenotemark{b}} &
\colhead{Resolution} &
\colhead{$\mathcal{M}$\tablenotemark{c}} &
\colhead{$\mathcal{M}_{A,peak}$\tablenotemark{d}} &
\colhead{$\mu$} & 
\colhead{$\rho_c$} &
\colhead{$r_c$} &
\colhead{$r_0$}  \\
\colhead{} &
\colhead{(K)} &
\colhead{AU} &
\colhead{} &
\colhead{} &
\colhead{} &
\colhead{(10$^{-19}$ g cm$^{-3}$)} &
\colhead{(pc)} &
\colhead{(pc)} 
}
\startdata
T10M1MU1 & 10 & 25 & 1 & 0.13 & 1.2 & 1.9 & 0.17 & 0.5 \\
T10M3MU1 & 10 & 25 & 3 & 0.38 & 1.2 & 1.9 & 0.17 & 0.5 \\
T10M3MU20 & 10 & 25 & 3 & 3.2 & 20 & 1.9 & 0.17 & 0.5 \\
T10M6MU2\tablenotemark{a} & 10 & 25 & 6.4 & 1.44 & 2 & 1.9 & 0.17 & 0.5 \\
T10M6MU200\tablenotemark{a} & 10 & 25 & 6.4 & 163 & 200 & 1.9 & 0.17 & 0.5 \\
T20M3MU2\tablenotemark{a} & 20 & 12.5 & 3 & 0.14 & 2 & 15 & 0.085 & 0.25 \\
T20M3MU5 & 20 & 12.5& 3 & 1.68 & 5 & 15 & 0.085 & 0.25 \\
T20M3MU200\tablenotemark{a} & 20 & 12.5& 3 & 63 & 200 & 15 & 0.085 & 0.25 \\
T20M6MU2\tablenotemark{a} & 20 & 12.5& 6.4 & 1.4 & 2 & 15 & 0.085 & 0.25 \\
T20M6MU5 & 20 & 12.5& 6.4 & 3.6 & 5 & 15 & 0.085 & 0.25 \\
T20M6MU200\tablenotemark{a} & 20 & 12.5& 6.4 & 144 & 200 & 15 & 0.085 & 0.25 \\
\enddata
\tablenotetext{a}{Adapted from \citet{2018AA...615A..94F}.}
\tablenotetext{b}{Initial gas and dust temperature.}
\tablenotetext{c}{Initial Mach number, which is the ratio between the turbulent and thermal velocity dispersions.}
\tablenotetext{c}{Initial peak Alfv\'{e}n Mach number, which is the ratio between the non-thermal velocity dispersion of the whole simulation and the Alfv\'{e}n velocity of the center magnetic field value (maximum amplitude). Since the initial amplitude of the magnetic field scales with the column density, the initial Alfv\'{e}n velocity is not uniform through the cloud. So we only present the value of the initial peak Alfv\'{e}n Mach number, which is the minimum value of the initial Alfv\'{e}n Mach number.}
\end{deluxetable*}

Based on the results of the MHD simulations, we perform radiative transfer modelings and recompute the temperature using the POLARIS code \citep{2016A&A...593A..87R}. The radiative torque (RAT) alignment theory is assumed to be the dominant mechanism for the dust grain alignment. The sinks as in the RAMSES output are set as the central heating sources of the POLARIS input. Other setups of the radiative transfer modelings are identical to those used in \citet{2019MNRAS.488.4897V}. An interstellar
radiation field of strength $G_0 = 1$ \citep{1983A&A...128..212M} is included as the external radiation field. The gas-to-dust ratio is assumed to be 100 and the dust is composed of 62.5\% astronomical silicates and 37.2\% graphite grains \citep{1977ApJ...217..425M}. The dust size distribution is described by d$n(a)\propto a^{-3.5}$d$a$ \citep{1977ApJ...217..425M}, where $a$ is the effective radius and $n(a)$ is the number of dust grains. The minimum and maximum of the grain size are set to be 5 nm and 1 $\mu$m, respectively. 

Based on the results of the radiative transfer modelings, we produce synthetic 1.3 mm Stokes $I$, $Q$, and $U$ maps of the simulated dust emission observed at three orthogonal directions (corresponding to xy, zy, and xz planes) at a distance of 2 kpc with POLARIS. We post-process these synthetic $I$, $Q$, and $U$ images through the CASA tasks \textit{simobserve} and \textit{simanalyze} to produce synthetic interferometer observations. We adopt the ALMA Cycle 7 configurations \footnote{https://almascience.eso.org/observing/prior-cycle-observing-and-configuration-schedule} (Configuration 1 to 5, hereafter AC1 to AC5) for these CASA tasks. Table \ref{tab:almaconfig} lists the synthesized beam size and the maximum recoverable scale of the ALMA configurations. Thermal noises are not considered in these synthetic ALMA observations. The polarized intensity $PI$, polarization percentage $P$, and polarization position angle $\theta$ are estimated with $PI = \sqrt{Q^2 + U^2}$, $P = PI/I$, and $\theta = \frac{1}{2} \arctan (U/Q)$, respectively. 

\begin{table}[]
\centering
\begin{tabular}{ccc}\hline

Configuration & Synthesized & Maximum recoverable\\
&beam  ($''$) & scale ($''$) \\\hline
AC1 & 1.5 & 12.6 \\
AC2 & 1.0 & 9.8 \\
AC3 & 0.61 & 7.0 \\
AC4 & 0.40 & 4.9 \\
AC5 & 0.23 & 2.9 \\
\end{tabular}
\caption{Beam size and maximum recoverable scale of different ALMA configurations at 1.3 mm \label{tab:almaconfig}}
\end{table}

\section{Results} \label{section:results}
\subsection{Results of simulations} \label{sec:paras}
Our simulation models are state-of-the-art and accurately reproduce the key outcomes of massive star formation. In Appendix \ref{app:imexam} we show examples of the synthetic images for a model initially with strong magnetic field and weak turbulence (T10M1MU1) and a model initially with weak magnetic field and strong turbulence (T10M6MU200) at the time when SFE = 15\%. In this work we focus on the ability to measure the magnetic field strength using the dust polarization signal as observed from realistic 3D simulation models. We refrain from analysing the difference of magnetic field morphology among different models in this work and will present them in detail in a separate study. Here we briefly describe the outcome of the simulations.

To investigate how the MHD simulations are affected by the the initial model parameter, we estimate the physical parameters within spheres (indicated by superscript ``sph'') of different radii with respect to the most massive sink for all simulation models for comparison. For the volume density $n_{\mathrm{H_2}}$, we adopt the volume-weighted value in the concerned space. Because the kinetic energy is proportional to the gas mass, we adopt the mass-weighted standard deviation of the velocity within the concerned space as the total kinetic velocity dispersion of the system: $\delta v = (v_{\mathrm{rms}}^2 - \overline{v}^2)^{1/2}$, where $v_{\mathrm{rms}} = (\sum (m^{\mathrm{cell}} (v^{\mathrm{cell}})^2) / \sum m^{\mathrm{cell}})^{1/2}$ is the mass-weighted rms velocity, $m^{\mathrm{cell}}$ is the mass of a cell, $v^{\mathrm{cell}}$ is the velocity of a cell, and $\overline{v}$ is the mass-weighted mean velocity. The equivalent total (i.e., rms) magnetic field strength $B^{\mathrm{tot}}$ within the concerned space is estimated as $B^{\mathrm{tot}} = (\sum (V^{\mathrm{cell}} (B^{\mathrm{cell}})^2) / \sum V^{\mathrm{cell}})^{1/2}$, where $B^{\mathrm{cell}}$ is the total magnetic field strength of a cell and $V^{\mathrm{cell}}$ is the volume of a cell. We also estimate the modified magnetic critical parameter as $\lambda_{\mathrm{mod}} = M/M_{\mathrm{B}}^{\mathrm{tot}}$, where $M$ is the total mass (including gas mass and sink mass) in the concerned space, and $M_{\mathrm{B}}^{\mathrm{tot}}$ is the total magnetic virial (critical) mass \citep{2020ApJ...895..142L}. $M_{\mathrm{B}}^{\mathrm{tot}}$ is given by
\begin{equation}
M_{\mathrm{B}}^{\mathrm{tot}} = \frac{\pi R^2 B^{\mathrm{tot}}}{\sqrt{\frac{3(3-i)}{2(5-2i)}\mu_0\pi G}} ,
\end{equation}
where $R$ is the radius, $\mu_0$ is the permeability of vacuum, $G$ is the gravitational constant, and $i$ is the power-law index of the radial density profile. $i$ is derived by fitting the radial density profile with a power law $\rho \propto r^{-i}$. We note that principally the magnetic critical parameter could be directly derived from the ratio between the gravitational energy and the total magnetic energy in the simulation space. However, directly estimating the gravitational energy is very time-consuming for 3D space with a large number of grids and is beyond our computation ability. So we refrain from directly deriving the magnetic critical parameter from energy ratios. 

\subsubsection{Properties of magnetic fields}

\begin{figure*}[!htbp]
 \gridline{\fig{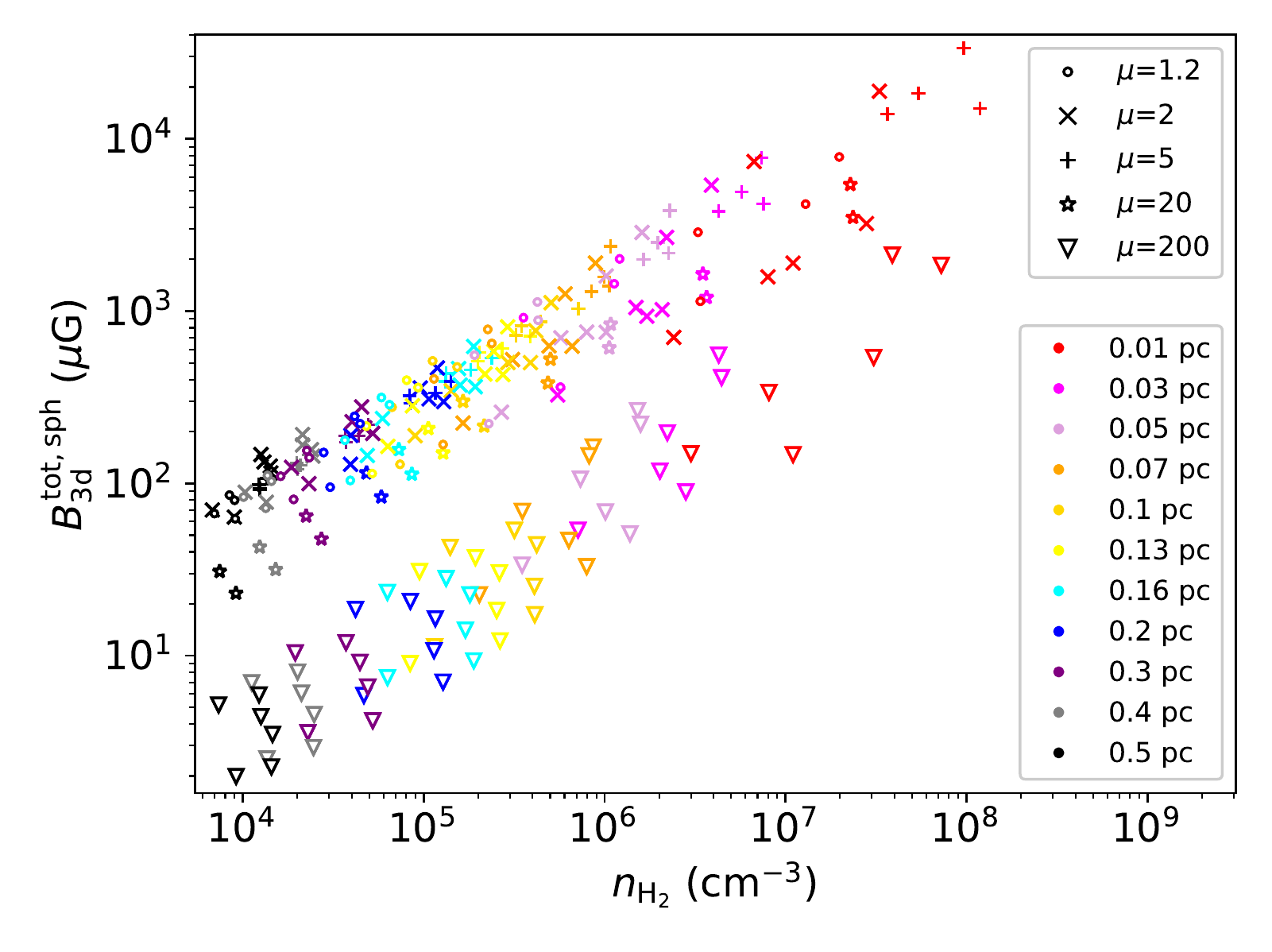}{0.45\textwidth}{}
  \fig{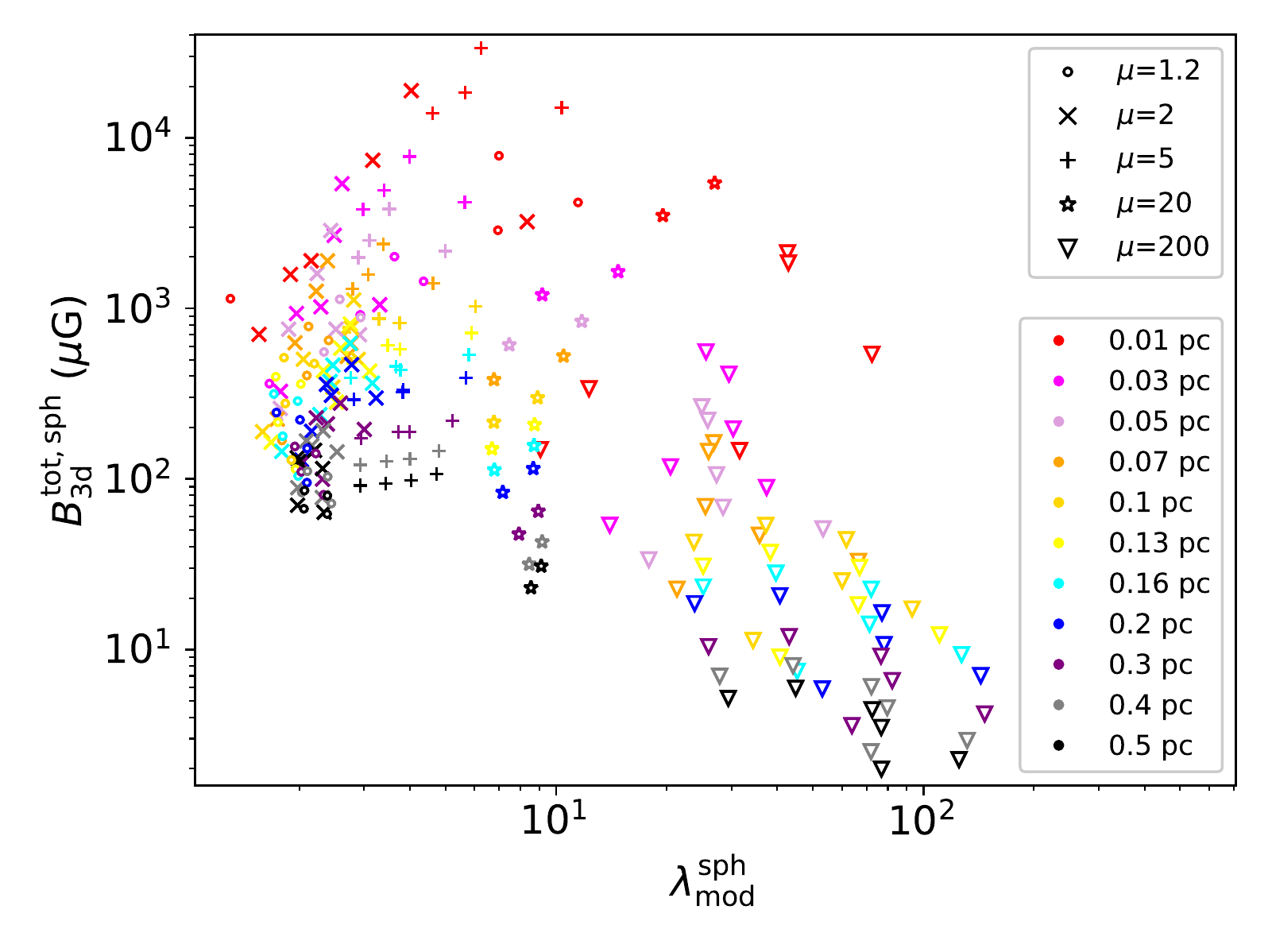}{0.45\textwidth}{}}
\caption{Left: The relation between the 3D total magnetic field strength and the volume density. Right: The relation between the 3D total magnetic field strength and the magnetic critical parameter. The estimations are within different radii in all simulation models at two different time snapshots. The magnetic field strength is the volume-weighted value.  Different colors correspond to different radii. Different symbols represent different initial $\mu$ values. Larger $\mu$ values correspond to weaker magnetic fields.   \label{fig:B_para}}
\end{figure*}

Figure \ref{fig:B_para}(a) shows the relation between the 3D total magnetic field strength within spheres ($B^{\mathrm{tot,sph}}_{\mathrm{3d}}$) and the volume density. It seems that models with smaller initial $\mu$ value (i.e., stronger initial magnetic field) tend to have stronger total magnetic field strength. The total magnetic field strength is positively correlated with the volume density and negatively correlated with the radius.

Figure \ref{fig:B_para}(b) shows the relation between the $B^{\mathrm{tot,sph}}_{\mathrm{3d}}$ and the magnetic critical parameter within spheres ($\lambda^{\mathrm{sph}}_{\mathrm{mod}}$). Models with smaller initial $\mu$ values have smaller $\lambda^{\mathrm{sph}}_{\mathrm{mod}}$, suggesting that the magnetic critical parameter of a simulation is determined by the initial magnetic critical parameter. In our simulations, the magnetic critical parameter does not show a strong relation with radius. We also find that if the sink mass is not included in the calculation of the magnetic critical parameter, there will be a fake relation that the magnetic critical parameter is positively correlated with radius for each simulation. 


\subsubsection{Properties of velocity fields}

The measured velocity dispersion contains the contributions from outflows, gravitational collapse (i.e., infall), rotation, and turbulence. To properly estimate the turbulent velocity dispersion, the contribution from other mechanisms must be removed from the total velocity dispersion.

We firstly consider the contribution from outflows. Since the resolution of our simulations (12.5 AU or 25 AU) does not allow to model the detailed process of disk accretion, our simulations do not produce the commonly observed parsec-scale jet-like or wide-angle outflows \citep{2007prpl.conf..245A} that are expected to be driven by accretion-related protostellar winds \citep[e.g., X-wind or disk winds:][]{2006ApJ...649..845S, 2006MNRAS.365.1131P}. However, there are still some materials ejected from the center that may be related to the some MHD processes in our simulations. These ``outflows'' would still contribute to the total velocity dispersion. Since we do not track the motion of particles, we cannot discern low-velocity outflows from other low-velocity gas motions in our simulations. Observationally, low-velocity outflowing gas usually cannot be distinguished from the environmental gas either. The distinguishable outflows usually have 1D outflowing velocities of $>$5 km/s in massive star-forming regions \citep[e.g.,][]{2002A&A...383..892B, 2005ApJ...625..864Z, 2018ApJ...860..106L}. So we only investigate the contribution from high-velocity outflows in the simulations.  To remove the contribution from the high-velocity outflows, we exclude grids with 3D velocity $>$10 km s$^{-1}$ (approximately with 1D velocity $>$5.8 km s$^{-1}$) and measure the mass-weighted velocity dispersion of the remaining grids as the non-outflow velocity dispersion. 

\begin{figure*}[!htbp]
 \gridline{\fig{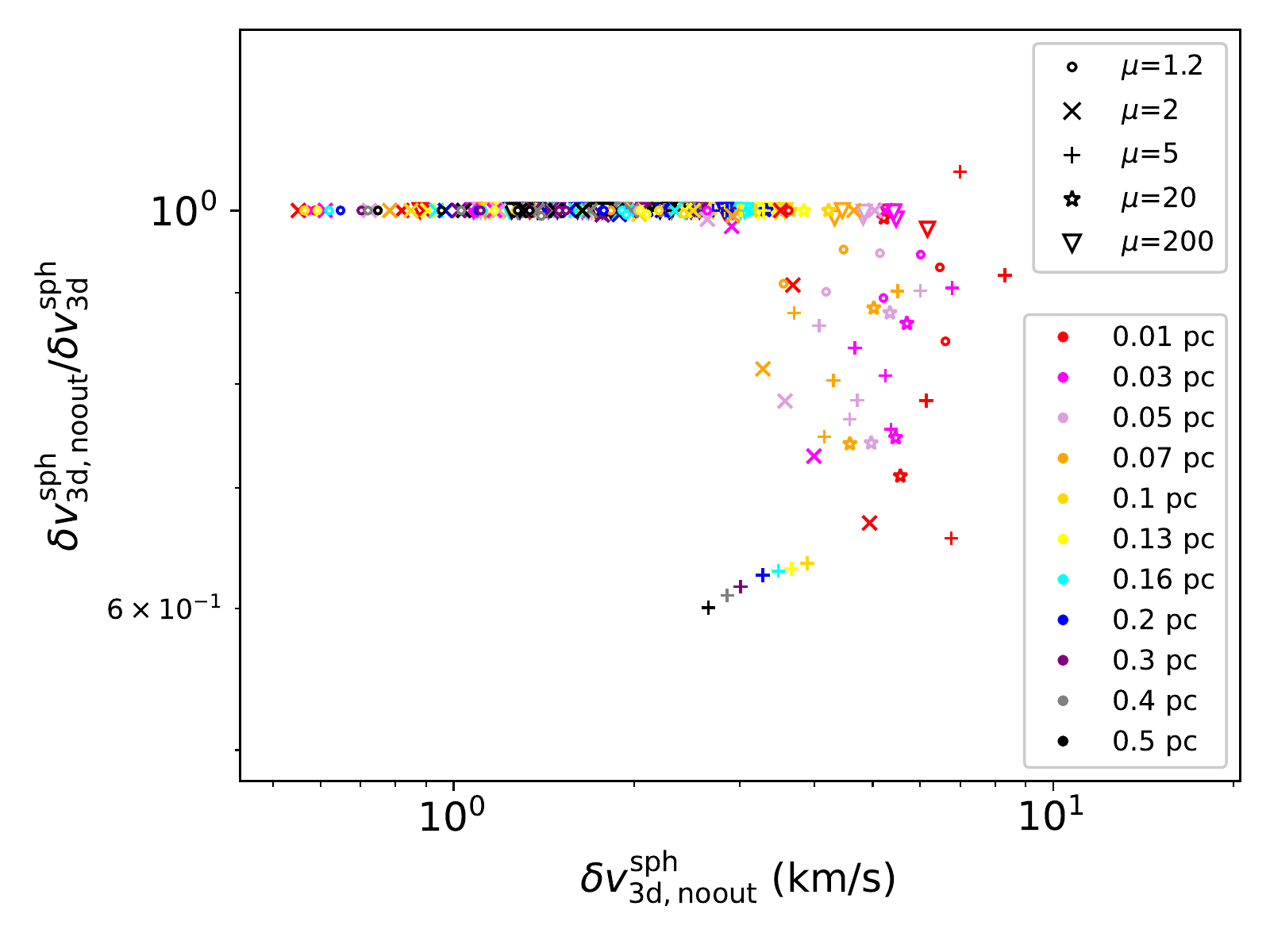}{0.45\textwidth}{}}
 \gridline{\fig{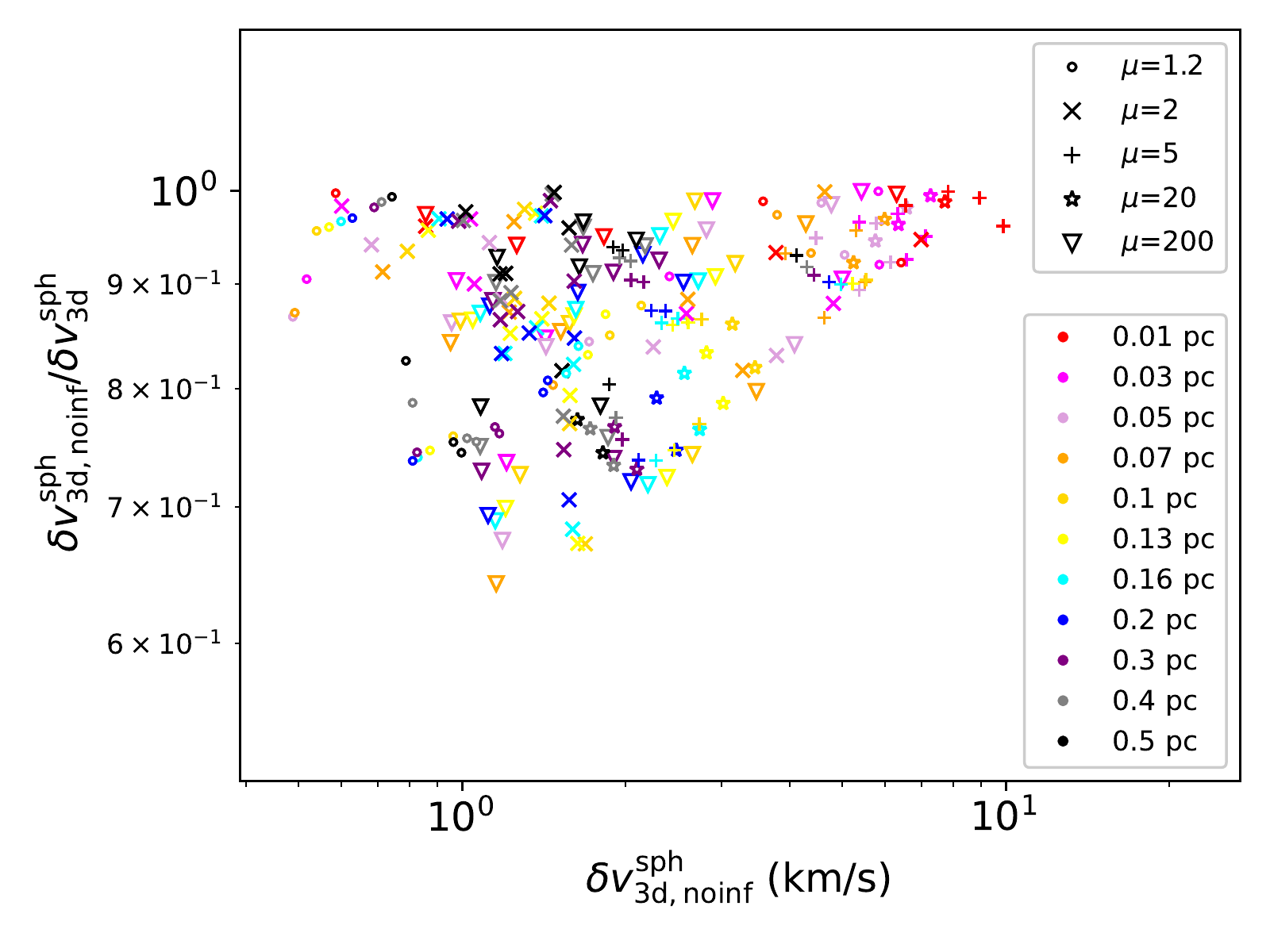}{0.45\textwidth}{}}
 \gridline{\fig{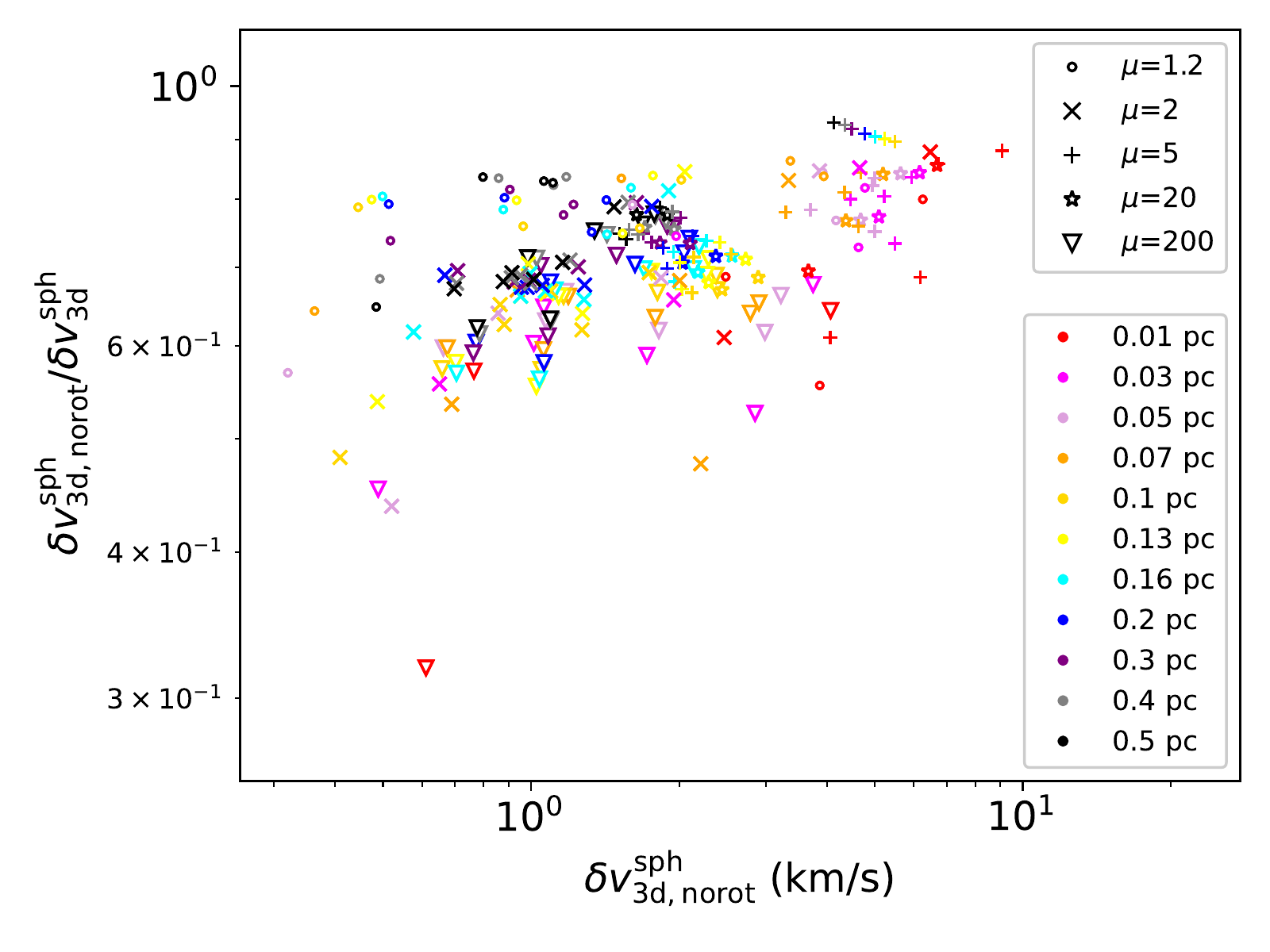}{0.45\textwidth}{}}
\caption{Upper: The ratio between the non-outflow velocity dispersion and the total velocity dispersion. Middle: The ratio between the non-infall velocity dispersion and the total velocity dispersion. Bottom: The ratio between the non-rotational velocity dispersion and the total velocity dispersion. The velocity dispersion is the mass-weighted value. Different symbols correspond to different initial $\mu$ values and different colors represents different radii.   \label{rdvnorot3ddv3d_dvnorot3d}}
\end{figure*}

Figure \ref{rdvnorot3ddv3d_dvnorot3d}(a) shows the ratio between the 3D non-outflow velocity dispersion ($\delta v_{\mathrm{3d,noout}}$) and the 3D total velocity dispersion ($\delta v_{\mathrm{3d}}$) within different radii in all simulation models. In most of the cases, the contribution from outflows to the total velocity dispersion is negligible. Only in some cases, mostly at small radii, the ratio between the non-outflow and total velocity dispersions could be considerably lower than 1. 

We then consider the contribution from infall. We adopt a simple spherical symmetric collapsing model for infall. The mass-weighted inward radial velocity with respect to the most massive sink is measured as the average radial infall velocity ($ v_{\mathrm{r,inf}}$) within spheres of different radii. We neglect the infall toward sinks of smaller mass. During the calculation, we also exclude grids with 3D velocity $>$10 km s$^{-1}$. The 3D non-infall velocity dispersion is estimated as $\delta v_{\mathrm{3d,noinf}} = (\delta v_{\mathrm{3d}}^2 - v_{\mathrm{r,inf}}^2)^{0.5}$. 

Figure \ref{rdvnorot3ddv3d_dvnorot3d}(b) shows the ratio between the 3D non-infall velocity dispersion and the 3D total velocity dispersion within different radii in all simulation models. In most of the cases, the ratio between the non-infall and total velocity dispersions are greater than 0.71 (corresponding to an energy ratio of 0.5), which suggests the infall energy is mostly smaller than half of the total kinetic energy. The average ratio between the two velocity dispersions is $\sim$0.87, which indicates that the infall does not affect the velocity dispersion too much. 

Finally, we consider the contribution from the rotation. We adopt a simple rotation model where the rotation is along the axis of the total angular momentum. The total angular momentum is estimated by $\boldsymbol{L} = \sum{\boldsymbol{r^{\mathrm{cell}}} \times m^{\mathrm{cell}} \boldsymbol{v^{\mathrm{cell}}}}$, where $\boldsymbol{r^{\mathrm{cell}}}$ is the position vector from the most massive sink to the position of an individual cell. The tangential velocity at the plane perpendicular to the axis of the total angular momentum is measured as the rotational velocity. We estimated the total angular momentum and mass-weighted rotational velocity within different radii for all simulation models. Similarly, we exclude grids with 3D velocity $>$10 km s$^{-1}$ in the calculations. Once the average rotational velocity ($ v_{\mathrm{t,rot}}$) is determined, the 3D non-rotational velocity dispersion can be estimated as $\delta v_{\mathrm{3d,norot}} = (\delta v_{\mathrm{3d}}^2 - v_{\mathrm{t,rot}}^2)^{0.5}$. 

Figure \ref{rdvnorot3ddv3d_dvnorot3d}(c) shows the ratio between the 3D non-rotational velocity dispersion and the 3D total velocity dispersion within different radii in all simulation models. The average ratio between the two velocity dispersions is $\sim$0.71 (corresponding to an energy ratio of $\sim$0.5), which indicates the rotation could significantly affect the measured velocity dispersion and contribute to the total kinetic energy. However, this may be because our simulations do not produce outflows that are comparable to the observed ones. So the excess angular momentum cannot be transferred out via outflows and the rotation in our simulations would be stronger than in the actual star-forming regions. 

As mentioned above, the infall and rotation in our simulations can both contribute to the total velocity dispersion. However, it is hard to include both mechanisms in a simple model to exclude their combined contributions. Since the contribution from the rotation is larger than that of the infall in the total velocity dispersion, we adopt the non-rotational velocity dispersion as the turbulent velocity dispersion $\delta v_{\mathrm{turb}}$ in the following analysis.

\begin{figure*}[!htbp]
 \gridline{\fig{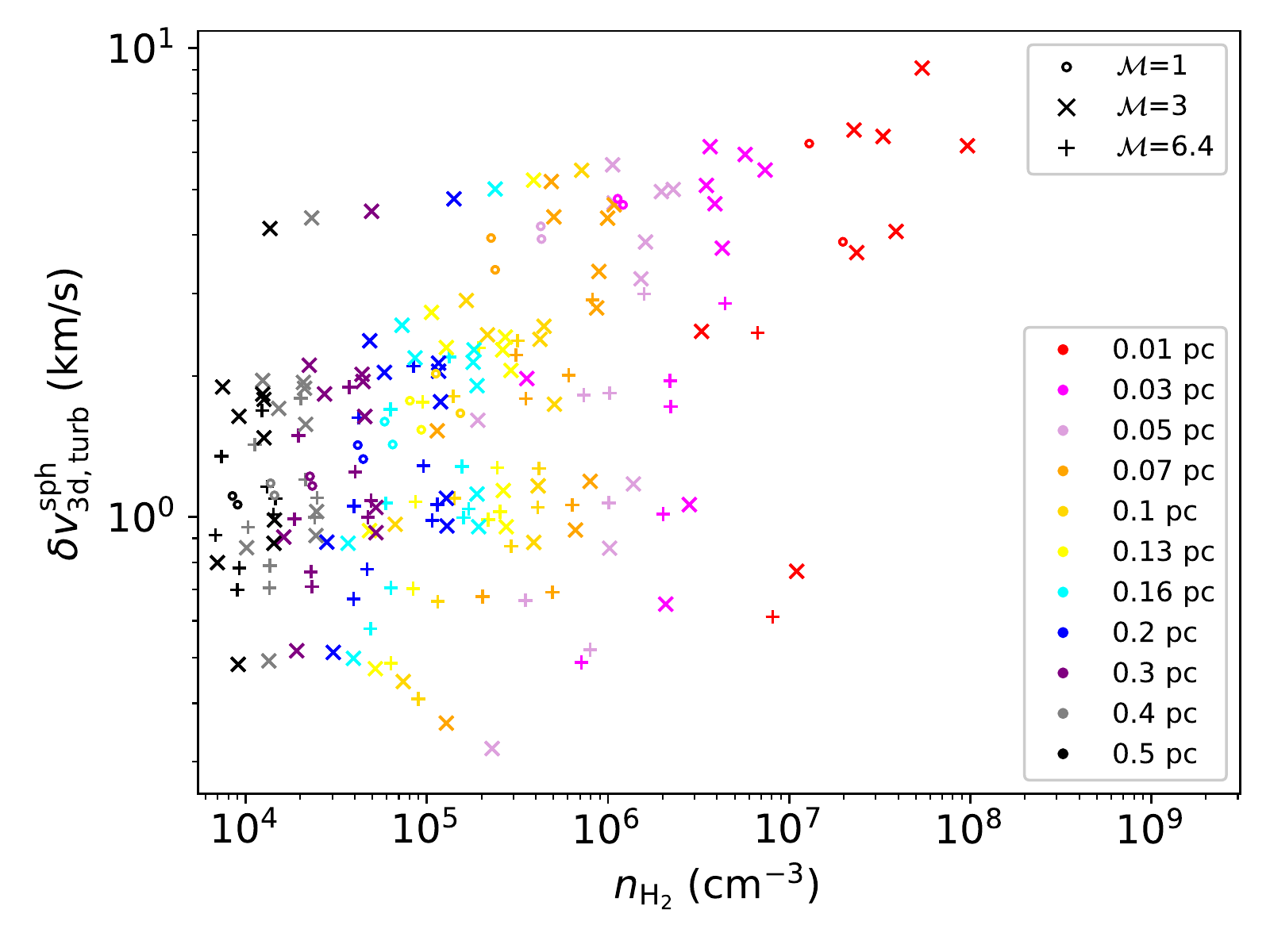}{0.45\textwidth}{}
 \fig{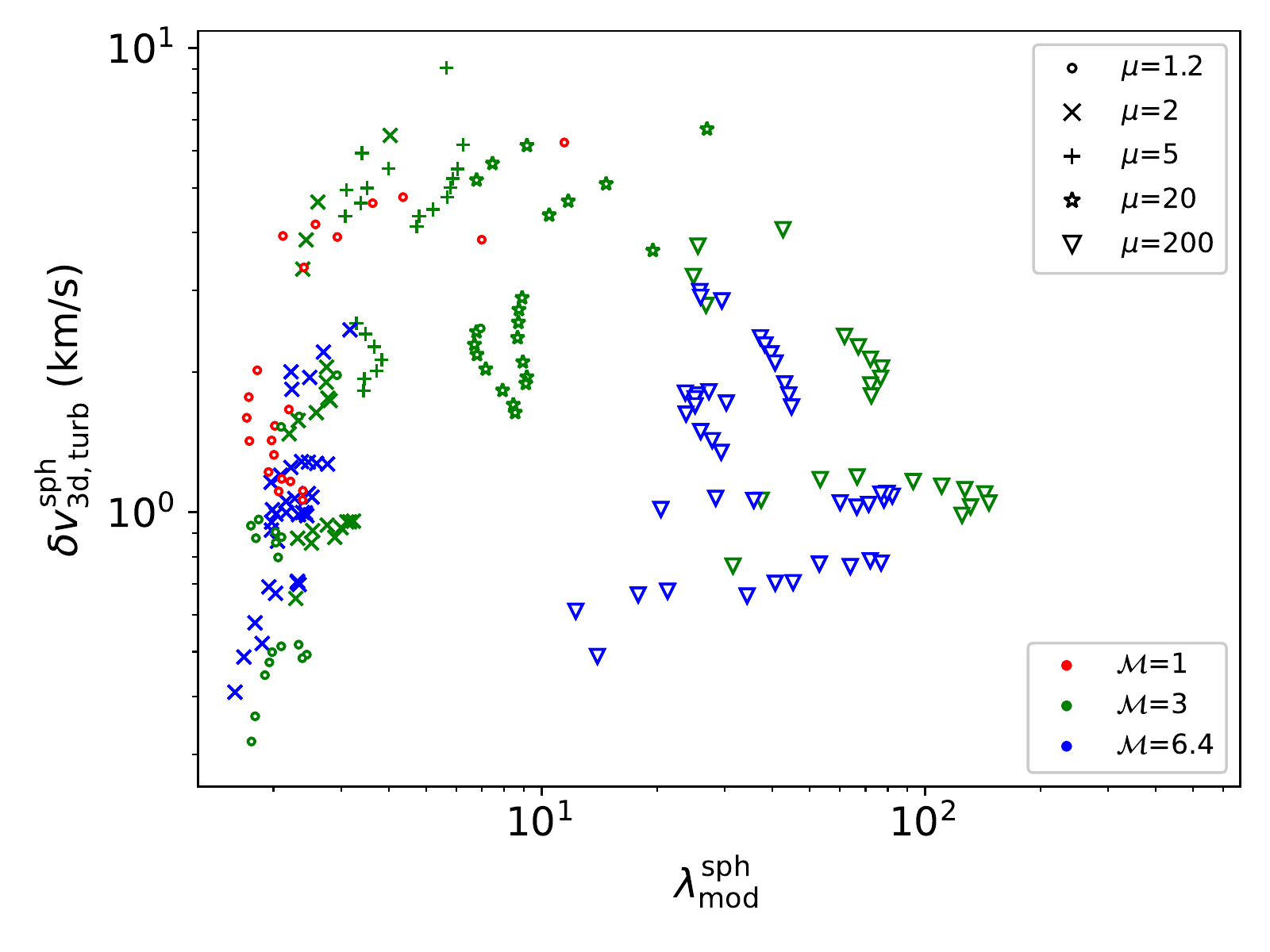}{0.45\textwidth}{}}
\caption{Left: The relation between the 3D turbulent velocity dispersion and the volume density. Different symbols correspond to different initial Mach numbers and different colors represent different radii. Right: The relation between the 3D turbulent velocity dispersion and the magnetic critical parameter. Different symbols represent different initial $\mu$ values and different colors represent different initial Mach numbers. The estimations are within different radii in all simulation models at two different time snapshots. The velocity dispersion is the mass-weighted value.  Larger initial $\mathcal{M}$ values correspond to stronger initial turbulence.  \label{fig:dv_para}}
\end{figure*}

Figure \ref{fig:dv_para}(a) shows the relation between the 3D turbulent velocity dispersion within spheres ($\delta v^{\mathrm{sph}}_{\mathrm{3d,turb}}$) and the volume density. Models with different initial Mach numbers do not show distinct difference in ranges of turbulent velocity dispersion, suggesting that the initial turbulence environment might not significantly affect the magnitude of turbulence in star forming regions if there is no continuously turbulence injection and/or when the self-gravity is dominant. In this situation, the turbulence is from the star formation activities. Generally, there is no strong relation between the turbulent velocity dispersion and the density/radius.

Figure \ref{fig:dv_para}(b) shows the relation between the $\delta v^{\mathrm{sph}}_{\mathrm{3d,turb}}$ and the magnetic critical parameter. With the same initial $\mu$ value, models with larger initial Mach number do not show larger velocity dispersion. This confirms the conclusion that the initial turbulence could decay quickly if there is no continuously turbulence injection. 

\subsection{The principle of the DCF method}\label{section:oridcf}
The DCF method is widely used to estimate the plane-of-sky magnetic field strength from polarized dust continuum observations. The basic assumption of the DCF method is that there is a mean (uniform/ordered) and prominent magnetic field component $\boldsymbol{B^{\mathrm{u}}} = \langle \boldsymbol{B^{\mathrm{loc}}} \rangle $, where $\boldsymbol{B^{\mathrm{loc}}}$ is the local plane-of-sky magnetic field vector. Then it assumes Alfv\'{e}nic magnetic field perturbations, i.e., the perturbation on the mean magnetic field is due to turbulence, or in other words, there is an equipartition between the turbulent magnetic energy\footnote{All the magnetic-related equations and terms in this work are in SI units or CGS units unless otherwise noted.} $E_B^{\mathrm{t}} = (B^{\mathrm{t}})^2 V / (2\mu_0)$  and the turbulent kinetic energy $E_\mathrm{K}^{\mathrm{t}} = \rho V \delta v^2/2$, where $B^{\mathrm{t}}$ is the equivalent turbulent magnetic field strength, $V$ is the volume, and $\rho$ is the gas density. The equivalent turbulent magnetic field strength $B^{\mathrm{t}}$ is estimated as $B^{\mathrm{t}} = ((B^{\mathrm{tot}})^2 - (B^{\mathrm{u}})^2)^{1/2}$, where the equivalent total (i.e., rms) magnetic field strength $B^{\mathrm{tot}}$ is estimated as $B^{\mathrm{tot}} = \langle (B^{\mathrm{loc}})^2 \rangle^{1/2}$. Thus, at the orientation perpendicular to ($\perp$) the mean plane-of-sky field and in the plane of sky, we have $B^{\mathrm{t}}_{\mathrm{pos\perp}} = \sqrt{\mu_0 \rho }\delta v_{\mathrm{pos\perp}}$, where ``pos'' stands for the plane of sky. If the turbulence is isotropic (see Appendix \ref{app:projection} for a discussion), i.e., $\delta v_{\mathrm{pos\perp}}^2 = \delta v_{\mathrm{pos\|}}^2 = \delta v_{\mathrm{los}}^2 = \delta v_{\mathrm{3d}}^2 /3$, where ``$\|$'' stands for the orientation parallel to the mean plane-of sky magnetic field, ``los'' stands for the line of sight, and ``3d'' stands for three Dimension (3D), we have
\begin{equation}\label{eq:eqequi}
B^{\mathrm{t}}_{\mathrm{pos\perp}} = \sqrt{\mu_0 \rho }\delta v_{\mathrm{los}}.
\end{equation}

To estimate $B^{\mathrm{u}}_{\mathrm{pos}}$, Equation \ref{eq:eqequi} can be transformed to 
\begin{equation}\label{eq:eqequitrans}
B^{\mathrm{u}}_{\mathrm{pos}} = \sqrt{\mu_0 \rho }\frac{\delta v_{\mathrm{los}}}{B^{\mathrm{t}}_{\mathrm{pos\perp}}/B^{\mathrm{u}}_{\mathrm{pos}}}.
\end{equation}
If we define angle $\phi$ as the difference between the local polarization angle $\theta$ and the mean polarization angle $\overline{\theta}$, the relation between the local $\phi$ and magnetic field components can be written as:
\begin{equation}\label{eq:tanphi}
\tan \phi = \frac{\vert\boldsymbol{B^{\mathrm{t,loc}}_{\mathrm{pos\perp}}}\vert}{\vert \boldsymbol{B^{\mathrm{t,loc}}_{\mathrm{pos\|}}} + \boldsymbol{B^{\mathrm{u}}_{\mathrm{pos}}} \vert},
\end{equation}
where $\boldsymbol{B^{\mathrm{t,loc}}_{\mathrm{pos\perp}}}$ and $\boldsymbol{B^{\mathrm{t,loc}}_{\mathrm{pos\|}}}$ are the local turbulent field vector components with orientations perpendicular to and parallel to the $\boldsymbol{B^{\mathrm{u}}_{\mathrm{pos}}}$, respectively. When the turbulent magnetic field is small compared to the ordered magnetic field ($B^{\mathrm{t}}_{\mathrm{pos\|}} \sim B^{\mathrm{t}}_{\mathrm{pos\perp}} \ll B^{\mathrm{u}}_{\mathrm{pos}}$), there is $B^{\mathrm{t}}_{\mathrm{pos\perp}}/B^{\mathrm{u}}_{\mathrm{pos}} \sim \delta(\tan \phi)$, where $\delta(\tan \phi)$ is the dispersion of $\tan \phi$. Under these assumptions, the uniform plane-of-sky magnetic field strength is given by
\begin{equation}\label{eq:eqdcf}
B^{\mathrm{u,dcf}}_{\mathrm{pos}} \sim \sqrt{\mu_0 \rho }\frac{\delta v_{\mathrm{los}}}{\delta (\tan \phi)}.
\end{equation}
When $\phi$ is small, there is $B^{\mathrm{t}}_{\mathrm{pos\perp}}/B^{\mathrm{u}}_{\mathrm{pos}} \sim \delta\phi$. The uniform plane-of-sky magnetic field strength in small angle approximation is given by
\begin{equation}\label{eq:eqdcf1}
B^{\mathrm{u,dcf,sa}}_{\mathrm{pos}} \sim \sqrt{\mu_0 \rho }\frac{\delta v_{\mathrm{los}}}{\delta\phi}.
\end{equation}

On the other hand, Equation \ref{eq:eqequi} can be rewritten as
\begin{equation}\label{eq:eqequitranstot}
B^{\mathrm{tot}}_{\mathrm{pos}} = \sqrt{\mu_0 \rho }\frac{\delta v_{\mathrm{los}}}{B^{\mathrm{t}}_{\mathrm{pos\perp}}/B^{\mathrm{tot}}_{\mathrm{pos}}}
\end{equation}
to estimate the equivalent plane-of-sky total magnetic field strength $B^{\mathrm{tot}}_{\mathrm{pos}}$. Similarly, we have
\begin{equation}\label{eq:sinphi}
\sin \phi = \frac{\vert\boldsymbol{B^{\mathrm{t,loc}}_{\mathrm{pos\perp}}}\vert}{\vert \boldsymbol{B^{\mathrm{loc}}_{\mathrm{pos}}}\vert},
\end{equation}
If $B^{\mathrm{loc}}_{\mathrm{pos}}$ does not vary too much, there is $B^{\mathrm{t}}_{\mathrm{pos\perp}}/B^{\mathrm{tot}}_{\mathrm{pos}} \sim \delta(\sin \phi)$. Thus, the $B^{\mathrm{tot,dcf}}_{\mathrm{pos}}$ could be estimated as
\begin{equation}\label{eq:eqdcftot}
B^{\mathrm{tot,dcf}}_{\mathrm{pos}} \sim \sqrt{\mu_0 \rho }\frac{\delta v_{\mathrm{los}}}{\delta (\sin \phi)}.
\end{equation}
Under small angle approximation, $B^{\mathrm{t}}_{\mathrm{pos\perp}}/B^{\mathrm{tot}}_{\mathrm{pos}} \sim \delta \phi$ and the plane-of-sky total magnetic field strength is given by
\begin{equation}\label{eq:eqdcftot1}
B^{\mathrm{tot,dcf,sa}}_{\mathrm{pos}} \sim \sqrt{\mu_0 \rho }\frac{\delta v_{\mathrm{los}}}{\delta \phi}.
\end{equation}




\subsection{Angular dispersions and ratios between magnetic field components}\label{sec:compangb}
\subsubsection{Comparing angular dispersions with ratios between magnetic field components in the model}\label{sec:compangbmodel}

The DCF method assumes the turbulent-to-ordered magnetic field strength ratio $B^{\mathrm{t}}_{\mathrm{pos\perp}}/B^{\mathrm{u}}_{\mathrm{pos}}$ and the turbulent-to-total magnetic field strength ratio $B^{\mathrm{t}}_{\mathrm{pos\perp}}/B^{\mathrm{tot}}_{\mathrm{pos}}$ are traced by the dispersion of magnetic field position angles. Here we test the reliability of this assumption with our simulation results.

We measure the angular dispersions and the strength ratios between magnetic field components within spheres of different radii with respect to the most massive sink and at three orthogonal projections (xy, xz, and zy). The orientation of plane-of-sky magnetic fields projected in the xy, xz, and zy planes in the simulation grids are gridded with a cell size of 0.0013 pc (see Appendix \ref{app:gridsize} for a discussion about grid size). In the calculation of the angular dispersion within a sphere, the direction of $\boldsymbol{B^{\mathrm{u}}_{\mathrm{pos}}}$ within the concerned space is adopted as the orientation of the mean plane-of-sky magnetic field position angle. We define the angle $\phi_{\mathrm{sim}}$ to represent the difference between the local and mean plane-of-sky magnetic field orientation in the simulation space. Then we measure the dispersion of $\phi_{\mathrm{sim}}$, $\sin \phi_{\mathrm{sim}}$, and $\tan \phi_{\mathrm{sim}}$. Because the magnetic energy is proportional to the volume, we adopt the volume-weighted value of the magnetic field strength. The $B^{\mathrm{u}}_{\mathrm{pos}}$ is estimated as $B^{\mathrm{u}}_{\mathrm{pos}} = \vert \langle \boldsymbol{B^{\mathrm{cell}}_{\mathrm{pos}}} \rangle \vert$, where $\boldsymbol{B^{\mathrm{cell}}_{\mathrm{pos}}}$ is the plane-of-sky magnetic field vector of a cell. At orientation in the plane-of-sky and perpendicular to the mean field orientation, we have $B^{\mathrm{u}}_{\mathrm{pos\perp}} = 0$ and  $B^{\mathrm{tot}}_{\mathrm{pos\perp}} = B^{\mathrm{t}}_{\mathrm{pos\perp}} = (\langle B^{\mathrm{cell}}_{\mathrm{pos\perp}})^2 \rangle^{1/2}$. It should be noted that because $B^{\mathrm{cell}}_{\mathrm{pos}}$ is the rms plane-of-sky magnetic field strength of a cell, $\langle B^{\mathrm{cell}}_{\mathrm{pos}} \rangle$ should not be used to represent the $B^{\mathrm{u}}_{\mathrm{pos}}$ as adopted by some previous simulation works. 



\begin{figure*}[!htbp]
 \gridline{\fig{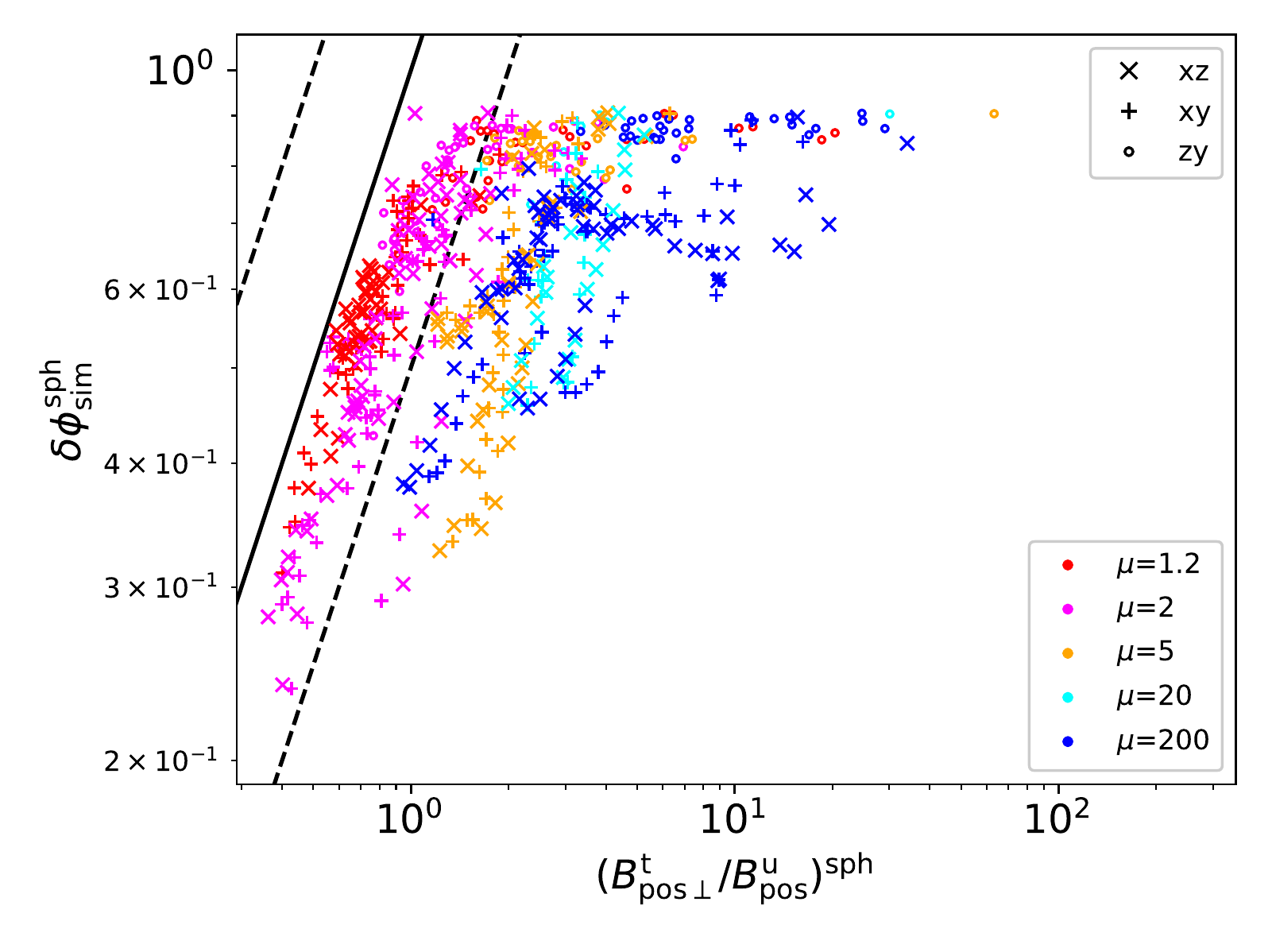}{0.45\textwidth}{a}
 \fig{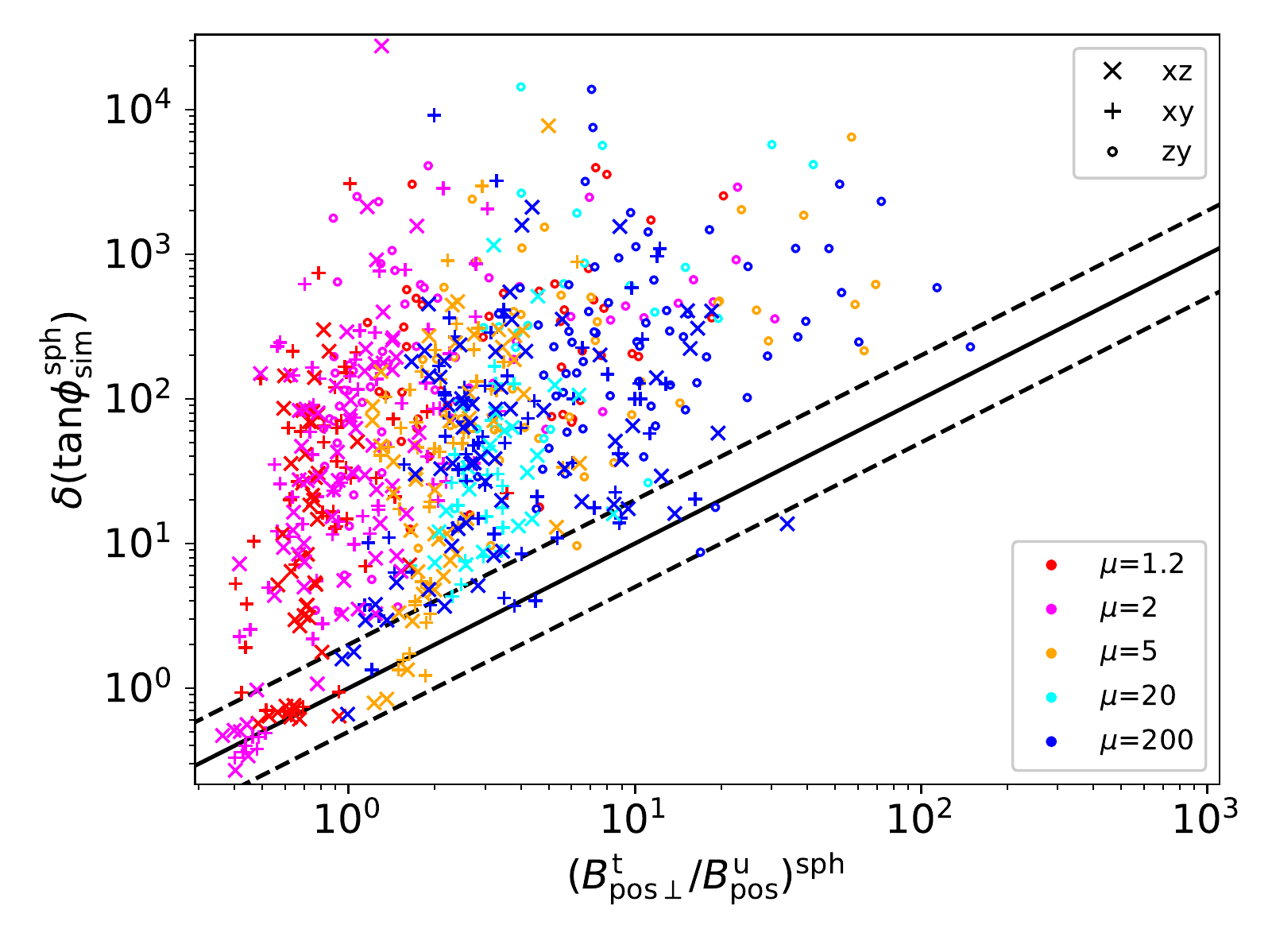}{0.45\textwidth}{b}}
\caption{Dispersions of (a) $ \phi^{\mathrm{sph}}_{\mathrm{sim}}$ and (b) $\tan \phi^{\mathrm{sph}}_{\mathrm{sim}}$ versus turbulent-to-ordered magnetic field strength ratio for all simulation models within different radii with respect to the most massive sink and at three orthogonal projections. Different symbols represent different projections. Different colors correspond to different initial $\mu$ values. Solid lines correspond to 1:1 relation. Dashed lines correspond to 1:2 and 2:1 relations.  \label{fig:disp_BtBu}}
\end{figure*}

Figure \ref{fig:disp_BtBu} shows the angular dispersions estimated within spheres of different radii in the simulation space ($\delta \phi^{\mathrm{sph}}_{\mathrm{sim}}$ and $\delta (\tan \phi^{\mathrm{sph}}_{\mathrm{sim}})$) versus the turbulent-to-ordered magnetic field strength ratio $(B^{\mathrm{t}}_{\mathrm{pos\perp}}/B^{\mathrm{u}}_{\mathrm{pos}})^{\mathrm{sph}}$ measured in the same space. The angular dispersions in the zy plane, where the initial magnetic field orientation is along the line of sight (i.e., the x-axis), are generally greater than those in the xy and xz plane, where the initial magnetic field orientation is in the plane of sky. This trend is consistent with what was found in \citet{2008ApJ...679..537F}. Figure \ref{fig:disp_BtBu} also indicates that models with stronger initial magnetic fields (i.e., smaller $\mu$) tend to have smaller turbulent-to-ordered magnetic field strength ratio and $\delta \phi^{\mathrm{sph}}_{\mathrm{sim}}$ are better correlated with $(B^{\mathrm{t}}_{\mathrm{pos\perp}}/B^{\mathrm{u}}_{\mathrm{pos}})^{\mathrm{sph}}$ for models with stronger initial magnetic fields. On the other hand, models with different initial $\mu$ values do not show significant difference in the value range of the angular dispersion. Figure \ref{fig:disp_BtBu}(a) shows that the ratio between $\delta \phi^{\mathrm{sph}}_{\mathrm{sim}}$ and  $(B^{\mathrm{t}}_{\mathrm{pos\perp}}/B^{\mathrm{u}}_{\mathrm{pos}})^{\mathrm{sph}}$ is from 1 to a factor of a few when $\delta \phi^{\mathrm{sph}}_{\mathrm{sim}} \lesssim 25\degr$ ($\sim$0.44), while $\delta \phi^{\mathrm{sph}}_{\mathrm{sim}}$ can underestimate  $(B^{\mathrm{t}}_{\mathrm{pos\perp}}/B^{\mathrm{u}}_{\mathrm{pos}})^{\mathrm{sph}}$ by up to more than two orders of magnitude for large $\delta \phi^{\mathrm{sph}}_{\mathrm{sim}}$ values. Figure \ref{fig:disp_BtBu}(b) shows that $\delta (\tan \phi^{\mathrm{sph}}_{\mathrm{sim}})$ are correlated with $(B^{\mathrm{t}}_{\mathrm{pos\perp}}/B^{\mathrm{u}}_{\mathrm{pos}})^{\mathrm{sph}}$ within a factor of $\sim$2 when $\delta (\tan \phi^{\mathrm{sph}}_{\mathrm{sim}}) \lesssim 1$. For large $\delta (\tan \phi^{\mathrm{sph}}_{\mathrm{sim}})$ values, there are significant scatters and $\delta (\tan \phi^{\mathrm{sph}}_{\mathrm{sim}})$ can greatly overestimate $(B^{\mathrm{t}}_{\mathrm{pos\perp}}/B^{\mathrm{u}}_{\mathrm{pos}})^{\mathrm{sph}}$ by several orders of magnitude. In summary, $\delta \phi^{\mathrm{sph}}_{\mathrm{sim}}$ and $\delta (\tan \phi^{\mathrm{sph}}_{\mathrm{sim}})$ are correlated with the turbulent-to-ordered magnetic field strength ratio only when the angular dispersions are small.

\begin{figure*}[!htbp]
 \gridline{\fig{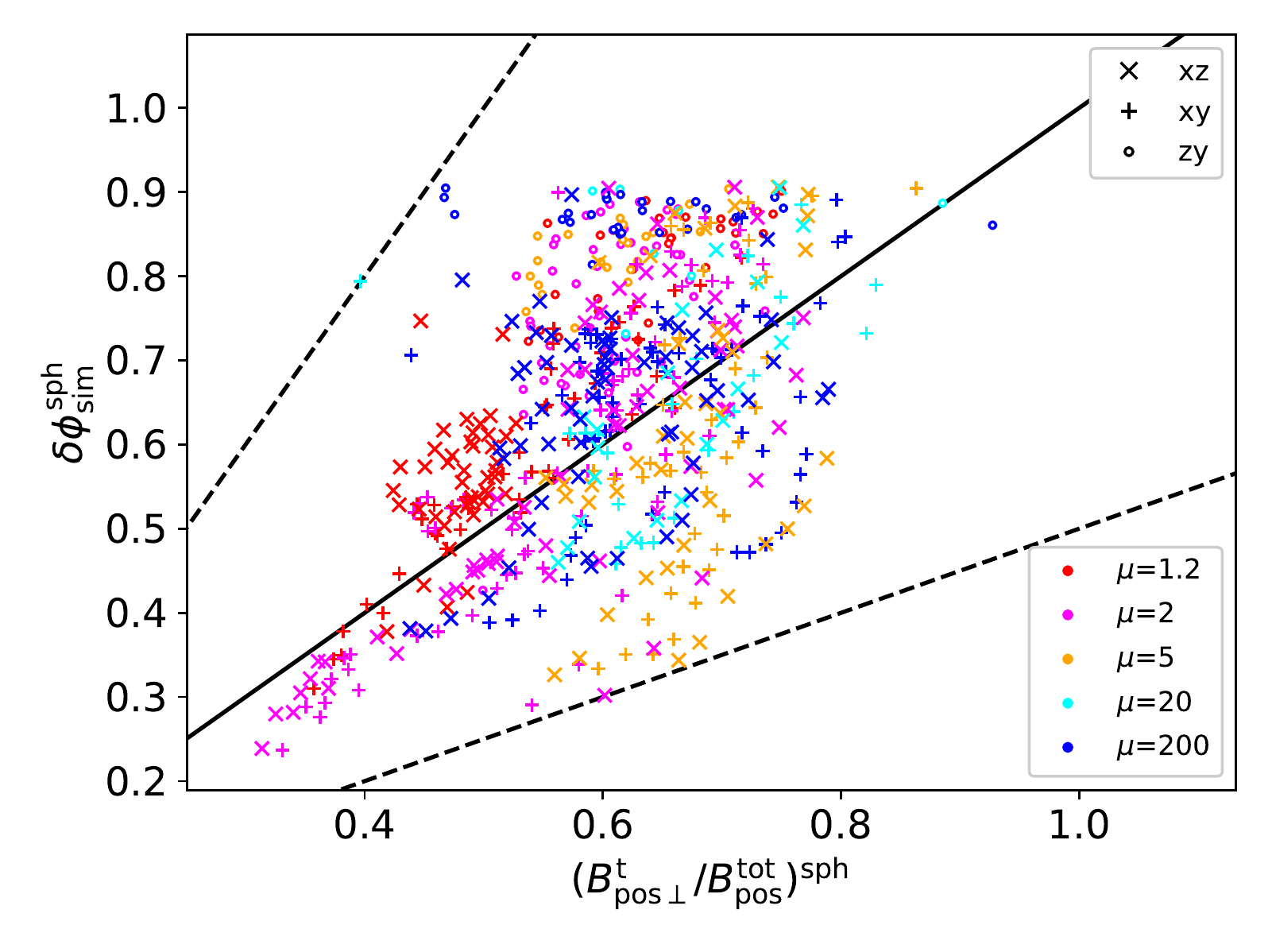}{0.45\textwidth}{a}
 \fig{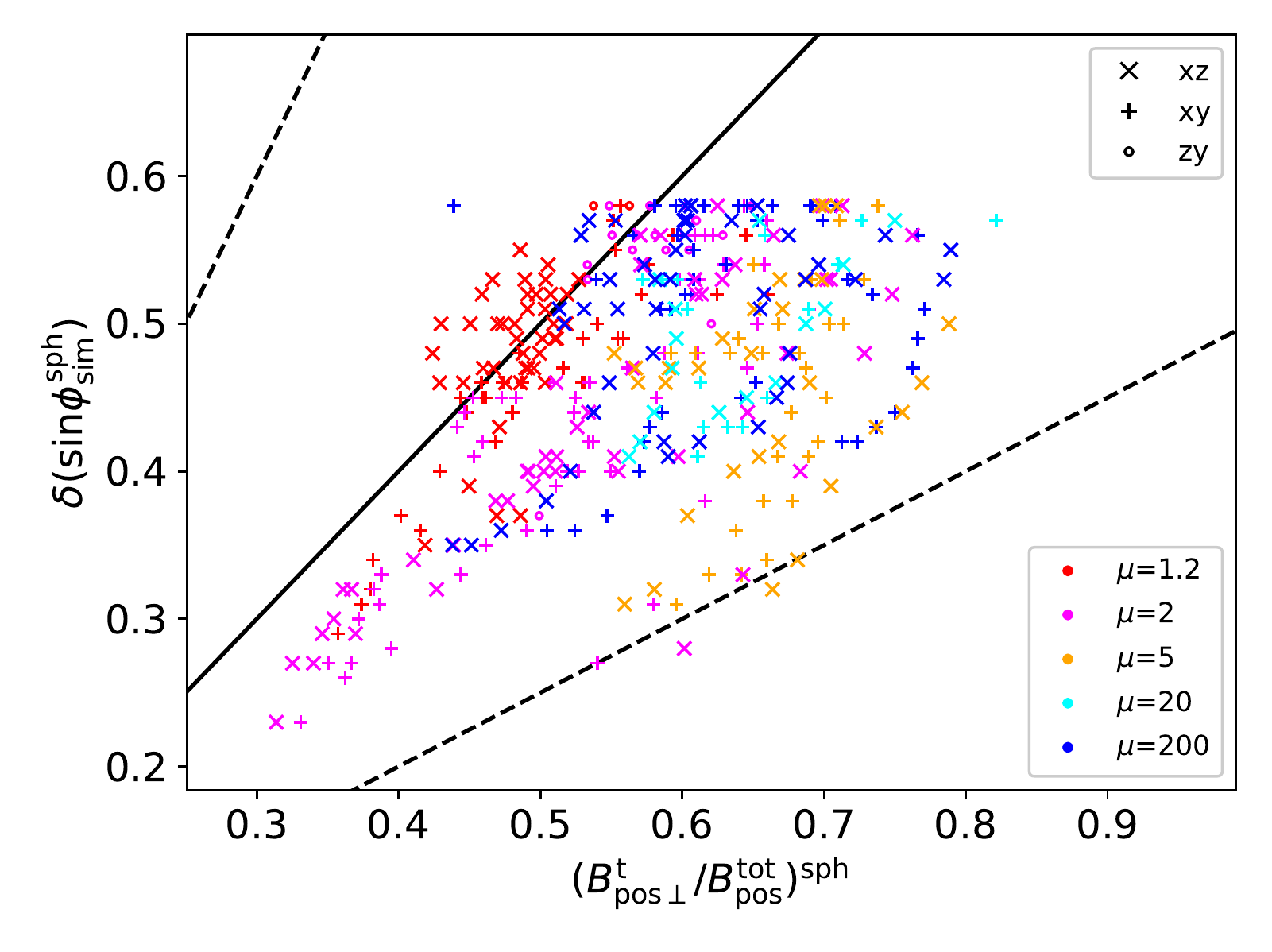}{0.45\textwidth}{b}}
\caption{Dispersions of (a) $ \phi^{\mathrm{sph}}_{\mathrm{sim}}$ and (b) $\sin \phi^{\mathrm{sph}}_{\mathrm{sim}}$ versus turbulent-to-total magnetic field strength ratio for all simulation models within different radii with respect to the most massive sink and at three orthogonal projections. Different symbols represent different projections. Different colors correspond to different initial $\mu$ values. Solid lines correspond to 1:1 relation. Dashed lines correspond to 1:2 and 2:1 relations.  \label{fig:disp_BtBtot}}
\end{figure*}

Figure \ref{fig:disp_BtBtot} shows the angular dispersions estimated within spheres in the simulation space ($\delta \phi^{\mathrm{sph}}_{\mathrm{sim}}$ and $\delta (\sin \phi^{\mathrm{sph}}_{\mathrm{sim}})$) versus the turbulent-to-total magnetic field strength ratio $(B^{\mathrm{t}}_{\mathrm{pos\perp}}/B^{\mathrm{tot}}_{\mathrm{pos}})^{\mathrm{sph}}$ measured in the same space. Both $\delta \phi^{\mathrm{sph}}_{\mathrm{sim}}$ and $\delta (\sin \phi^{\mathrm{sph}}_{\mathrm{sim}})$ correlate well with $(B^{\mathrm{t}}_{\mathrm{pos\perp}}/B^{\mathrm{tot}}_{\mathrm{pos}})^{\mathrm{sph}}$ within a factor of 2. Models with different initial $\mu$ values do not show significant difference in the value range of the turbulent-to-total magnetic field strength ratio or the angular dispersion. Models with different initial $\mu$ values also do not show significant difference in the correlation between the angular dispersion and the turbulent-to-total magnetic field strength ratio.

\subsubsection{Comparing angular dispersions in the model with angular dispersions in synthetic polarization maps}\label{sec:angobs_angsim}

The DCF method assumes the plane-of-sky angular dispersions of magnetic fields in 3D spaces are represented by the angular dispersions obtained from 2D polarization maps. As firstly proposed by \citet{1991ApJ...373..509M} and \citet{1990ApJ...362..545Z}, the effect of signal integration along the line of sight may underestimate the angular dispersion because of complex ordered magnetic field structure or perturbation from multiple independent turbulent cells. Here we investigate the signal integration effect on the measured angular dispersions.  

We measure the observed angular dispersions, $\delta\phi_{\mathrm{obs}}$, $\delta (\tan \phi_{\mathrm{obs}})$, and $\delta (\sin \phi_{\mathrm{obs}})$, within circles with different radii with respect to the projected position of the most massive sink on the synthetic polarization maps at three orthogonal projections. There are more than 200 uniformly sampled pixels within the smallest analysing radius ($\sim$0.01 pc) in the polarization map, so the estimated angular dispersion does not suffer from the uncertainty from sparse sampling. The way to estimate the observed angular dispersion from 2D synthetic polarization maps is similar to that in \citet{2001ApJ...559.1005P}, which is to shift the origin of the observed position angle $\phi_{\mathrm{obs}}$ to the approximate minimum of the angular distribution and measure the standard deviations of the $\phi_{\mathrm{obs}}$, $\tan \phi_{\mathrm{obs}}$, or $\sin \phi_{\mathrm{obs}}$. In principle, the observed polarization emissions arise from all the polarized dust grains in the volume along the line of sight if the emission is optically thin. In practice, the polarized gas in the outer low-density space only has negligible contribution to the synthetic polarized dust emission due to mass concentration, so the observed polarized emission is usually correlated with condensed structures (i.e., clumps and cores). It is unclear whether the synthetic polarized emission is dominated by the contribution from the central condensed structures or the material at outer low-density space also has nonnegligible contribution, so we compare the observed angular dispersions within circles in polarization maps with the angular dispersions within both cylinders (indicated by superscript ``cyl'') and spheres in the simulation space. All the concerned space and area are centered at the position or projected position of the most massive sink.

\begin{figure*}[!htbp]
 \gridline{\fig{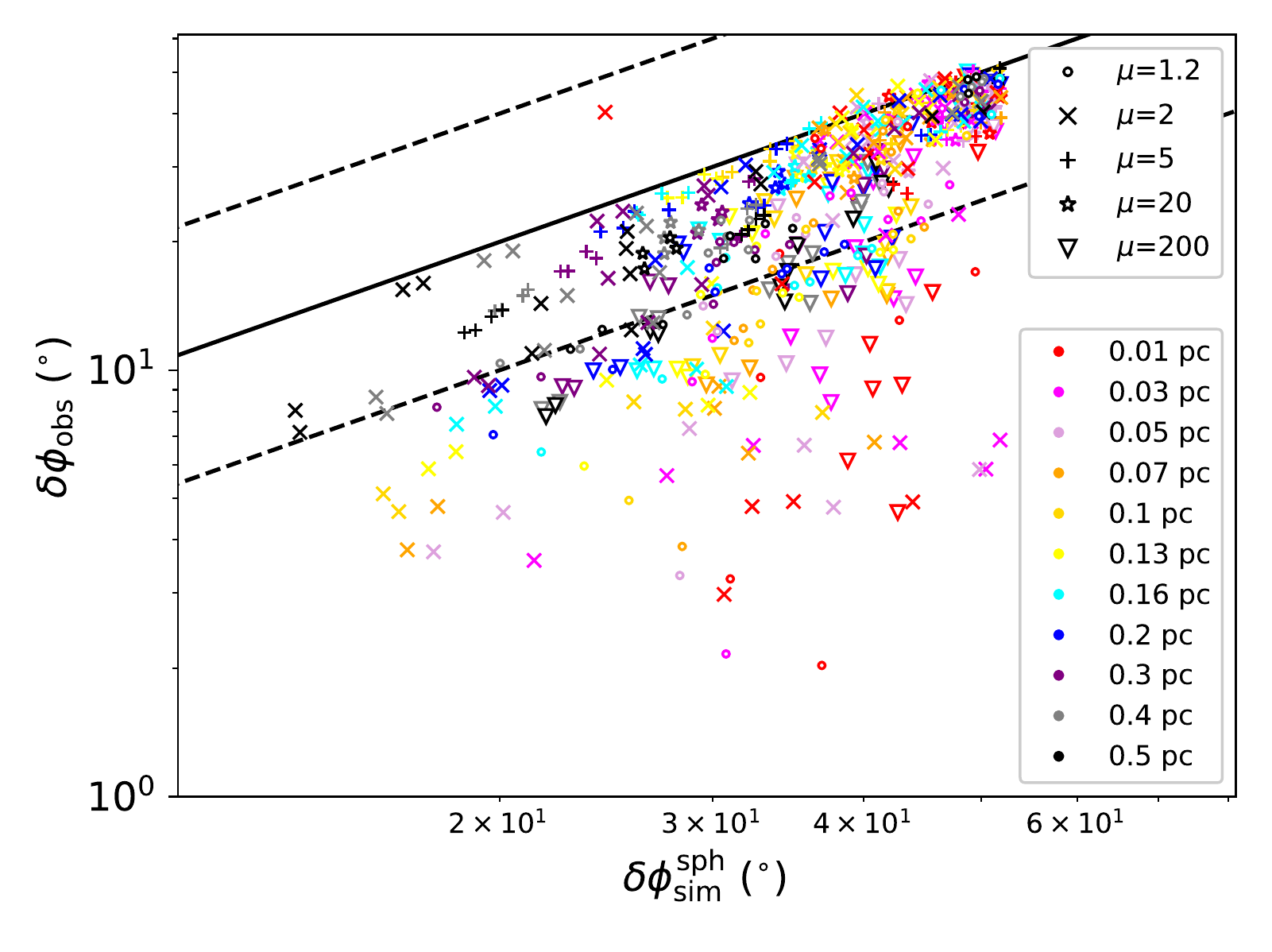}{0.45\textwidth}{a}
 \fig{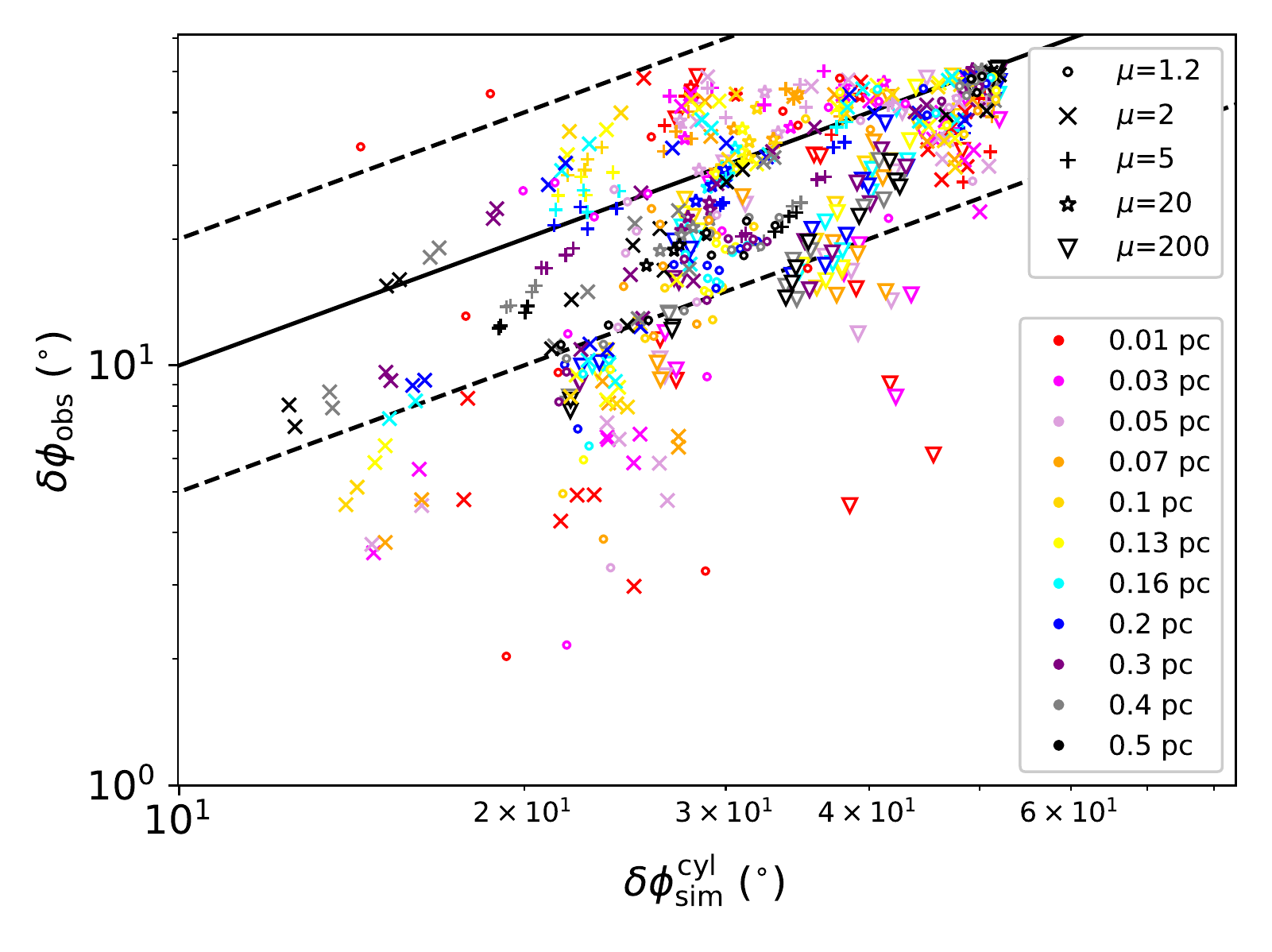}{0.45\textwidth}{b}}
  \gridline{\fig{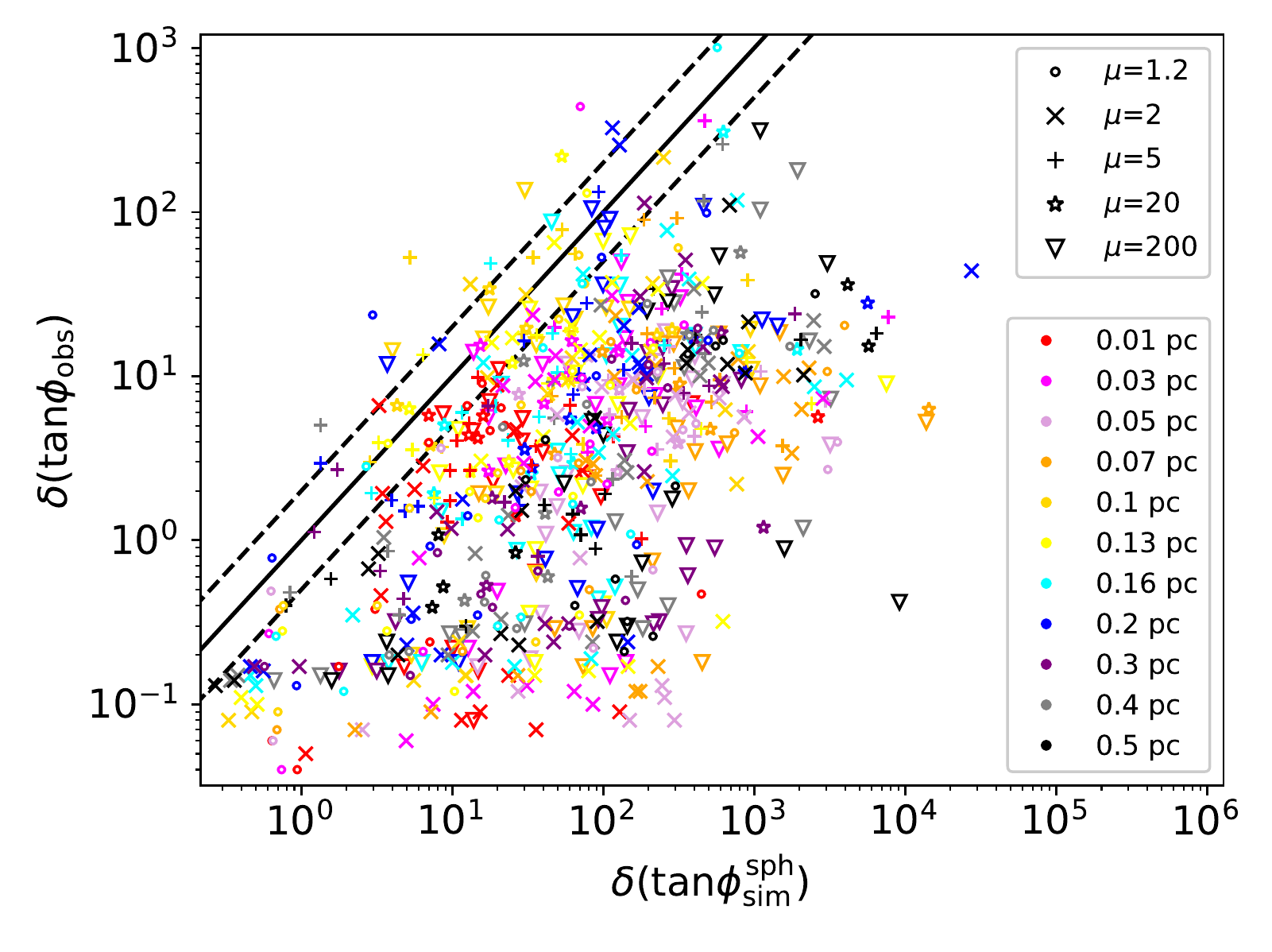}{0.45\textwidth}{c}
 \fig{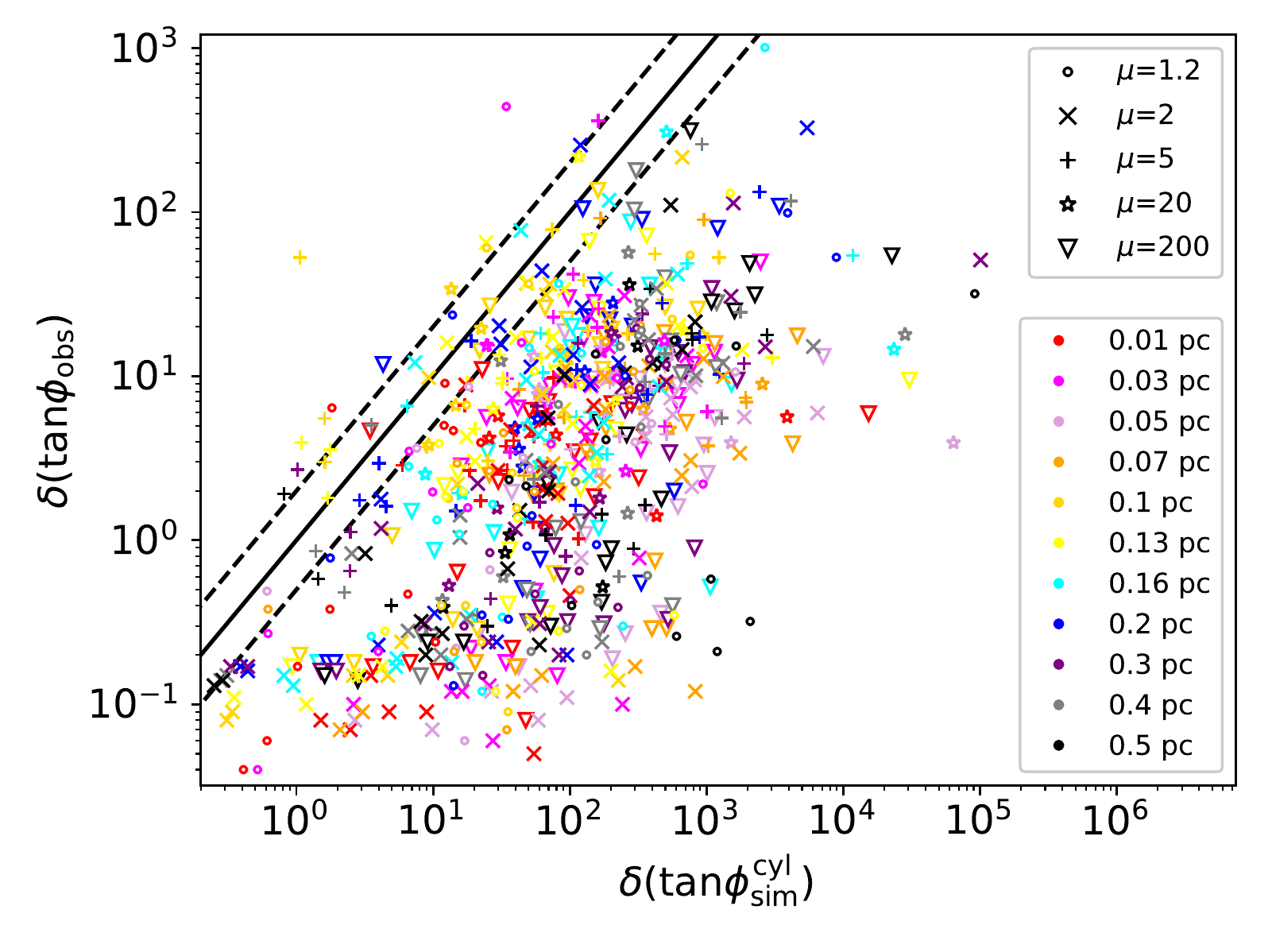}{0.45\textwidth}{d}}
  \gridline{\fig{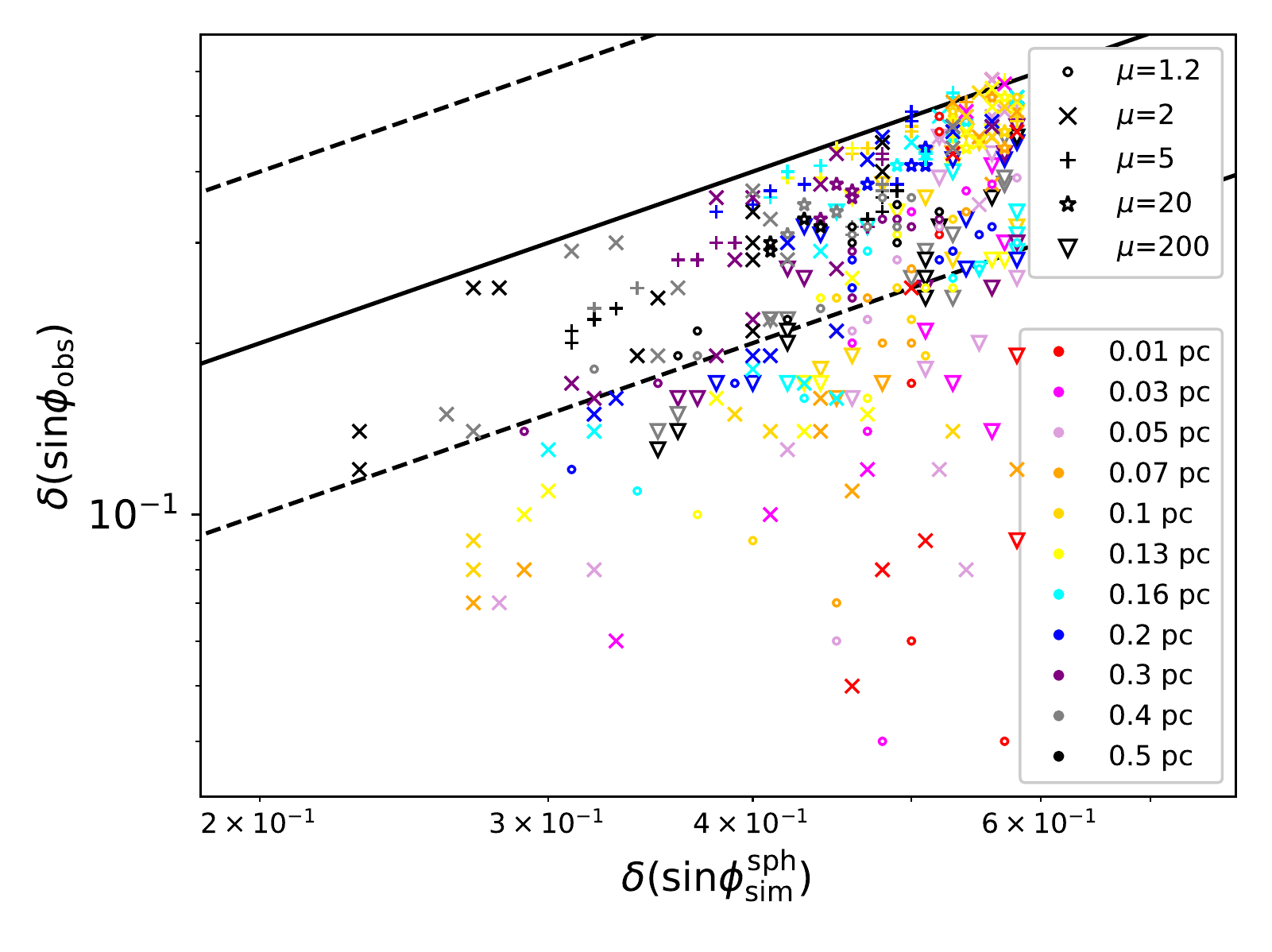}{0.45\textwidth}{e}
 \fig{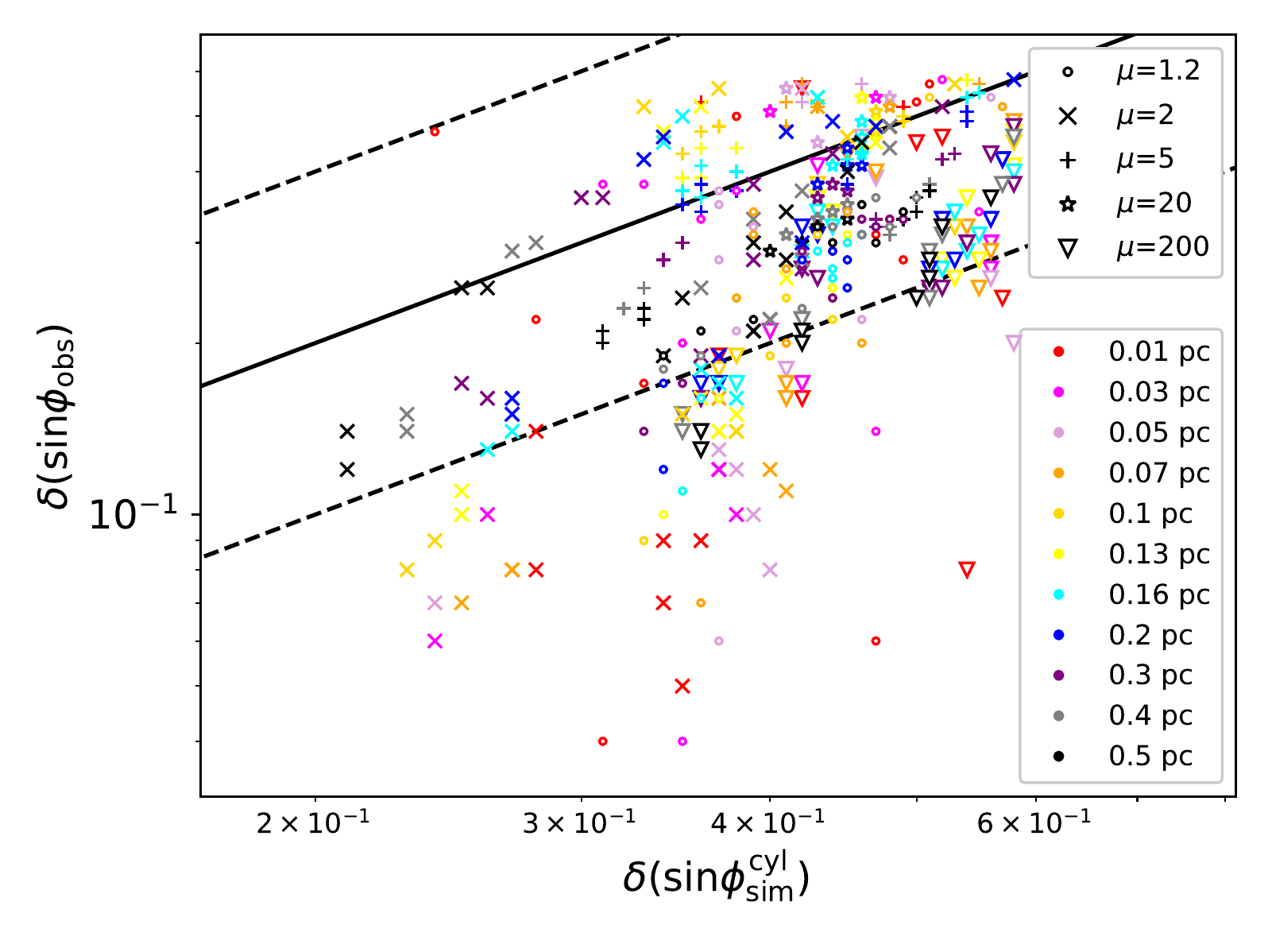}{0.45\textwidth}{f}}
\caption{Observed angular dispersion within circles versus angular dispersions in the simulation space obtained in spheres (left panels) and cylinders (right panels) for all simulation models within different radii with respect to the most massive sink and at three orthogonal projections. Different symbols represent different initial $\mu$ values. Different colors correspond to different radii. Solid lines correspond to 1:1 relation. Dashed lines correspond to 1:2 and 2:1 relations.  \label{fig:ang_angsim}}
\end{figure*}

Figure \ref{fig:ang_angsim} shows the relation between the angular dispersions estimated within spheres and cylinders in the simulation space and the observed angular dispersion within the projected circles in polarization maps. As shown in Figures \ref{fig:ang_angsim}(c) and (d), the angular dispersion in the form of $\delta (\tan \phi)$ has large scatters. Thus, we avoid using $\delta (\tan \phi)$ in the following analysis. As shown in Figures \ref{fig:ang_angsim}(a), (b), (e), and (f), the observed angular dispersions and angular dispersions in the simulation space are positively correlated with each other and the observed angular dispersions are overall smaller than the angular dispersions in the simulation space. This underestimation of the angular dispersion could be due to line-of-sight signal integrations through more than one turbulent cells or due to field tangling of complex field geometry along the line of sight. For some data points within small radii (radii$\lesssim$0.1 pc), we observe a significant underestimation of the angular dispersion in the model 3D space, in some of our synthetic observations. Generally, the ratio between the angular dispersion in polarization maps and the angular dispersion in simulation space at radii$>$0.1 pc is $\gtrsim$0.5. There is no significant difference in the correlation between the angular dispersions in the simulation space and in the polarization map for models with different initial $\mu$ values.

Table \ref{tab:angangsim} summarises the average ratios between the angular dispersions in polarization maps and the angular dispersion in the simulation space for data points with radii$>0.1$ pc. These average ratios statistically take into account the contribution from the ordered field structure and should be regarded as the correction factor for the effect of signal integration along the line-of-sight.

\begin{deluxetable}{ccc}[t!]
\tablecaption{Average ratios between the raw angular dispersions in polarization maps and angular dispersions in the simulation space for data points at radii$>0.1$ pc. Values in the parenthesis are the relative uncertainty. \label{tab:angangsim}}
\tablecolumns{3}
\tablewidth{0pt}
\tablehead{
\colhead{Average ratios} &
\colhead{Spheres\tablenotemark{a}} &
\colhead{Cylinders\tablenotemark{a}} 
}
\startdata
$\delta \phi_{\mathrm{obs}}$ /$\delta \phi_{\mathrm{sim}}$  &   0.70(28\%) &   0.77(31\%)\\ 
$\delta (\sin \phi_{\mathrm{obs}})$/$\delta (\sin \phi_{\mathrm{sim}})$ &   0.68(26\%) &   0.73(31\%)\\\hline
\enddata
\tablenotetext{a}{Indicates whether the angular dispersion in the simulation space are estimated in spheres or cylinders.}
\end{deluxetable}

\subsubsection{Comparing angular dispersions in polarization maps with ratios between magnetic field components in the model} \label{sec:compangBtB}

\begin{figure*}[!htbp]
 \gridline{\fig{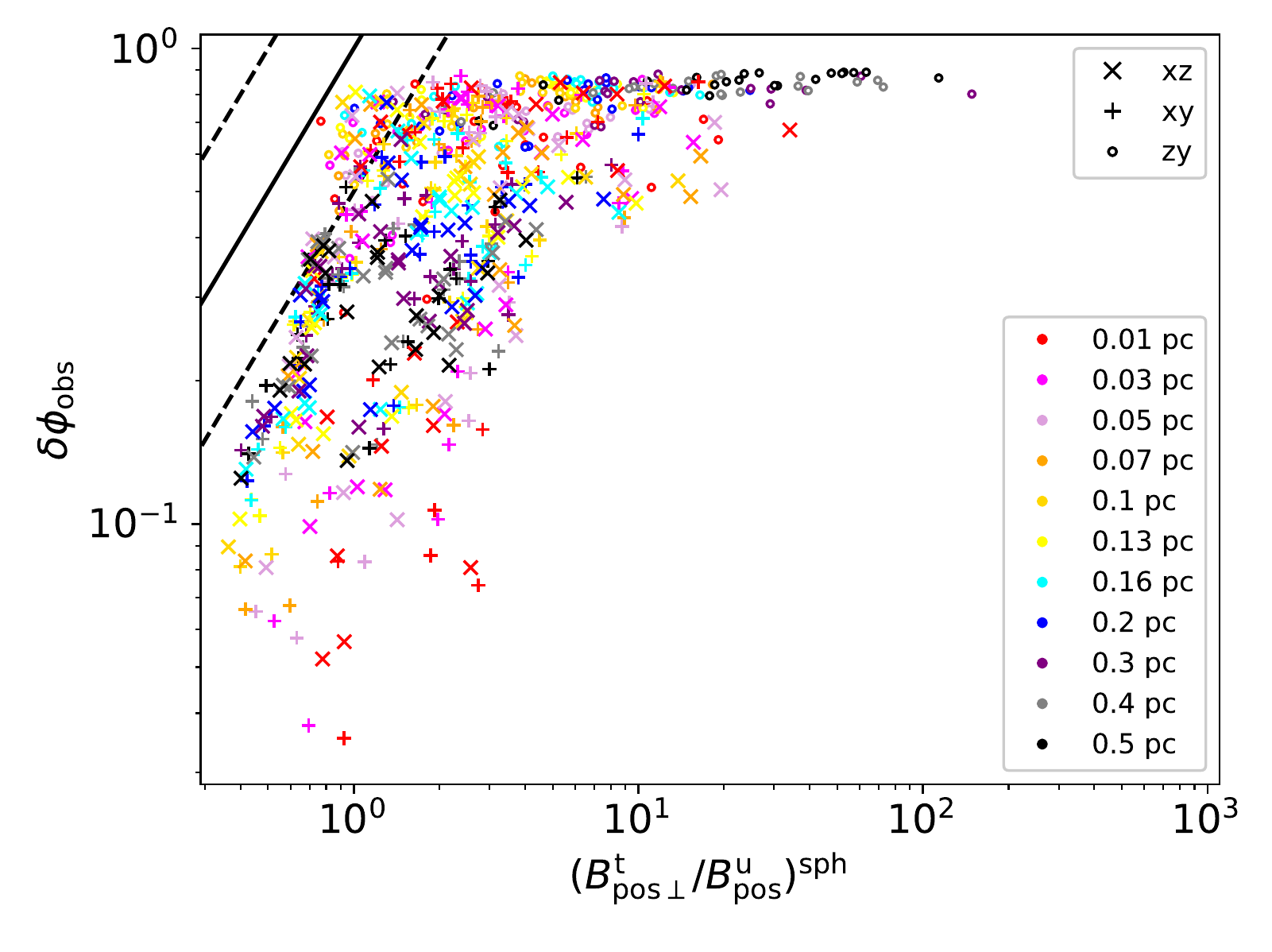}{0.45\textwidth}{a}
 \fig{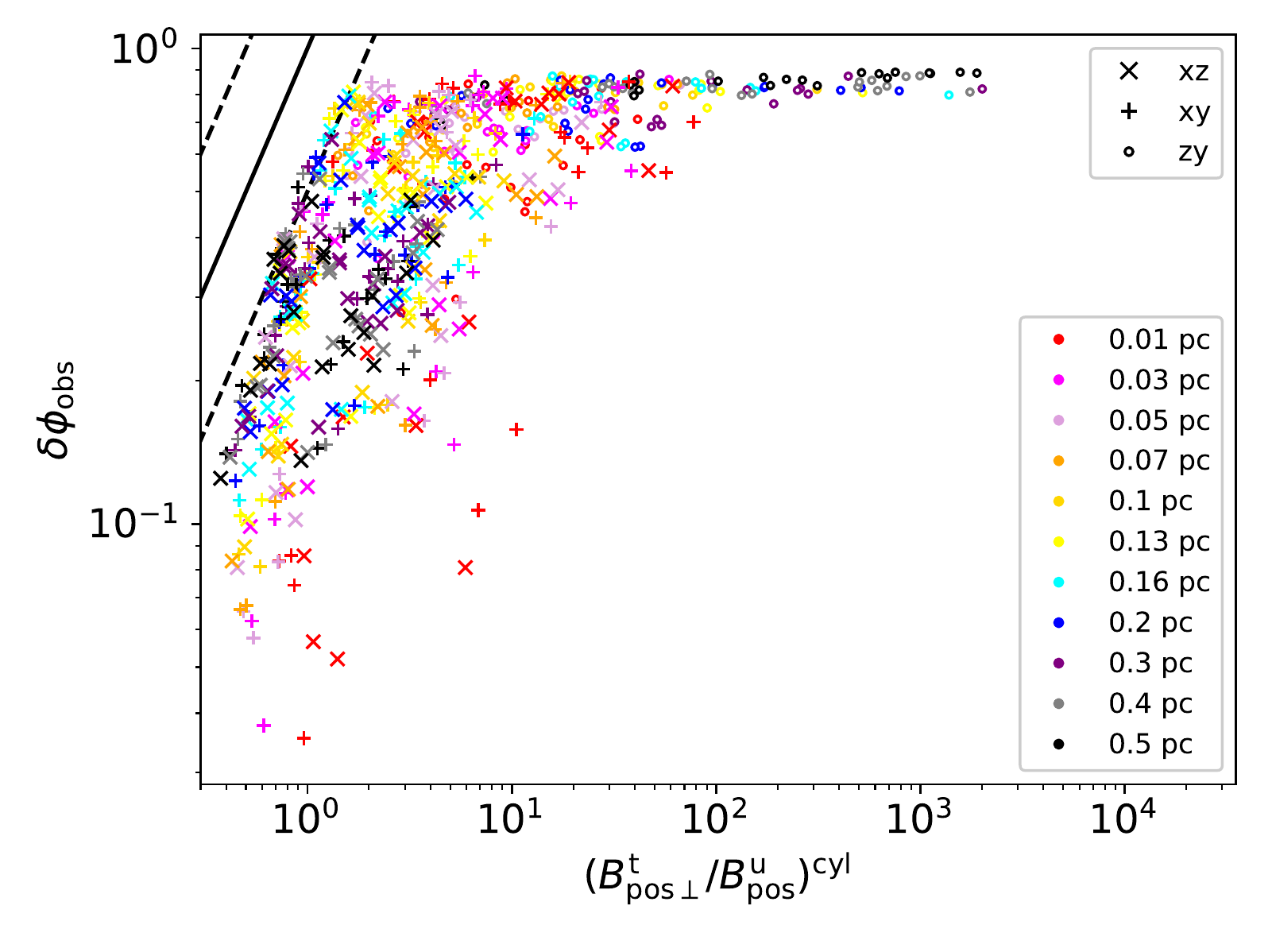}{0.45\textwidth}{b}}
 \gridline{\fig{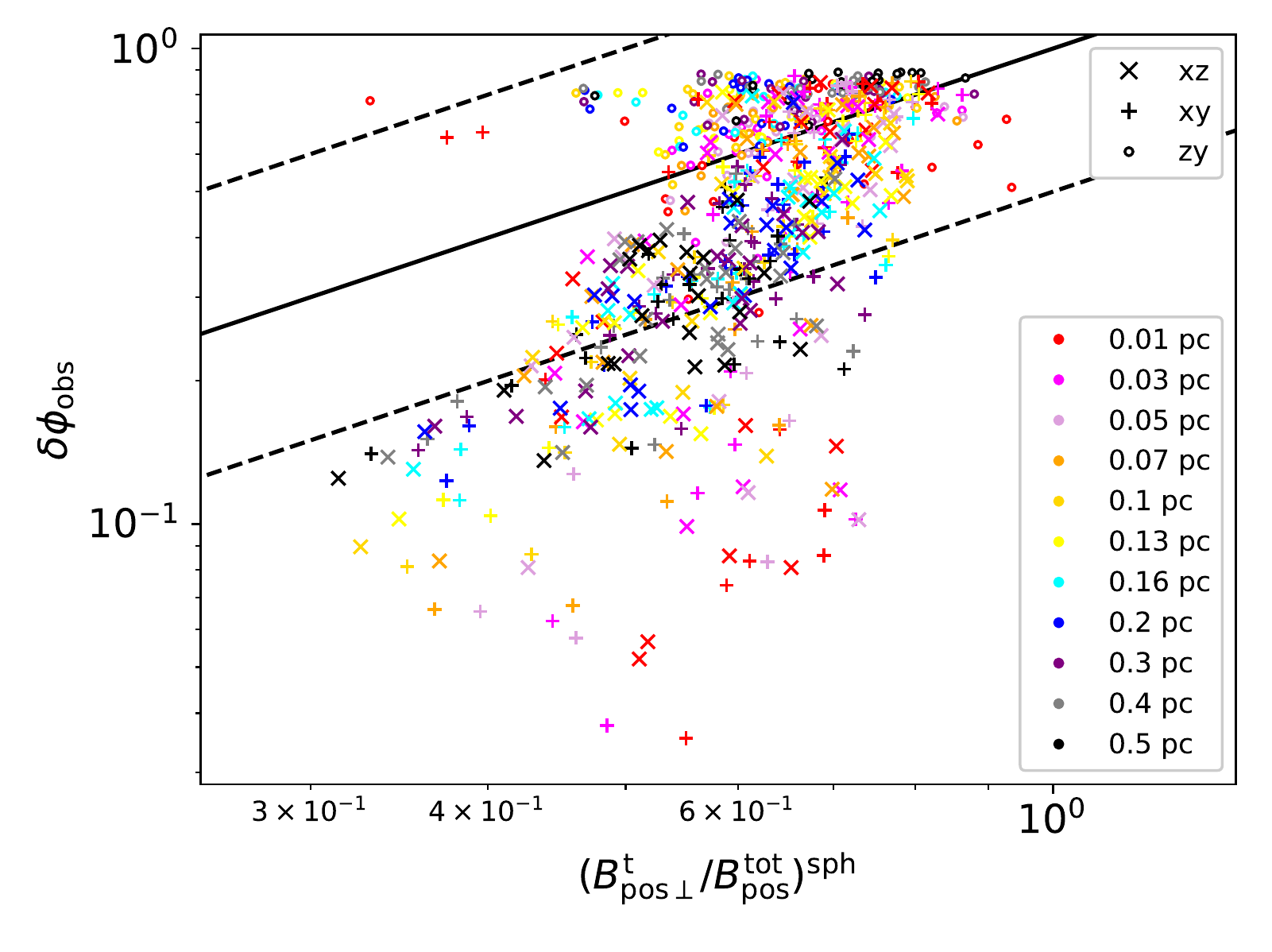}{0.45\textwidth}{c}
 \fig{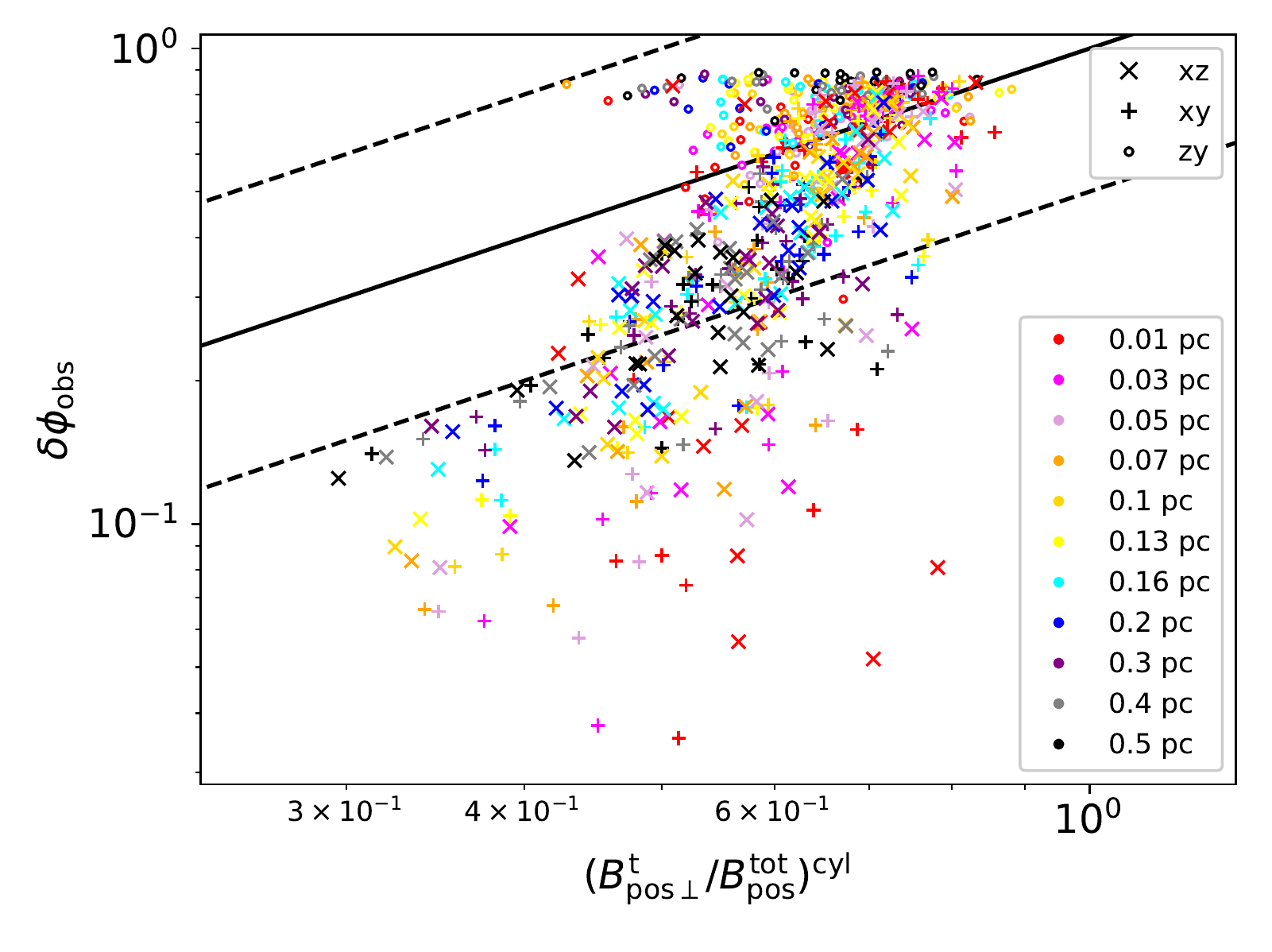}{0.45\textwidth}{d}}
 \gridline{\fig{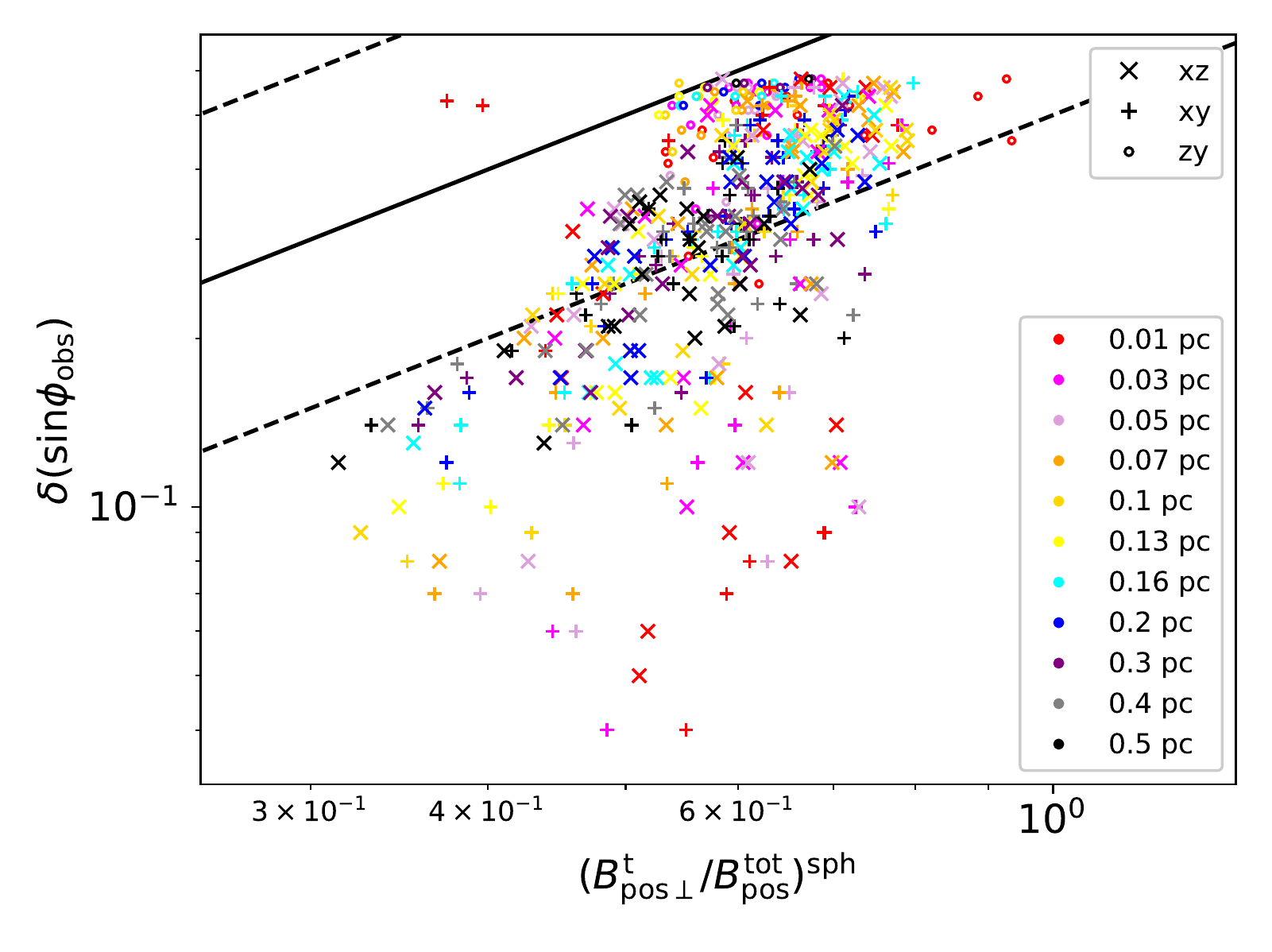}{0.45\textwidth}{e}
 \fig{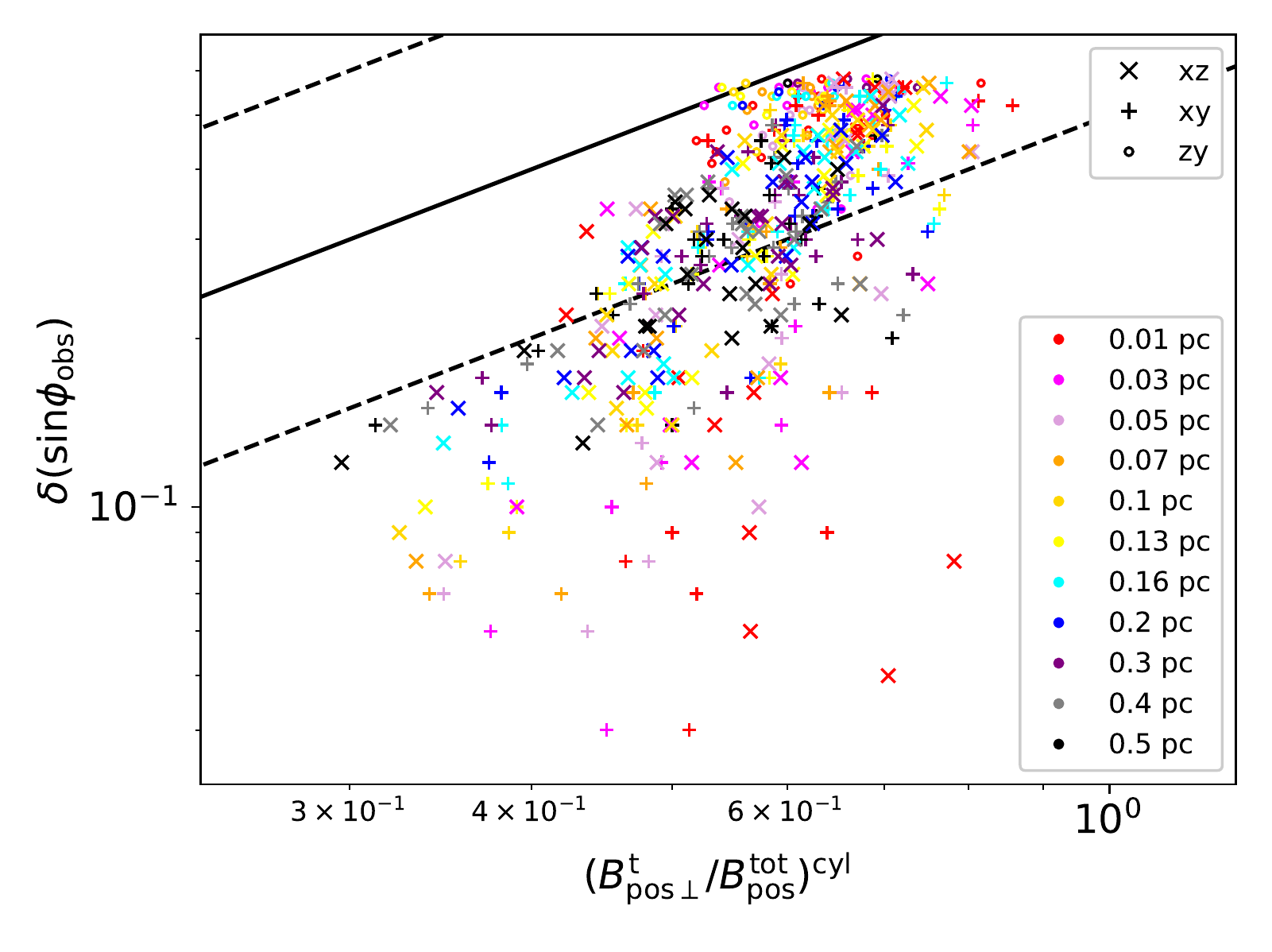}{0.45\textwidth}{f}}
\caption{Angular dispersion within circles in polarization maps versus turbulent-to-ordered or -total magnetic field strength ratios in the simulation space obtained in spheres (left panels) and cylinders (right panels) for all simulation models within different radii with respect to the most massive sink and at three orthogonal projections. Different symbols represent different projections. Different colors correspond to different radii. Solid lines correspond to 1:1 relation. Dashed lines correspond to 1:2 and 2:1 relations.  \label{fig:ang_BtB}}
\end{figure*}

We combine the results in Sections \ref{sec:compangbmodel} and \ref{sec:angobs_angsim} and compare the angular dispersion in polarization maps with the turbulent-to-ordered or -total magnetic field strength ratio in the simulation space (see Figure \ref{fig:ang_BtB}). In general, the measured angular dispersions in polarization maps are positively correlated with the turbulent-to-ordered or -total magnetic field strength ratio. Figures \ref{fig:ang_BtB}(a) and (b) indicate that $\delta\phi_{\mathrm{obs}}$ would overall underestimate the turbulent-to-ordered magnetic field strength ratio. Data points with large $\delta \phi_{\mathrm{obs}}$ (e.g., $> 0.44$, $\sim25\degr$) or small radii (radii$\lesssim$0.1 pc) show large scatters. Most of $\delta \phi_{\mathrm{obs}}$ in the zy plane are greater than $\sim25\degr$ and are poorly correlated with the turbulent-to-ordered magnetic field strength ratio. We also find that $\delta\phi_{\mathrm{obs}}$ is better correlated with $B^{\mathrm{t}}_{\mathrm{pos\perp}}/B^{\mathrm{u}}_{\mathrm{pos}}$ for models with stronger initial magnetic fields. Figures \ref{fig:ang_BtB}(c), (d), (e), and (f) indicate that the measured angular dispersion would be significantly lower than the turbulent-to-total mangetic field strength ratio for some data points with a small radius (radii$\lesssim$0.1 pc) and small angular dispersion. For data points with radii$>$0.1 pc, the measured angular dispersion is comparable to or only slightly lower than the turbulent-to-total mangetic field strength ratio. Overall, $\delta \phi_{\mathrm{obs}}$ in the zy plane are slightly greater than $B^{\mathrm{t}}_{\mathrm{pos\perp}}/B^{\mathrm{tot}}_{\mathrm{pos}}$, while $\delta \phi_{\mathrm{obs}}$ in the xz and xy planes are slightly smaller than $B^{\mathrm{t}}_{\mathrm{pos\perp}}/B^{\mathrm{tot}}_{\mathrm{pos}}$. On the other hand, $\delta (\sin \phi_{\mathrm{obs}})$ are generally smaller than the $B^{\mathrm{t}}_{\mathrm{pos\perp}}/B^{\mathrm{tot}}_{\mathrm{pos}}$ and $\delta (\sin \phi_{\mathrm{obs}})$ in the zy plane are better correlated with $B^{\mathrm{t}}_{\mathrm{pos\perp}}/B^{\mathrm{tot}}_{\mathrm{pos}}$ than in the xz and xy planes. There is no significant difference in the correlation between the angular dispersion and the turbulent-to-total magnetic field strength ratio for models with different initial $\mu$ values.

Table \ref{tab:angBtB} summarises the average ratios between the angular dispersions in polarization maps and turbulent-to-ordered or -total magnetic field strength ratios for data points with radii$>0.1$ pc. Due to large scatters, we exclude data points with $\delta \phi_{\mathrm{obs}} > 25\degr$ ($\sim$0.44) in the estimation of $\delta \phi_{\mathrm{obs}}$/$(B^{\mathrm{t}}_{\mathrm{pos\perp}}/B^{\mathrm{u}}_{\mathrm{pos}})$. If other assumptions (i.e., isotropic turbulence and energy equipartition) of the DCF method are satisfied, the ratios reported in Table \ref{tab:angBtB} should be regarded as the correction factors for the plane-of-sky uniform or total magnetic field strength derived from the DCF method. These correction factors statistically take into account the contribution from the ordered field structure to the measured angular dispersion at clump and core scales. 

\begin{deluxetable}{ccc}[t!]
\tablecaption{Average ratios between the angular dispersions in polarization maps and turbulent-to-ordered or -total magnetic field strength ratios for data points at radii$>0.1$ pc. The ratios are averaged over 3 different planes (xy, xz, and zy). These ratios should be regarded as the correction factor for the estimated magnetic field strength if assumptions of energy equipartition and isotropic turbulence are satisfied. Values in the parenthesis are the relative uncertainty. \label{tab:angBtB}}
\tablecolumns{3}
\tablewidth{0pt}
\tablehead{
\colhead{Average ratios} &
\colhead{Spheres\tablenotemark{a}} &
\colhead{Cylinders\tablenotemark{a}} 
}
\startdata
$\delta \phi_{\mathrm{obs}}$/$(B^{\mathrm{t}}_{\mathrm{pos\perp}}/B^{\mathrm{u}}_{\mathrm{pos}})$ \tablenotemark{b} &   0.25(46\%) &   0.25(49\%)\\ 
$\delta \phi_{\mathrm{obs}}$/$(B^{\mathrm{t}}_{\mathrm{pos\perp}}/B^{\mathrm{tot}}_{\mathrm{pos}})$ &   0.81(43\%) &   0.82(41\%)\\ 
$\delta (\sin \phi_{\mathrm{obs}})$/$(B^{\mathrm{t}}_{\mathrm{pos\perp}}/B^{\mathrm{tot}}_{\mathrm{pos}})$ &   0.57(29\%) &   0.58(28\%)\\ 
\enddata
\tablenotetext{a}{Indicates whether the turbulent-to-ordered or -total magnetic field strength ratios are estimated in spheres or cylinders in the simulation space.}
\tablenotetext{b}{Additional selection criteria of $\delta \phi_{\mathrm{obs}} \lesssim 25\degr$ ($\sim$0.44) is adopted.}
\end{deluxetable}


\subsection{Uncertainties in estimating the angular dispersion}
\subsubsection{The angular dispersion function method}\label{sec:adf}

There are several effects that may affect the measured angular dispersion. The effect of measuring angular dispersion at scales smaller than the turbulent correlation scale \citep[hereafter the turbulent correlation effect,][]{2009ApJ...696..567H}, the beam-smoothing effect of telescopes, and the effect of line-of-sight signal integration can underestimate the angular dispersion and thus overestimate the magnetic field strength. On the other hand, the contribution from the underlying ordered (large-scale) field structure could overestimate the angular dispersion and thus underestimate the magnetic field strength. The spatial filtering effect of interferometers could also affect the measured angular dispersion and magnetic field strength.

There were several approaches to quantitatively account for these effects that may affect the measured angular dispersion. The angular dispersion function method \citep[ADF.][]{2008ApJ...679..537F, 2009ApJ...696..567H, 2009ApJ...706.1504H, 2016ApJ...820...38H} is the only method to analytically and simultaneously account for the ordered field contribution, the beam-smoothing effect, the turbulent correlation effect, and the inteferometric filtering effect. Here we investigate the accuracy of the ADF method with our simulation results.


In Appendix \ref{app:ADForder}, we tested the ADF only accounting for the ordered field structure with simple Monte Carlo simulations and the ADF works well on taking into account this effect. Further, the ADF accounting for the ordered field structure and the plane-of-sky turbulent correlation effect is given by \citep{2009ApJ...696..567H, 2009ApJ...706.1504H, 2016ApJ...820...38H}: 
\begin{equation} \label{eq:adfnobeam}
1 - \langle \cos \lbrack \Delta \Phi (l)\rbrack \rangle \simeq a_2\arcmin l^2 +  (\frac{\langle B_{\mathrm{t}}^2 \rangle}{\langle B^2 \rangle})_{\mathrm{or,tc}}^{\mathrm{adf}}  \times \lbrack 1 - e^{-l^2/2l_{\delta}^2}\rbrack,
\end{equation}
where $\Delta \Phi (l)$ is the angular difference of two magnetic field line segments separated by a distance $l$, $a_2\arcmin l^2$ is the first term of the Taylor expansion of the ordered component of ADF, $l_{\delta}$ is the turbulent correlation length, $(\langle B_{\mathrm{t}}^2 \rangle/  \langle B^2\rangle)_{\mathrm{or,tc}}^{\mathrm{adf}} = (\langle B_{\mathrm{t}}^2 \rangle/ \langle B_0^2\rangle)_{\mathrm{or,tc}}^{\mathrm{adf}} / ((\langle B_{\mathrm{t}}^2 \rangle/ \langle B_0^2\rangle)_{\mathrm{or,tc}}^{\mathrm{adf}}+1) $ is the turbulent-to-total magnetic energy ratio and $(\langle B_{\mathrm{t}}^2 \rangle/  \langle B_0^2\rangle)_{\mathrm{or,tc}}^{\mathrm{adf}}$ is the turbulent-to-ordered energy ratio. 

We derive the ADFs within different radii in the polarization maps for all simulation models. Calculation of the ADFs can be very time-consuming if a large number of angle samples are concerned. Thus, we bin the synthetic Q and U maps with an appropriate bin size and regenerate the binned position angles for some of the calculations to make sure the number of angle sample is smaller than 3000. Using the first term of the Taylor expansion to represent the ordered magnetic field is only valid when the scale $l$ is smaller than an appropriate displacement $d$, while the displacement $d$ varies case by case. Thus, we fit each specific ADF over different maximum spatial scales and obtain the best fit for each scale via reduced $\chi^2$ minimization. We compare the reduced $\chi^2$ of the best fit for different spatial scales and adopt the one with the smallest reduced $\chi^2$ as the best fit for each specific ADF.

\begin{figure}[!htbp]
 \gridline{\fig{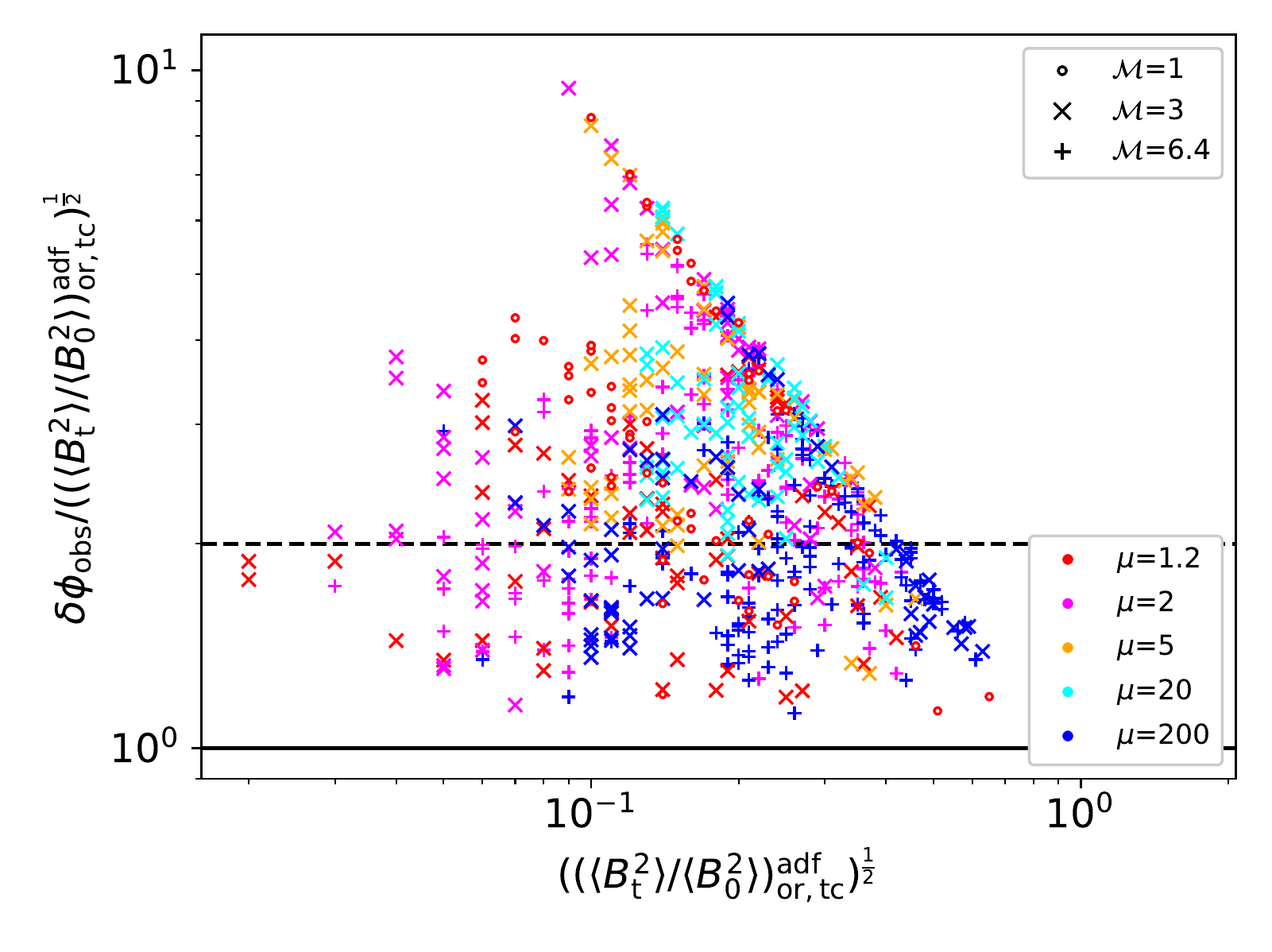}{0.45\textwidth}{}}
\gridline{\fig{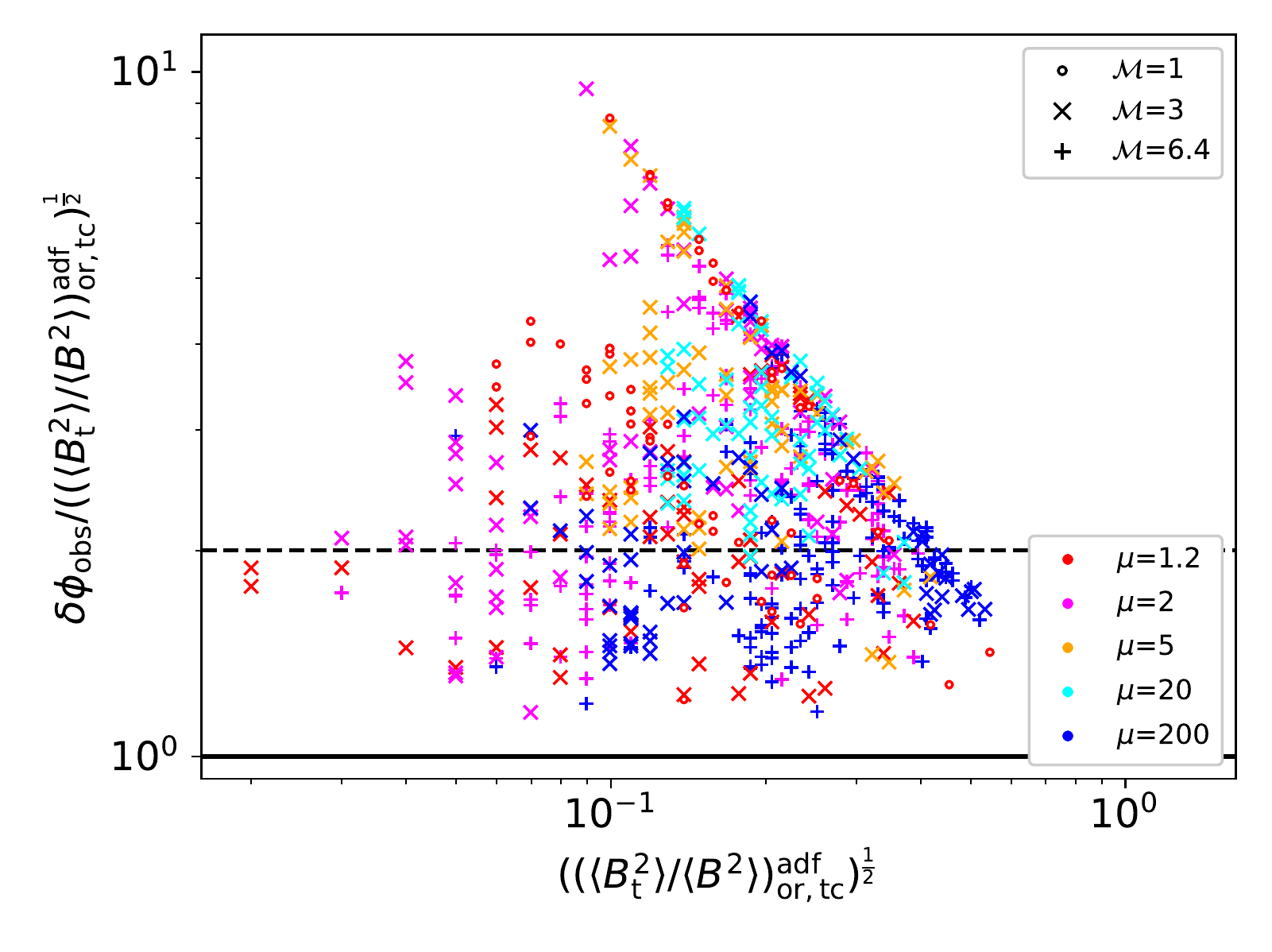}{0.45\textwidth}{}}
\gridline{\fig{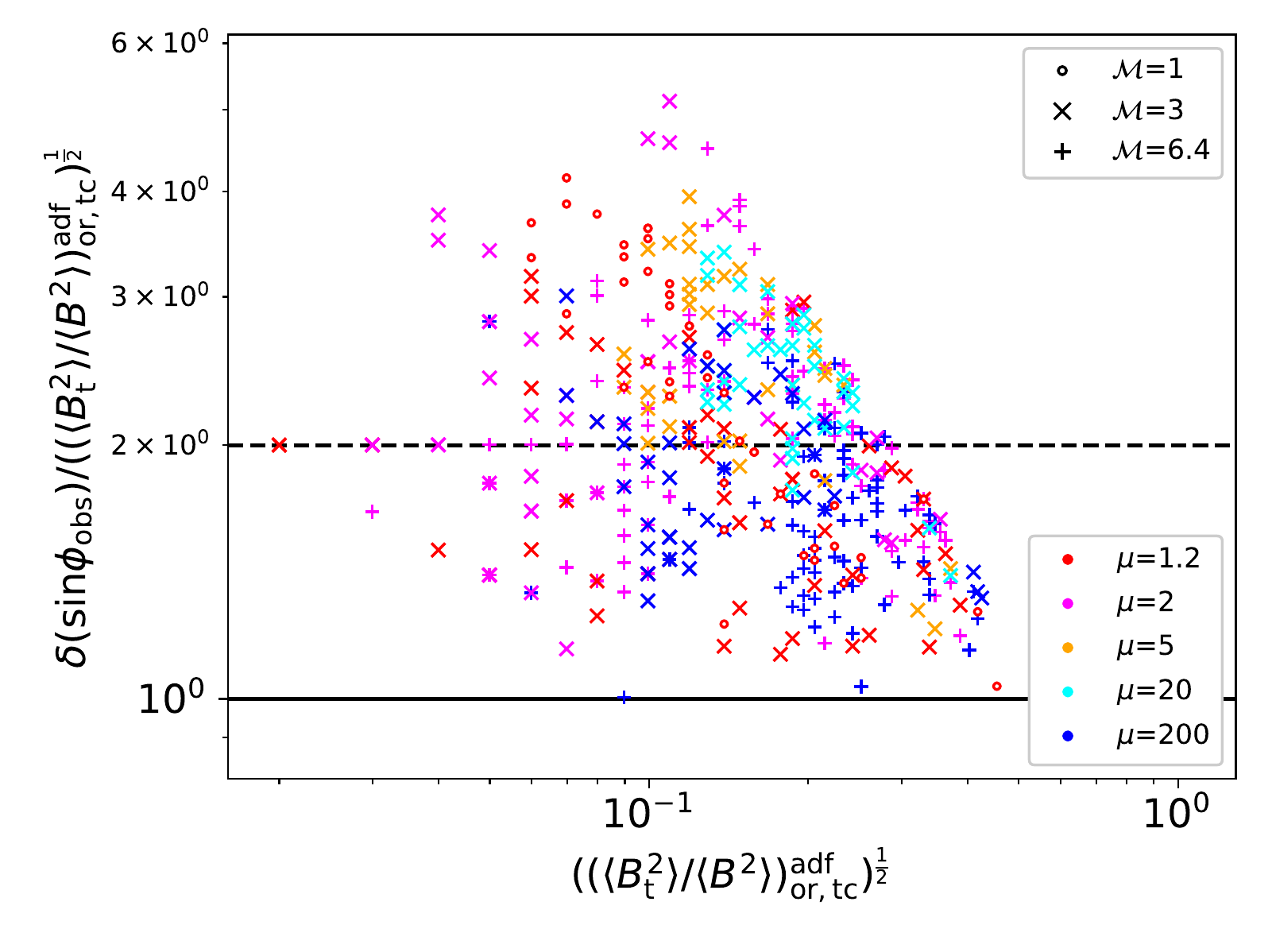}{0.45\textwidth}{}}
\caption{Ratio between the directly estimated angular dispersions in synthetic polarization maps and the turbulent-to-ordered or -total magnetic field strength ratio derived by the ADF method. Different symbols represent different initial Mach numbers. Different colors correspond to different initial $\mu$ values. \label{fig:adf_ang_btb}}
\end{figure}

Figure \ref{fig:adf_ang_btb} shows the relation between the directly estimated angular dispersions (see Section \ref{sec:angobs_angsim}) in synthetic polarization maps and the turbulent-to-ordered or -total magnetic field strength ratio derived from the ADF method. Generally, the directly estimated angular dispersions are greater than the turbulent-to-ordered or -total magnetic field strength ratio derived from the ADF method, which can be explained by the contribution from large-scale field structure. If the turbulent-to-ordered or -total magnetic field strength ratio correctly trace the true angular dispersion in polarization maps, Figure \ref{fig:adf_ang_btb} indicates that the contribution from ordered field structure can overestimate the angular dispersion by a few factors when the intrinsic angular dispersion in polarization maps is small. The average $\delta \phi_{\mathrm{obs}}/(\langle B_{\mathrm{t}}^2 \rangle/ \langle B_0^2\rangle)_{\mathrm{or,tc}}^{\mathrm{adf}}$, $\delta \phi_{\mathrm{obs}}/(\langle B_{\mathrm{t}}^2 \rangle/ \langle B^2\rangle)_{\mathrm{or,tc}}^{\mathrm{adf}}$, and $\delta (\sin \phi_{\mathrm{obs}})/(\langle B_{\mathrm{t}}^2 \rangle/ \langle B^2\rangle)_{\mathrm{or,tc}}^{\mathrm{adf}}$ are $\sim$2.7, $\sim$2.8, and $\sim$2.1, respectly, suggesting that the contribution from the ordered field may overestimate the angular dispersion by a factor of 2-3 on average. 


If the signal integration along the line of sight is considered, the turbulent-to-ordered energy ratio is given by $(\langle B_{\mathrm{t}}^2 \rangle/ \langle B_0^2\rangle)_{\mathrm{or,tc,si}}^{\mathrm{adf}} = (\langle B_{\mathrm{t}}^2 \rangle/ \langle B_0^2\rangle)_{\mathrm{or,tc}}^{\mathrm{adf}} \times N_{\mathrm{adf}}'$, where $N_{\mathrm{adf}}'$ is the number of turbulent cells along the line of sight. $N_{\mathrm{adf}}'$ is given by
\begin{equation}
N_{\mathrm{adf}}' = \frac{l_\Delta}{\sqrt{2\pi}l_{\delta}},
\end{equation}
where $l_\Delta$ is the effective thickness along the line of sight. Here an assumption is adopted that the line-of-sight turbulent correlation scale is identical to the plane-of-sky turbulent correlation scale. Similarly, the turbulent-to-total energy ratio $(\langle B_{\mathrm{t}}^2 \rangle/ \langle B^2\rangle)_{\mathrm{or,tc,si}}^{\mathrm{adf}}$ can be derived from the turbulent-to-ordered energy ratio taking into account this signal integration effect. 


Following \citet{2009ApJ...706.1504H}, we adopt the width at half of the maximum of the normalized auto-correlation function of the integrated normalized polarized flux as the effective thickness $l_\Delta$. However, for $\sim$35\% of our estimations, the minimum of the normalized auto-correlation function of the integrated normalized polarized flux is greater than half of the maximum, so the effective thickness cannot be derived in this way. Thus, these data are not included in the following ADF analysis. 

The turbulent-to-ordered or -total magnetic field strength ratio derived from the ADF method should be compared to the turbulent-to-ordered or -total magnetic field strength ratio without the contribution from ordered field structure in the 3D simulation space. However, due to the lack of an appropriate method to account for the 3D field geometry, this cannot be done in our analysis. In principle, the ADF method can be applied on the 3D field to remove the 3D field geometry and derive the 3D turbulent-to-ordered or -total field strength ratio. In 3D space, there is no need to take into account the line-of-sight signal integration. However, the upper limit of the derivable turbulent-to-ordered or -total field strength ratio without accounting for the siginal-integration effect are \textbf{$\sim$0.76 and $\sim$0.6}, respectively, for the ADF method (see Appendix \ref{app:ADForder}). So the 3D turbulent-to-total field strength ratio could be underestimated, and the 3D turbulent-to-ordered field strength ratio could be greatly underestimated if there are significantly turbulent field. Application of the ADF method on 3D space with large grids could also be very time-consuming. On the other hand, the unsharp masking method \citep{2017ApJ...846..122P} may also be applied to remove the 3D field structure, but its correctness is still uncertain in 3D space because this method does not consider the turbulent correlation length. Thus, we refrain from deriving the 3D turbulent-to-ordered or -total field strength ratio without the contribution from ordered field structure.

\begin{figure*}[!htbp]
 \gridline{\fig{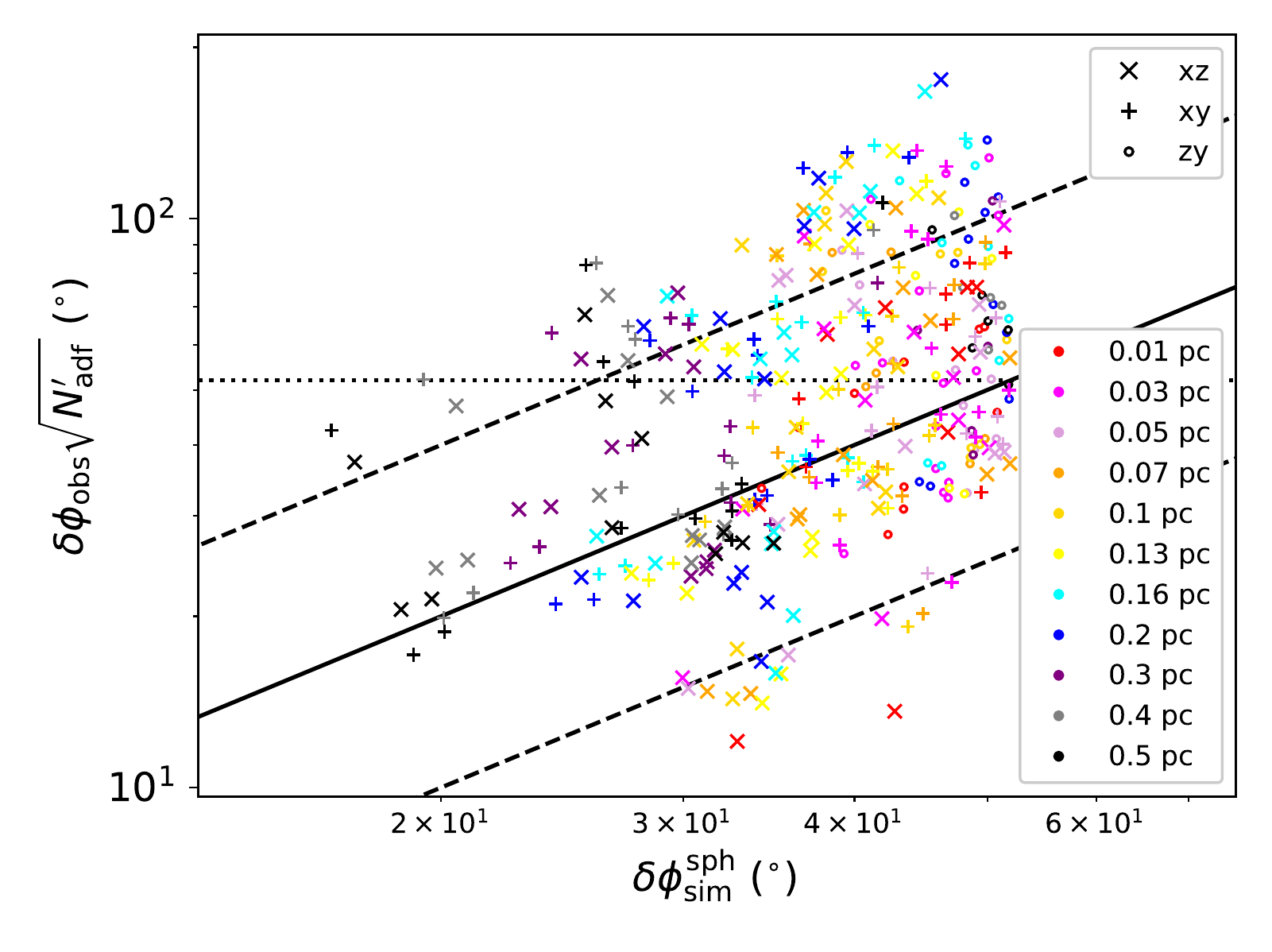}{0.45\textwidth}{a}
 \fig{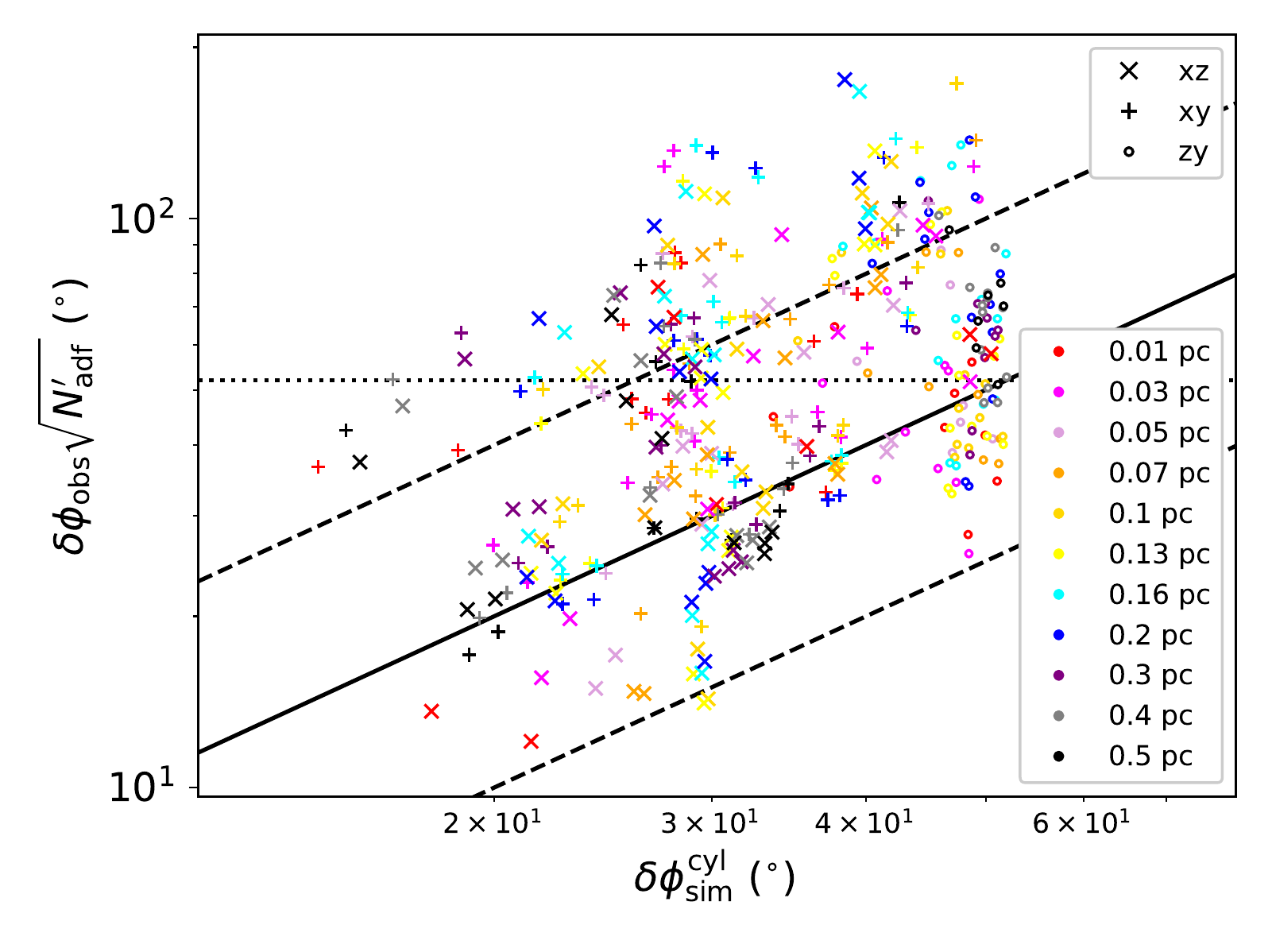}{0.45\textwidth}{b}}
  \gridline{\fig{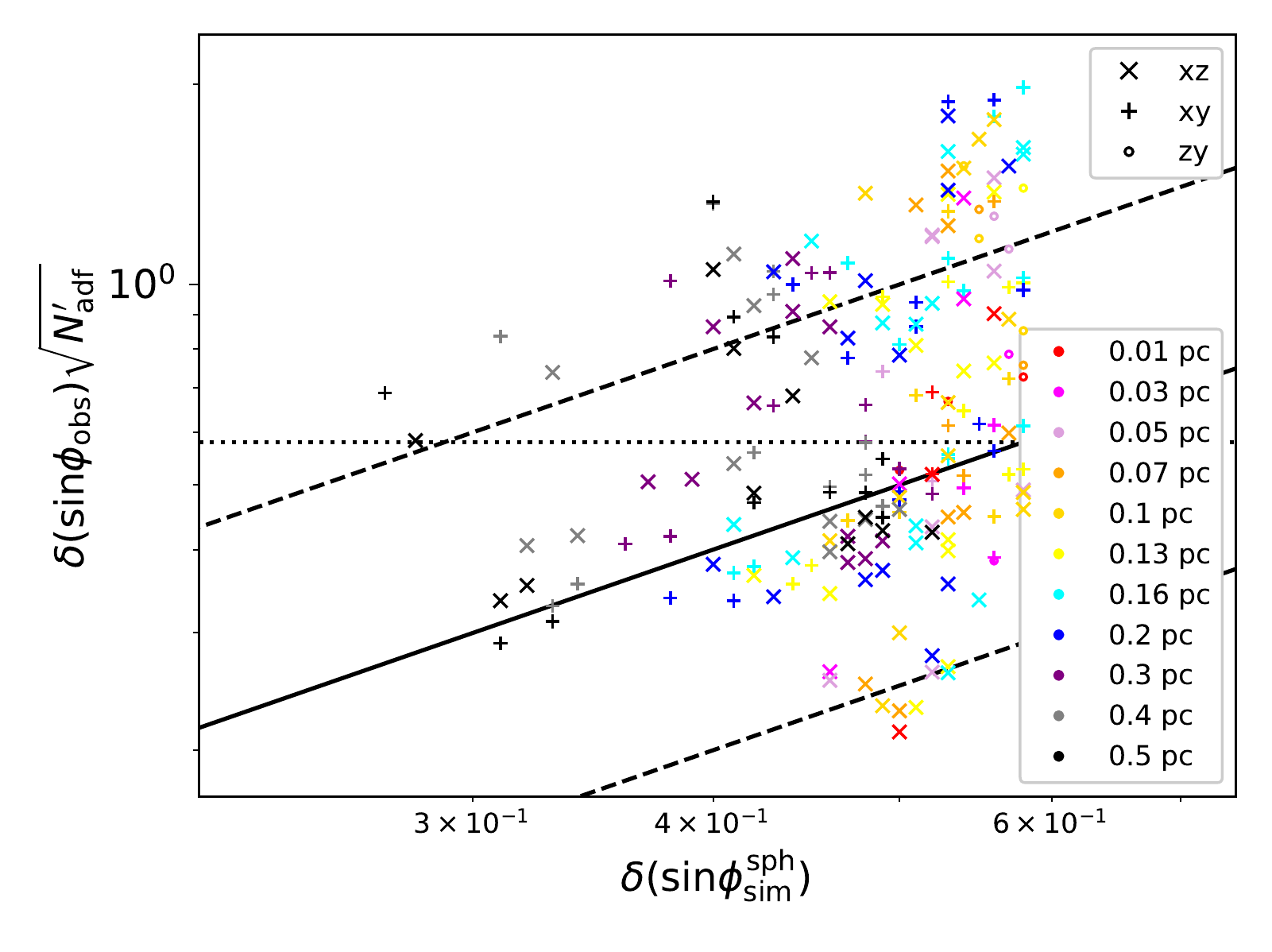}{0.45\textwidth}{c}
 \fig{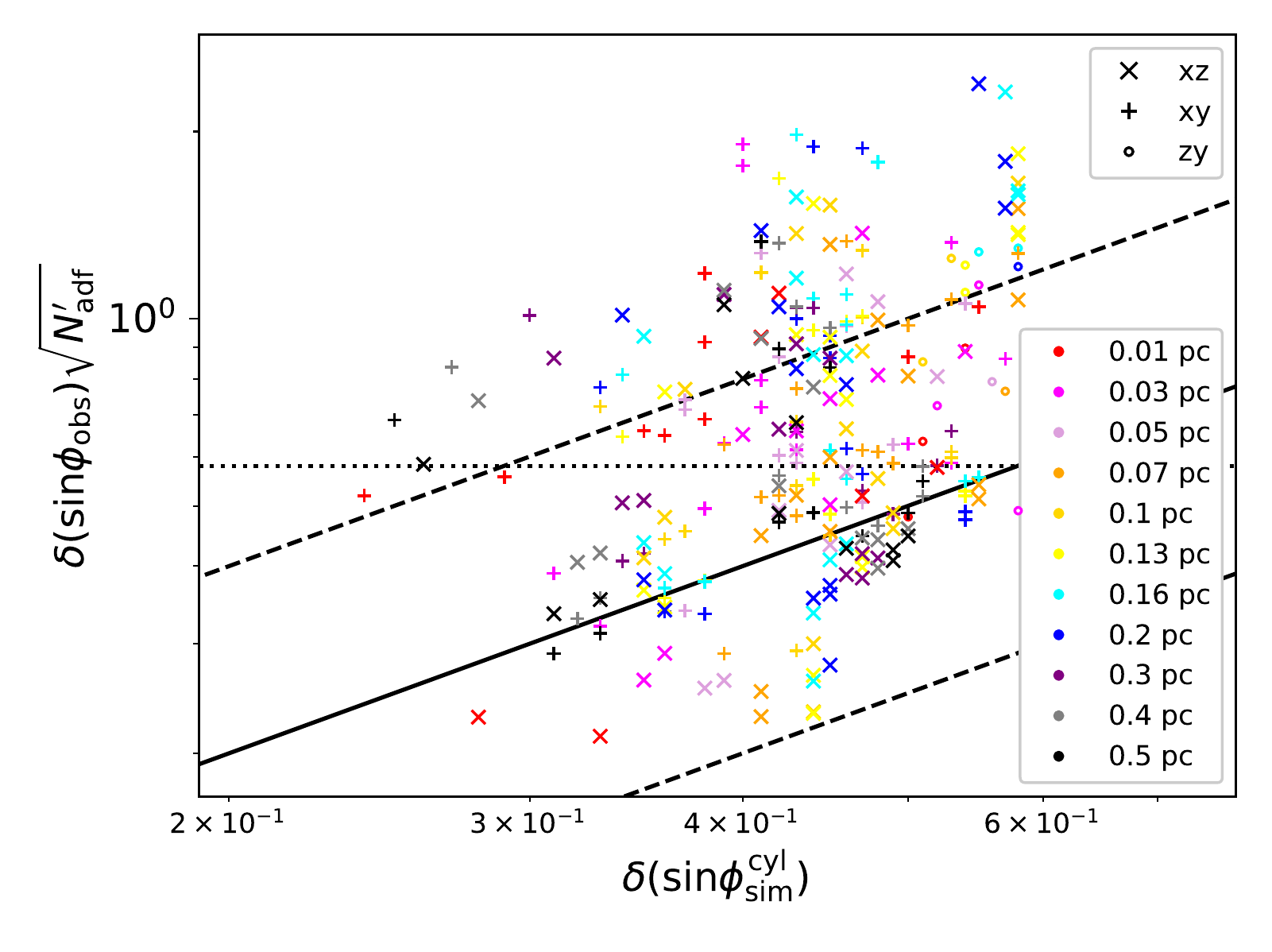}{0.45\textwidth}{d}}
\caption{Observed angular dispersion within circles in polarizaton maps corrected for $\sqrt{N_{\mathrm{adf}}'}$ versus angular dispersions in the simulation space obtained in spheres (left panels) and cylinders (right panels) for all simulation models within different radii with respect to the most massive sink and at three orthogonal projections. Different symbols represent different projections. Different colors correspond to different radii. Solid lines correspond to 1:1 relation. Dashed lines correspond to 1:2 and 2:1 relations. The horizonal dotted lines indicate the expected angular dispersion for random fields ($\sim$52$\degr$ for $\delta \phi$ and $\sim$0.58 for $\delta (\sin \phi)$). \label{fig:angcoradf_angsim}}
\end{figure*}

Alternatively, to investigate whether the ADF method can correctly account for the effect of signal integration along the line of sight, we correct the directly measured angular dispersion in polarization maps by a factor of $\sqrt{N_{\mathrm{adf}}'}$ and compare the corrected angular dispersion with the angular dispersion in simulation grids (see Figure \ref{fig:angcoradf_angsim}). For some data points, the angular dispersion after correction exceeds the value expected for a random field, suggesting that the number of turbulent cells along the line of sight is overestimated by the ADF method and the correction of $\sqrt{N_{\mathrm{adf}}'}$ is inappropriate for these data points. For other data points, the angular dispersion after correction generally agree with the angular dispersions in simulation grids within a factor of $\sim$2. The average ratios between the corrected angular dispersion and the angular dispersion in simulation space for these ``good'' data points are listed in Table \ref{tab:corangangsim}. The average ratios are $\sim$1, suggesting that the ADF method can correctly account for the line-of-sight signal integration effect. However, these good data points only make up 20\%-30\% of our total estimations, which suggests the ADF method might not be applicable to account for the line-of-sight signal integration in the majority of cases. 

Once $(\langle B_{\mathrm{t}}^2 \rangle/ \langle B_0^2\rangle)_{\mathrm{or,tc,si}}^{\mathrm{adf}}$ and $(\langle B_{\mathrm{t}}^2 \rangle/ \langle B^2\rangle)_{\mathrm{or,tc,si}}^{\mathrm{adf}}$ are derived, the plane-of-sky uniform magnetic field strength can be derived by \footnote{It should be noted that there is an approximation in Eq. 26 in \citet{2016ApJ...820...38H}, $B_0 \simeq \sqrt{4\pi \rho} \sigma(v) (\langle B_{\mathrm{t}}^2 \rangle / \langle B^2\rangle)^{-1/2}$ in Gaussian units, where the turbulent-to-total magnetic field strength ratio is used to derive the uniform magnetic field strength. The approximation is only valid when the turbulent field is relatively small compared to the uniform field. The form without approximation is $B_0 \simeq \sqrt{4\pi \rho} \sigma(v) (\langle B_{\mathrm{t}}^2 \rangle / \langle B_0^2\rangle)^{-1/2}$ or $B \simeq \sqrt{4\pi \rho} \sigma(v) (\langle B_{\mathrm{t}}^2 \rangle / \langle B^2\rangle)^{-1/2}$ in Gaussian units. }
\begin{equation}\label{eq:eqbadfu}
B^{\mathrm{u,adf}}_{\mathrm{pos,or,tc,si}} \sim \sqrt{\mu_0 \rho }\delta v_{\mathrm{los}}((\frac{\langle B_{\mathrm{t}}^2 \rangle}{ \langle B_0^2\rangle})_{\mathrm{or,tc,si}}^{\mathrm{adf}})^{-\frac{1}{2}}.
\end{equation}
and the plane-of-sky total magnetic field strength can be derived by
\begin{equation}\label{eq:eqbadftot}
B^{\mathrm{tot,adf}}_{\mathrm{pos,or,tc,si}} \sim \sqrt{\mu_0 \rho }\delta v_{\mathrm{los}}((\frac{\langle B_{\mathrm{t}}^2 \rangle}{ \langle B^2\rangle})_{\mathrm{or,tc,si}}^{\mathrm{adf}})^{-\frac{1}{2}}.
\end{equation}
The plane-of-sky uniform magnetic field strength $B^{\mathrm{u,adf}}_{\mathrm{pos,or,tc}}$ and the plane-of-sky total magnetic field strength $B^{\mathrm{tot,adf}}_{\mathrm{pos,or,tc}}$ without accounting for the line-of-sight signal integration effect can be derived in similar ways.

\subsubsection{The \citet{2016ApJ...821...21C} method}\label{sec:cy16}

As mentioned before, the amount of angular dispersion measured in polarization maps would be reduced by a factor of $\sqrt{N}$ compared to the intrinsic angular dispersion, where $N$ is the number of independent turbulent cells along the line of sight \citep{2009ApJ...696..567H, 2016ApJ...821...21C}. The ADF method analytically solves for the plane-of-sky turbulent correlation scale from the distribution of plane-of-sky polarization angles and derives the number of line-of-sight turbulent cells with the assumption that the line-of-sight turbulent correlation scale is identical to the plane-of-sky turbulent correlation scale. We test the ADF method in Section \ref{sec:adf} and find that the ADF method only correctly account for this effect for a small portion of the total estimations. Alternatively, \citet{2016ApJ...821...21C} proposed an approach (hereafter the CY16 method) to estimate $N$ from the ratio between the average line-of-sight velocity dispersion ($\delta v_{\mathrm{los}}$) and the standard deviation of centroid velocities ($\delta V_c$) in the considered region: 
\begin{equation}
    N_{\mathrm{cy16}} = (\frac{\delta v_{\mathrm{los}}}{\delta V_c})^2.
\end{equation}
We apply the CY16 method on our simulation data at different radii at three different projections for all simulation models. We adopt the mass-weighted standard deviation of centroid velocities within cylinders as $\delta V_c$. The $\delta v_{\mathrm{los}}$ is also estimated within cylinders. We derive $N_{\mathrm{cy16}}$ and use it to correct for the angular dispersion measured in polarization maps. 

\begin{figure*}[!htbp]
 \gridline{\fig{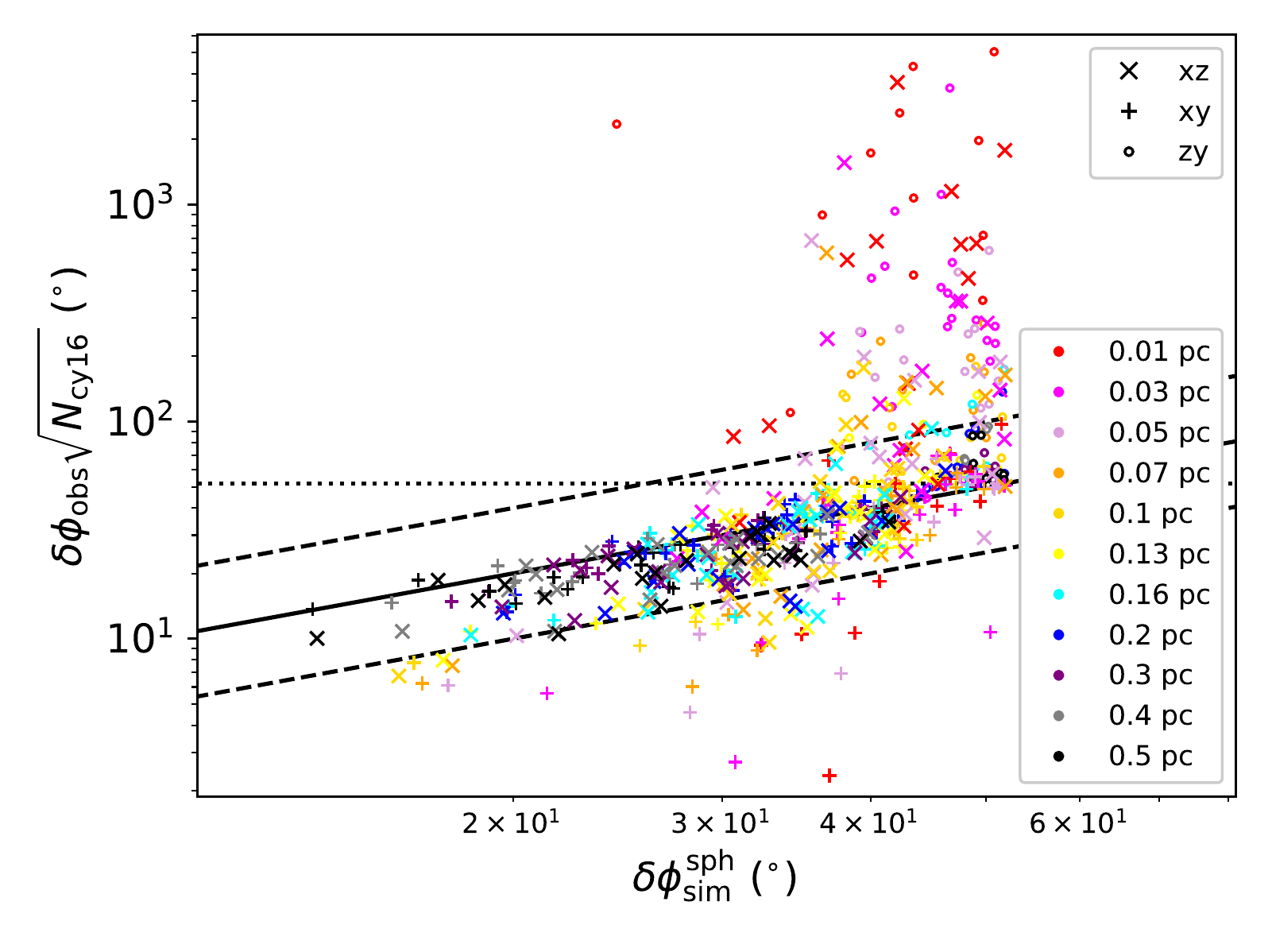}{0.45\textwidth}{a}
 \fig{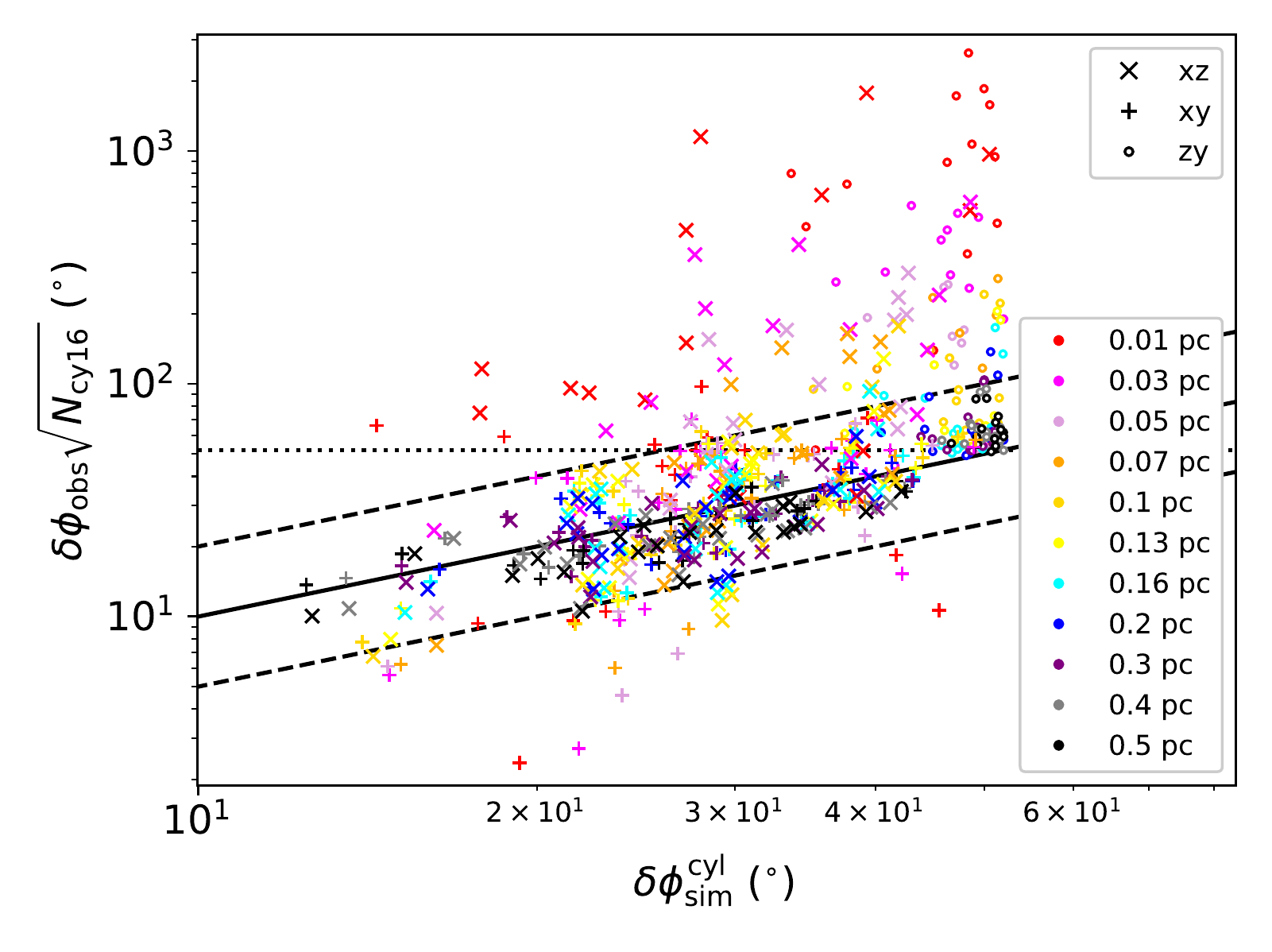}{0.45\textwidth}{b}}
  \gridline{\fig{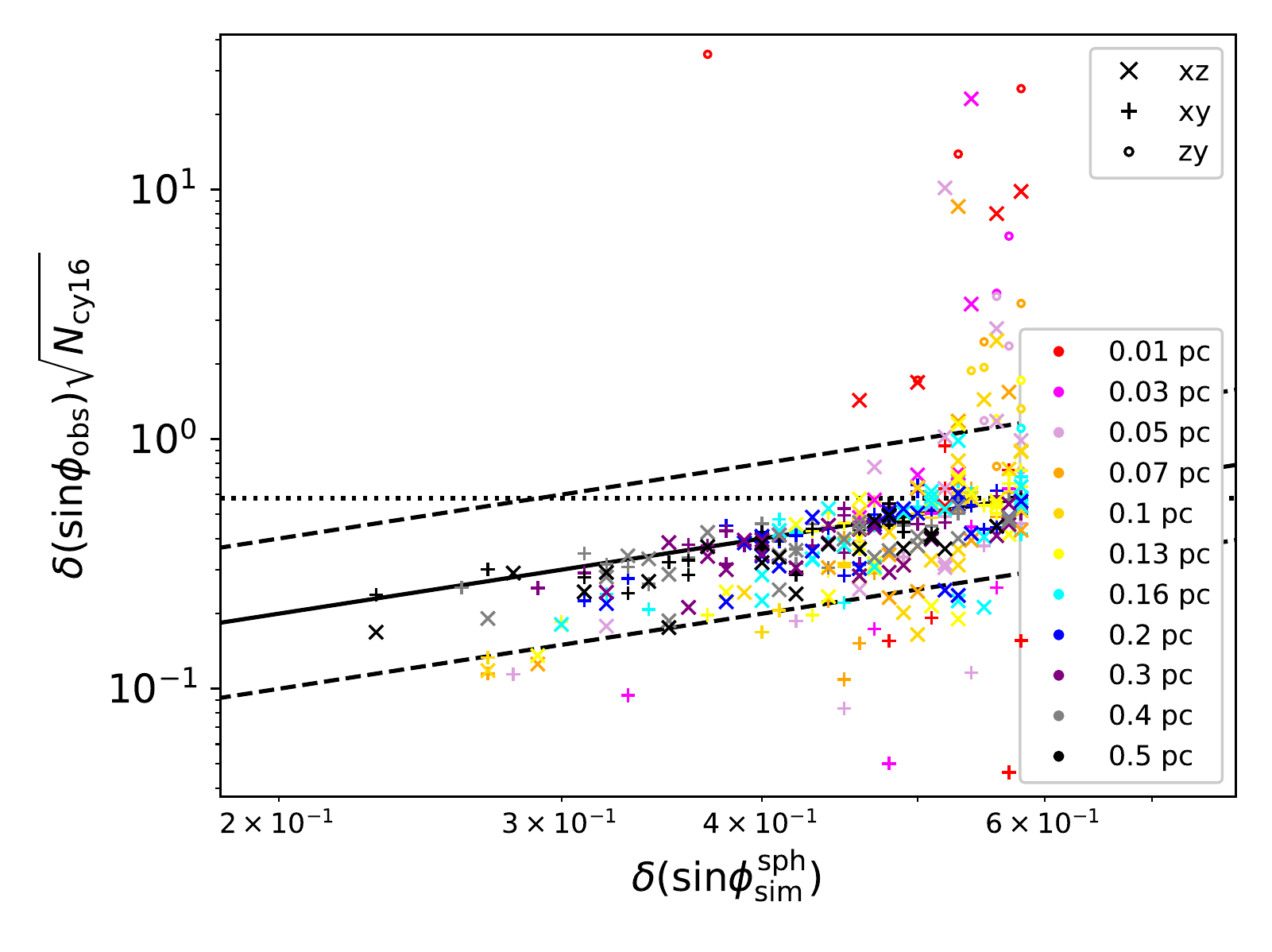}{0.45\textwidth}{c}
 \fig{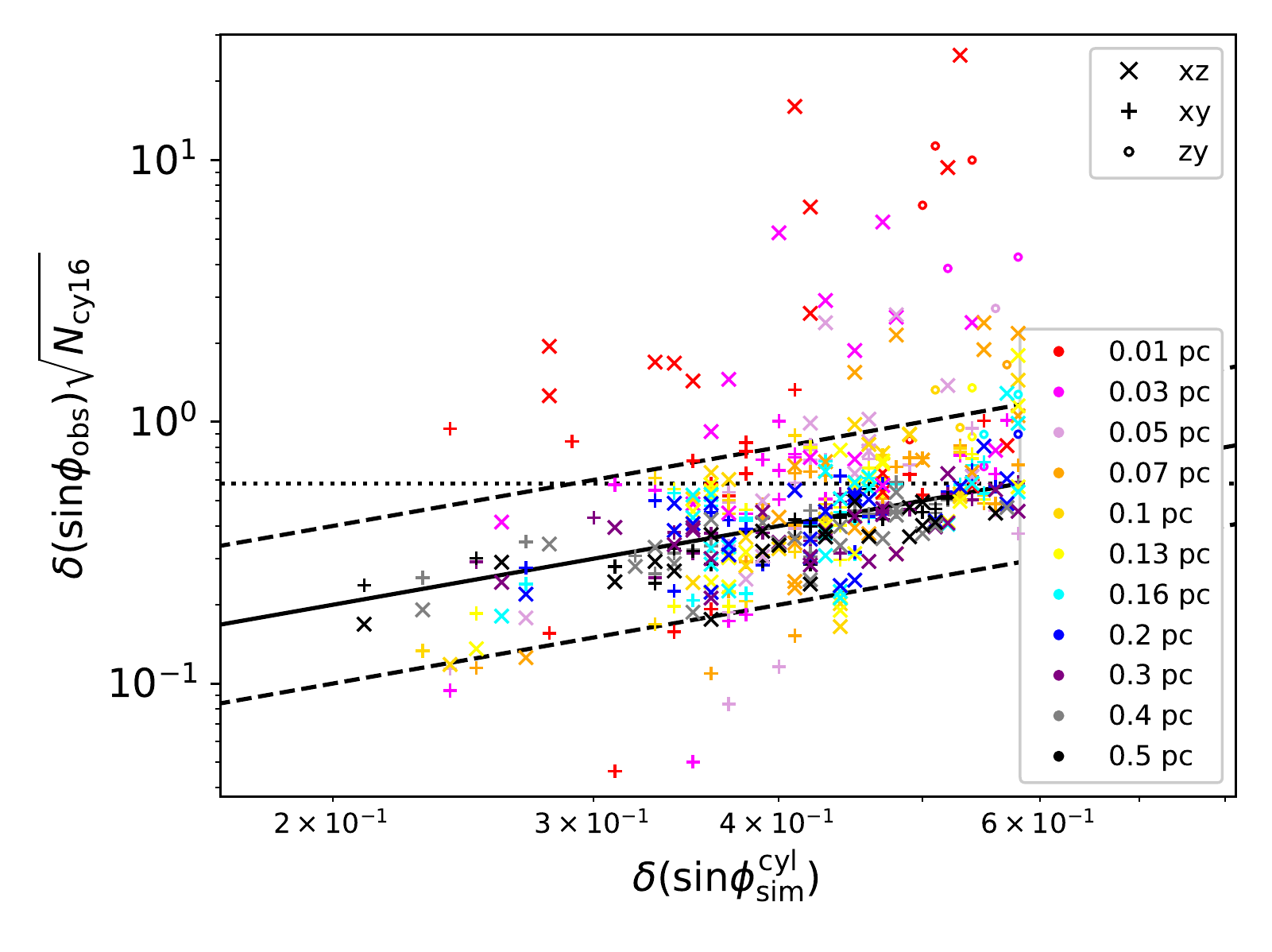}{0.45\textwidth}{d}}
\caption{Observed angular dispersion within circles in polarization maps corrected for $\sqrt{N_{\mathrm{cy16}}}$ versus angular dispersions in the simulation space obtained in spheres (left panels) and cylinders (right panels) for all simulation models within different radii with respect to the most massive sink and at three orthogonal projections. Different symbols represent different projections. Different colors correspond to different radii. Solid lines correspond to 1:1 relation. Dashed lines correspond to 1:2 and 2:1 relations. The horizonal dotted lines indicate the expected angular dispersion for random fields ($\sim$52$\degr$ for $\delta \phi$ and $\sim$0.58 for $\delta (\sin \phi)$). \label{fig:angcor_angsim}}
\end{figure*}

Figure \ref{fig:angcor_angsim} shows the relation between the corrected angular dispersions and the angular dispersions in the simulation space. The data points show large scatters at radii $<$0.1 pc. At radii $>$0.1 pc, the corrected angular dispersion and the angular dispersion in simulation grids are mostly consistent with each other within a factor of 2 and most corrected angular dispersions are below the value expected for random fields. 

\begin{deluxetable}{ccc}[t!]
\tablecaption{Average ratios between the corrected angular dispersions in polarization maps and angular dispersions in the simulation space for data points at radii$>0.1$ pc. Values in the parenthesis are the relative uncertainty. \label{tab:corangangsim}}
\tablecolumns{3}
\tablewidth{0pt}
\tablehead{
\colhead{Average ratios} &
\colhead{Spheres\tablenotemark{a}} &
\colhead{Cylinders\tablenotemark{a}} 
}
\startdata
$\sqrt{N_{\mathrm{adf}}'} \delta \phi_{\mathrm{obs}}$ /$\delta \phi_{\mathrm{sim}}$\tablenotemark{b}  &   1.00(35\%) &   1.06(37\%)\\ 
$\sqrt{N_{\mathrm{adf}}'} \delta (\sin \phi_{\mathrm{obs}})$/$\delta (\sin \phi_{\mathrm{sim}})$\tablenotemark{c} &   0.94(20\%) &   1.00(18\%)\\ \hline
$\sqrt{N_{\mathrm{cy16}}} \delta \phi_{\mathrm{obs}}$ /$\delta \phi_{\mathrm{sim}}$\tablenotemark{b}  &   0.86(23\%) &   0.94(27\%)\\ 
$\sqrt{N_{\mathrm{cy16}}} \delta (\sin \phi_{\mathrm{obs}})$/$\delta (\sin \phi_{\mathrm{sim}})$\tablenotemark{c} &   0.87(20\%) &   0.94(23\%)\\ 
\enddata
\tablenotetext{a}{Indicates whether the angular dispersion in the simulation space are estimated in spheres or cylinders.}
\tablenotetext{b}{Data points with corrected angular dispersion$>52\degr$ are not included.}
\tablenotetext{c}{Data points with corrected angular dispersion$>0.58$ are not included.}
\end{deluxetable}

The average ratios between the angular dispersion corrected by the CY16 method and the angular dispersion in simulation space at radii $>$0.1 are listed in Table \ref{tab:corangangsim}. We notice that the relative uncertainty of the average ratios corrected by the CY16 method is slightly lower than the relative uncertainty of the average ratios without correction (see Table \ref{tab:angangsim}), suggesting that the CY16 method provides a slightly better estimation than without correction. 

\begin{figure}[!htbp]
 \gridline{\fig{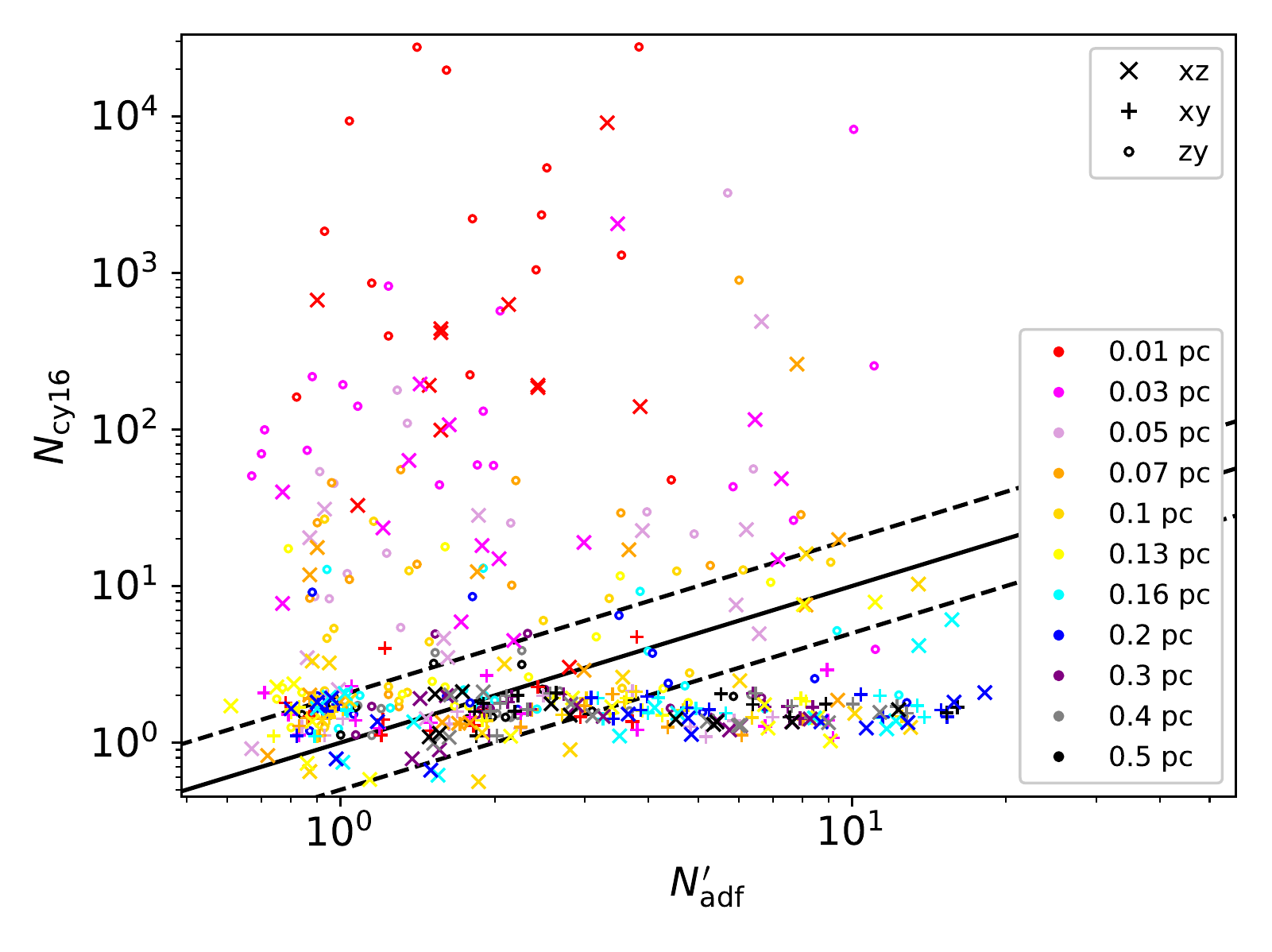}{0.45\textwidth}{}}
\caption{Comparison between $N_{\mathrm{cy16}}$ and $N_{\mathrm{adf}}'$. Different symbols represent different projections. Different colors correspond to different radii. Solid lines correspond to 1:1 relation. Dashed lines correspond to 1:2 and 2:1 relations.   \label{fig:ncy16_nadf}}
\end{figure}

In Figure \ref{fig:ncy16_nadf} we compare the number of line-of-sight turbulent cells derived from the ADF method and the CY16 method. For most data points at radii $<$0.1 pc, $N_{\mathrm{cy16}}$ could be significantly larger than $N_{\mathrm{adf}}'$. On the other hand, the value of $N_{\mathrm{cy16}}$ shows smaller variations than $N_{\mathrm{adf}}'$ at radii $>$0.1 pc. 

\subsubsection {Smoothing and filtering effects of synthetic interferometric observations}

As revealed by some previous numerical studies \citep{2001ApJ...561..800H, 2008ApJ...679..537F}, the effect of finite beam resolution of telescopes tends to underestimate the angular dispersion and thus overestimate the magnetic field strength. When it comes to inteferometric observations, there arise additional issues about the spatial filtering of large-scale structures. Here we investigate how the combination of the beam-smoothing effect and the filtering effect of interferometers affect the measured angular dispersion. It should be noted that the analyses carried out in previous subsections (Sections \ref{sec:compangb}, \ref{sec:adf}, and \ref{sec:cy16}) are on perfect synthetic skies, while the analyses in this subsection are carried on synthetic ALMA polarizaton observations. The synthetic ALMA polarizaton maps are gridded with pixel sizes of 1/5 of the beam sizes. 

\begin{figure*}[!htbp]
 \gridline{\fig{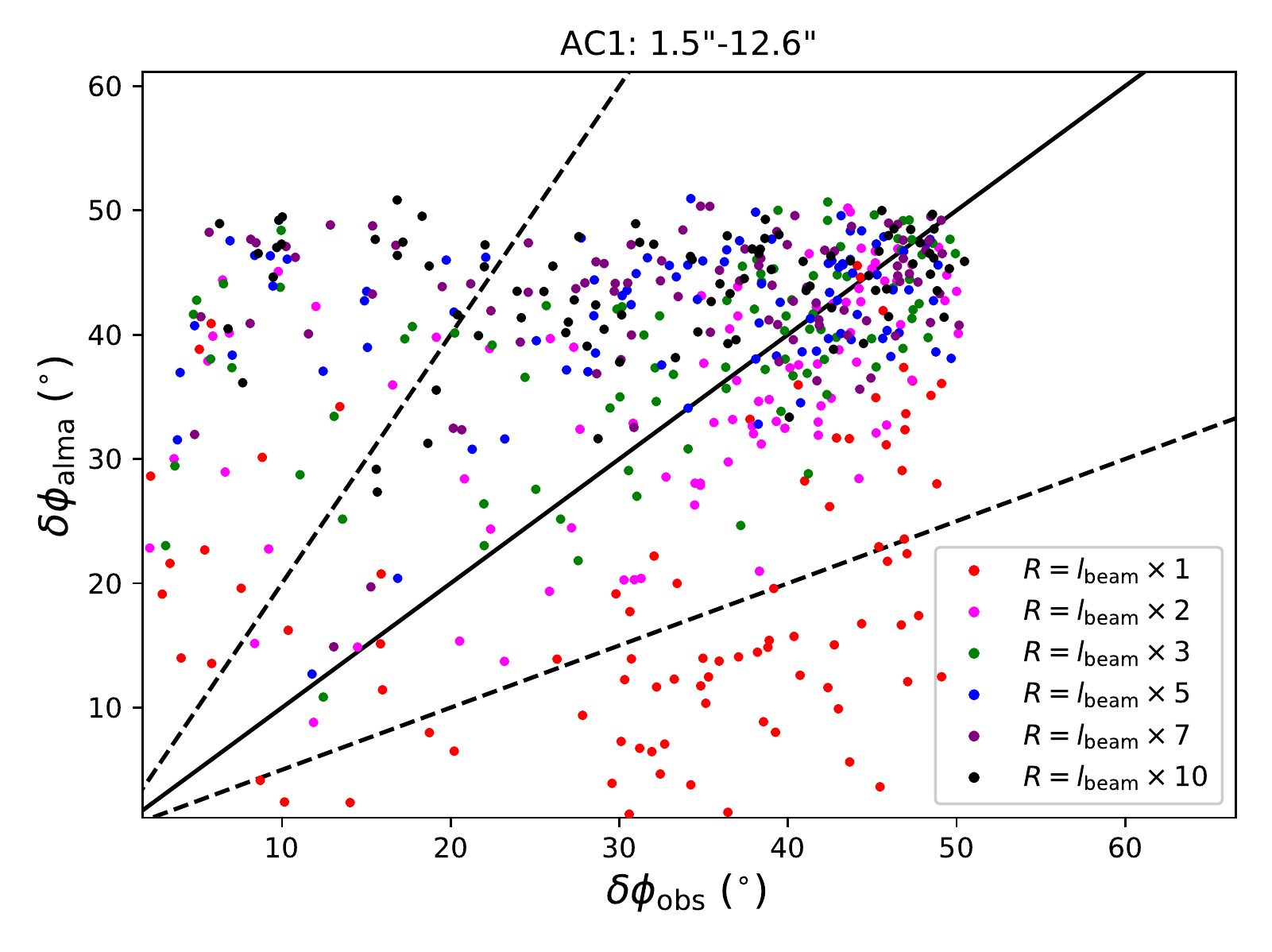}{0.45\textwidth}{a}
 \fig{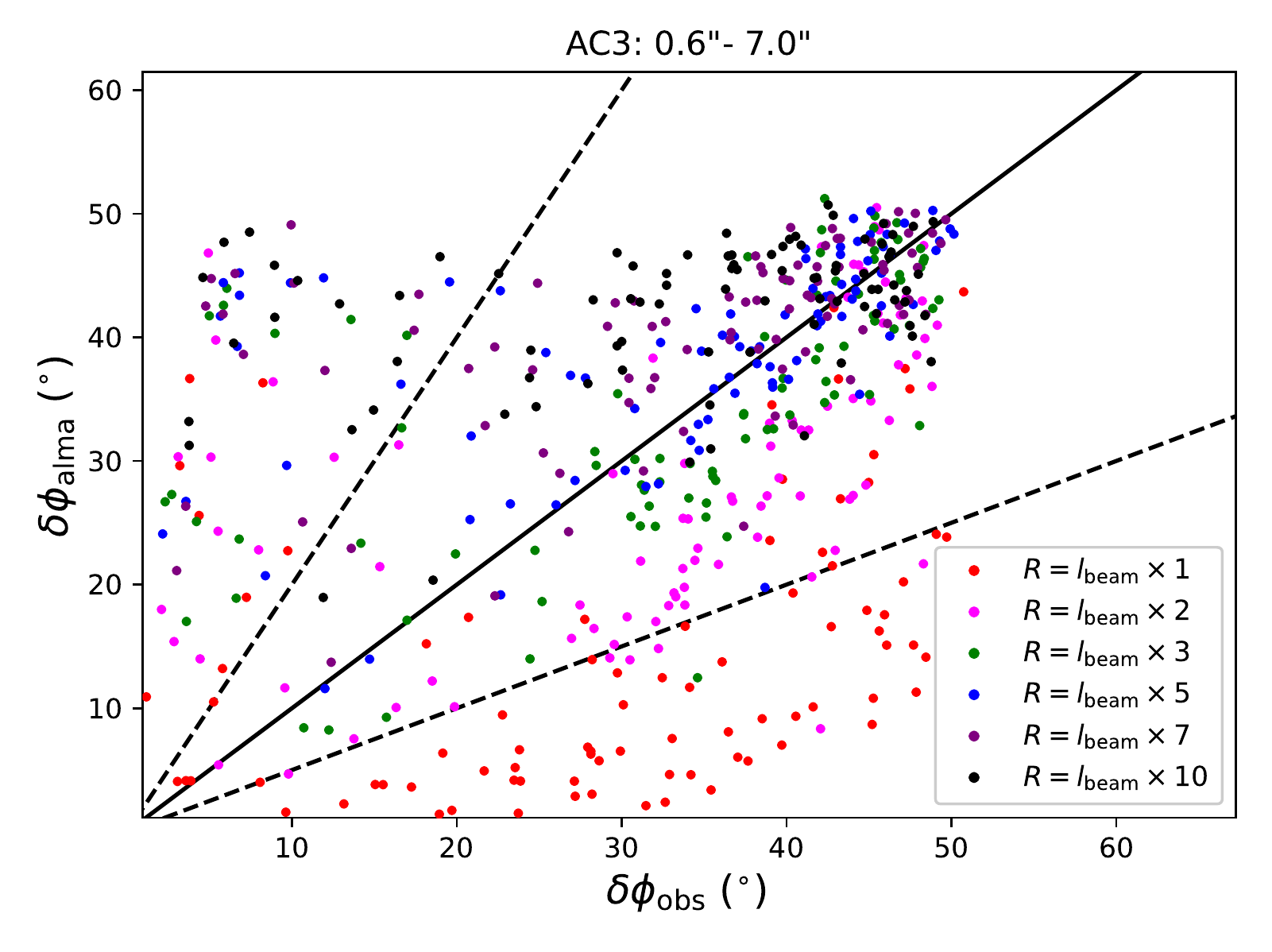}{0.45\textwidth}{b}}
  \gridline{  \fig{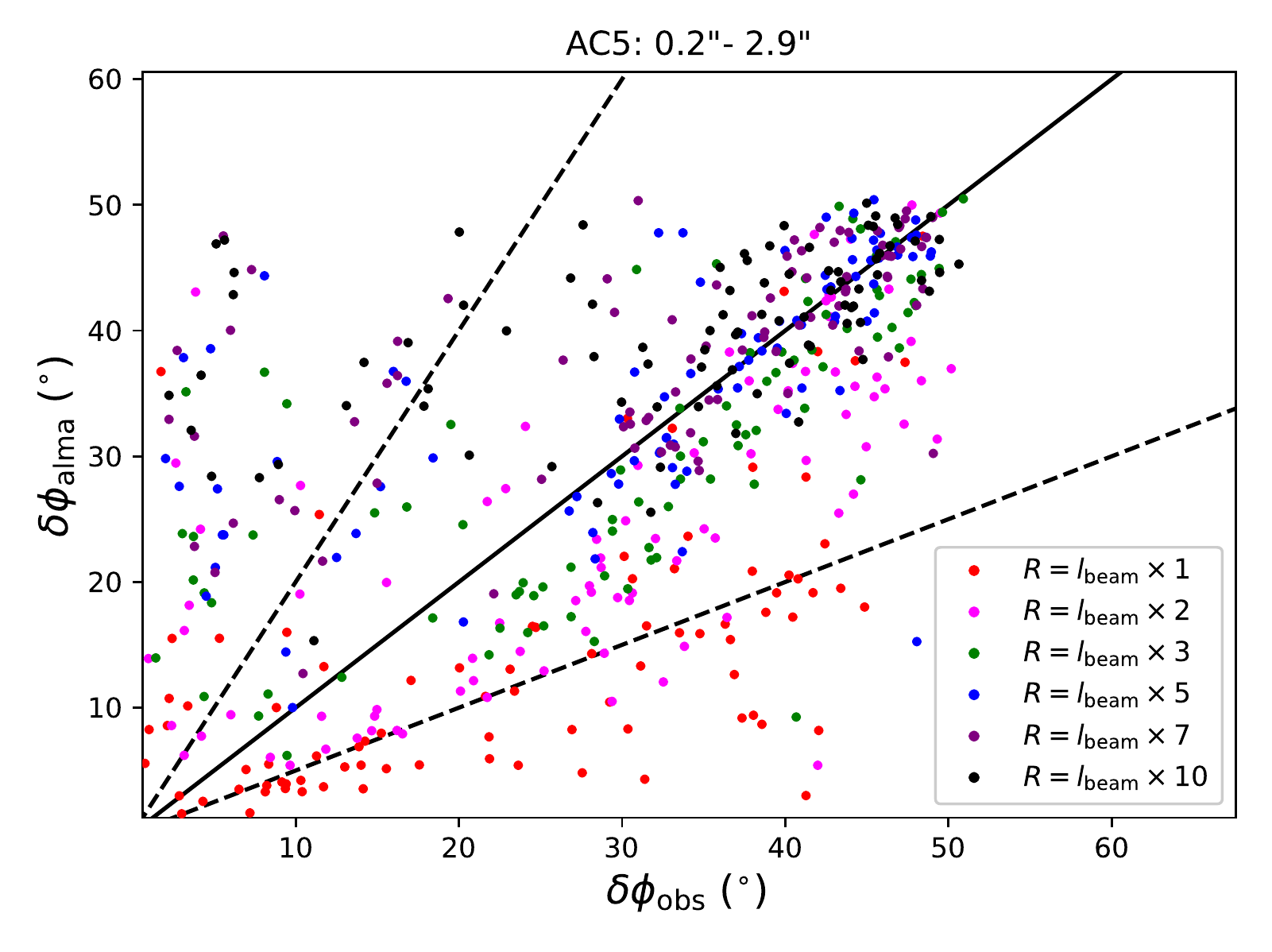}{0.45\textwidth}{c}}
\caption{Comparison between the directly measured angular dispersion in raw synthetic polarization maps, $\delta\phi_{\mathrm{obs}}$, and the directly measured angular dispersion in synthetic ALMA polarization maps, $\delta\phi_{\mathrm{alma}}$. Different colors correspond to different radii in units of multiples of the synthesized beam size. Solid lines correspond to 1:1 relation. Dashed lines correspond to 1:2 and 2:1 relations. The beam size and the filtering scale of each configuration of ALMA are indicated in the title.  \label{fig:ang_angalma}}
\end{figure*}

Firstly, we directly measure the dispersion of magnetic field position angles in synthetic ALMA maps ($\delta\phi_{\mathrm{alma}}$) in a similar way to the estimation of $\delta \phi_{\mathrm{obs}}$ in Section \ref{sec:angobs_angsim}. Only position angles at locations where $I>0$ are considered in the estimation. Figure \ref{fig:ang_angalma} shows the relation between $\delta\phi_{\mathrm{alma}}$ and $\delta \phi_{\mathrm{obs}}$. When the radius of the concerned area, $R$, is comparable to the synthesized beam size $l_{\mathrm{beam}}$ (i.e., $R \sim l_{\mathrm{beam}}$), $\delta\phi_{\mathrm{alma}}$ is overall smaller than $\delta\phi_{\mathrm{obs}}$, which means the synthetic ALMA observations would underestimate the angular dispersion when the concerned area is barely resolved. This could be easily explained by the lost of perturbation information because of beam-smoothing. On the other hand, for a significant amount of data points, the $\delta\phi_{\mathrm{alma}}$ are larger than $\delta\phi_{\mathrm{obs}}$. When $\theta_{\mathrm{obs}}$ is small, the synthetic ALMA observations could significantly overestimate the angular dispersion. As $R$ approaches the interferometer filtering scale $l_{\mathrm{filter}}$ ($l_{\mathrm{filter}} \sim $ 8.5, 9.8, 11.6, 12.2, and 12.4 $\times l_{\mathrm{beam}}$ for AC1, AC2, AC3, AC4, and AC5, respectively), the overestimation of angular dispersion is more obvious. This overestimation might be because the filtering effect of interferometric observations has altered the large-scale magnetic field structure (see Figures \ref{fig:mhdmapmu1} and \ref{fig:mhdmapmu1zyalma} for examples), or because there is not enough number of samples within the area to achieve statistical significance due to the excluding of some angle samples at locations where $I<0$. 

The ADF method is the only method that analytically takes into account the two observational effects. In the most complicated form of the ADF \citep{2016ApJ...820...38H}, it takes into account the contribution from ordered field structure, the turbulent correlation effect, the signal integration effect, the beam-smoothing effect, and the interferometer filtering effect. The ADF is given by \citep{2016ApJ...820...38H}:
\begin{align}
1 - \langle \cos \lbrack \Delta \Phi (l)\rbrack \rangle \simeq a_2\arcmin l^2 + \left[ \frac{N_{\mathrm{adf}}}{1+\frac{N_{\mathrm{adf}}}{(\langle B_{\mathrm{t}}^2 \rangle/ \langle B_0^2\rangle)_{\mathrm{or,tc,si,bs,if}}^{\mathrm{adf}}} } \right] \nonumber \\
\times \Bigg\{ \frac{1}{N_1} \left[ 1 - e^{-l^2/2(l_\delta^2+2W_1^2)}\right] \nonumber \\+ \frac{1}{N_2} \left[ 1 - e^{-l^2/2(l_\delta^2+2W_2^2)}\right] \nonumber \\- \frac{2}{N_{12}} \left[ 1 - e^{-l^2/2(l_\delta^2+W_1^2+W_2^2)}\right] \Bigg\},
\end{align}
where $W_1$ and $W_2$ are the standard deviation of the Gaussian profiles of the synthesized beam and the large-scale filtering (i.e., the FWHM divided by $\sqrt{8 \ln{2}}$), respectively. $N_{\mathrm{adf}}$ is the number of turbulent cells along the line of sight probed by the telescope beam given by: 
\begin{equation}
N_1 = \frac{(l_\delta^2 + 2W_1^2)l_\Delta}{\sqrt{2\pi}l_\delta^3},
\end{equation}
\begin{equation}
N_2 = \frac{(l_\delta^2 + 2W_2^2)l_\Delta}{\sqrt{2\pi}l_\delta^3},
\end{equation}
\begin{equation}
N_{12} = \frac{(l_\delta^2 + W_1^2 + W_2^2)l_\Delta}{\sqrt{2\pi}l_\delta^3},
\end{equation}
\begin{equation}
N_{\mathrm{adf}} = \left( \frac{1}{N_1} + \frac{1}{N_2} - \frac{2}{N_{12}} \right)^{-1}.
\end{equation}

Similar to the approach in Section \ref{sec:adf}, we derive the ADFs within different radii in the synthetic ALMA maps, fit the ADFs, and obtain the best fit via reduced $\chi^2$ minimization. When $R\sim l_{\mathrm{beam}}$ (i.e., the concerned area is barely resolved), the fitting of the ADFs does not converge. Once the turbulent-to-ordered magnetic field energy ratio taking into all the effects, $(\langle B_{\mathrm{t}}^2 \rangle/ \langle B_0^2\rangle)_{\mathrm{or,tc,si,bs,if}}^{\mathrm{adf}}$, is derived, the turbulent-to-ordered magnetic field energy ratio without accounting for the signal integration along the line of sight can be derived by $(\langle B_{\mathrm{t}}^2 \rangle/ \langle B_0^2\rangle)_{\mathrm{or,tc,bs,if}}^{\mathrm{adf}} = (\langle B_{\mathrm{t}}^2 \rangle/ \langle B_0^2\rangle)_{\mathrm{or,tc,si,bs,if}}^{\mathrm{adf}} / N_{\mathrm{adf}}$. 

\begin{figure*}[!htbp]
 \gridline{\fig{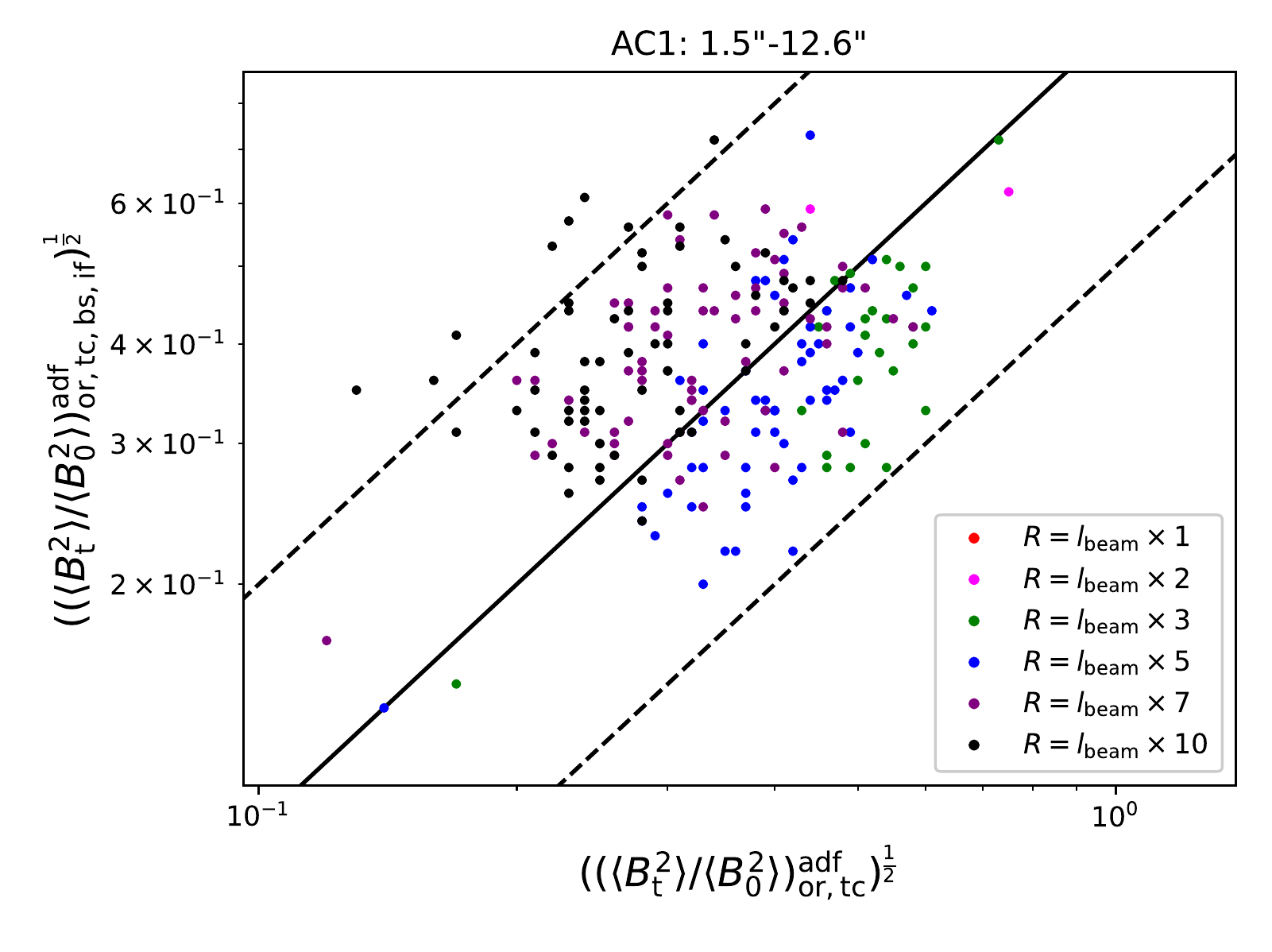}{0.45\textwidth}{a}
 \fig{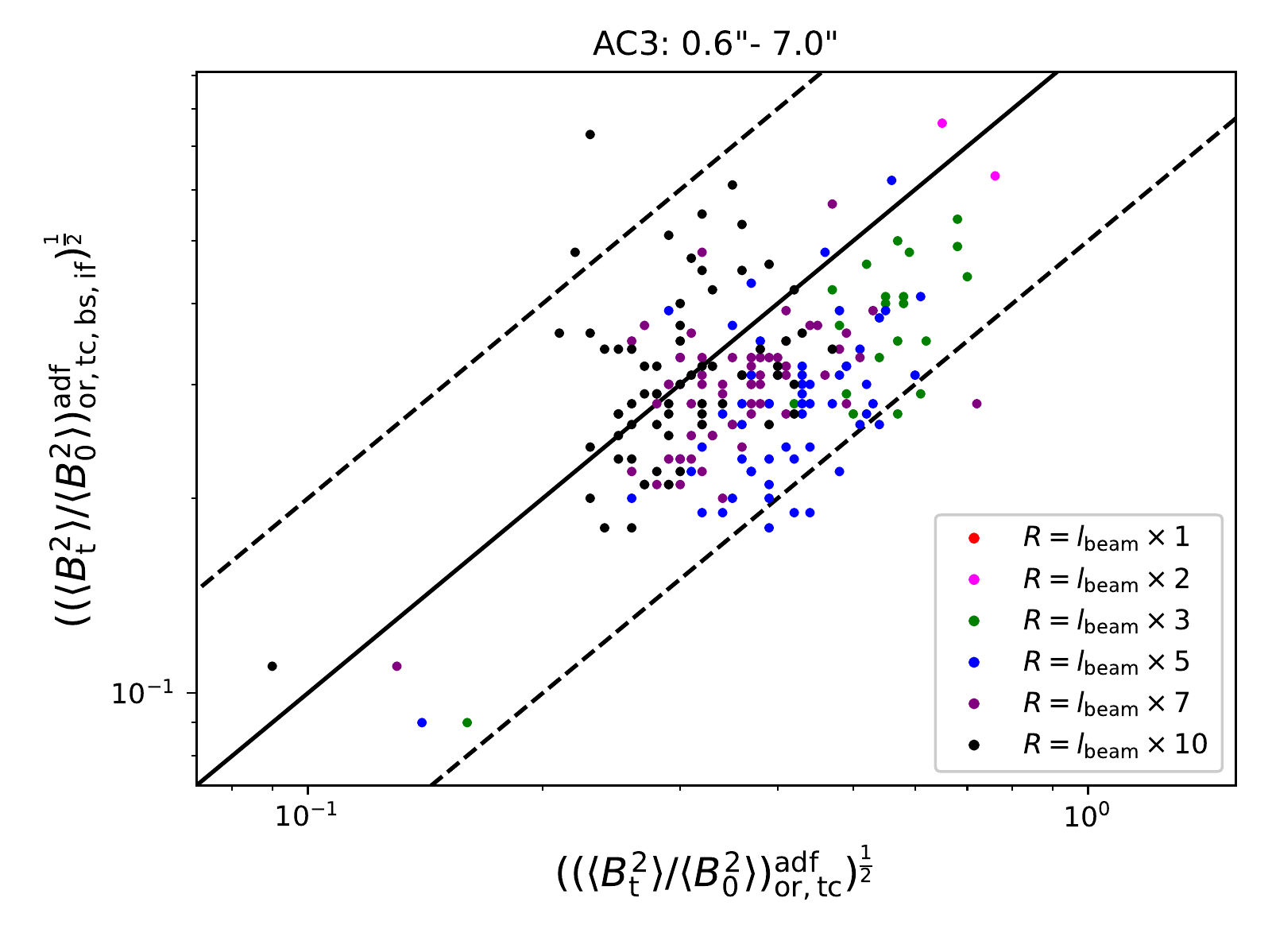}{0.45\textwidth}{b}}
 \gridline{  \fig{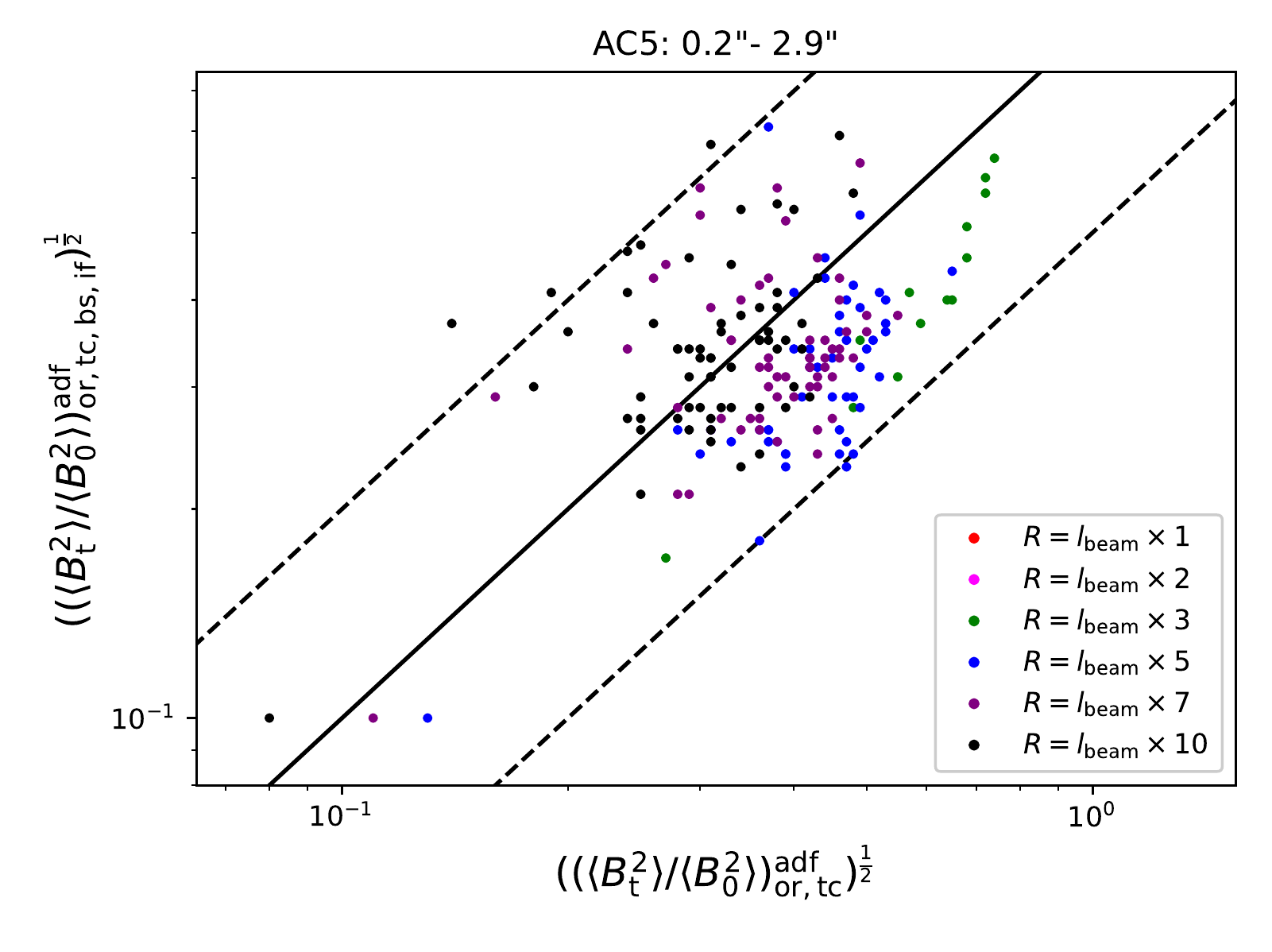}{0.45\textwidth}{c}
  \fig{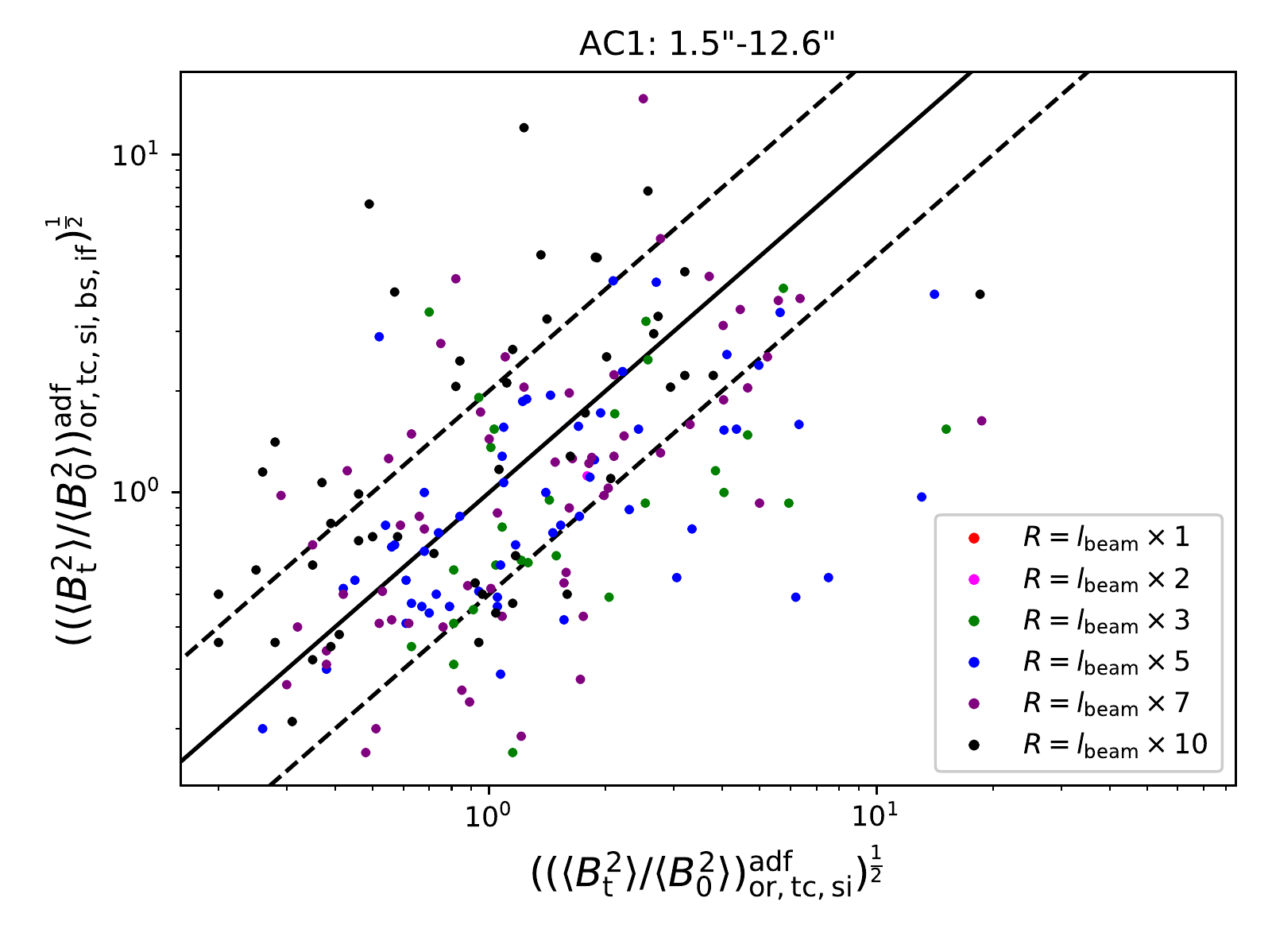}{0.45\textwidth}{d}}
 \gridline{\fig{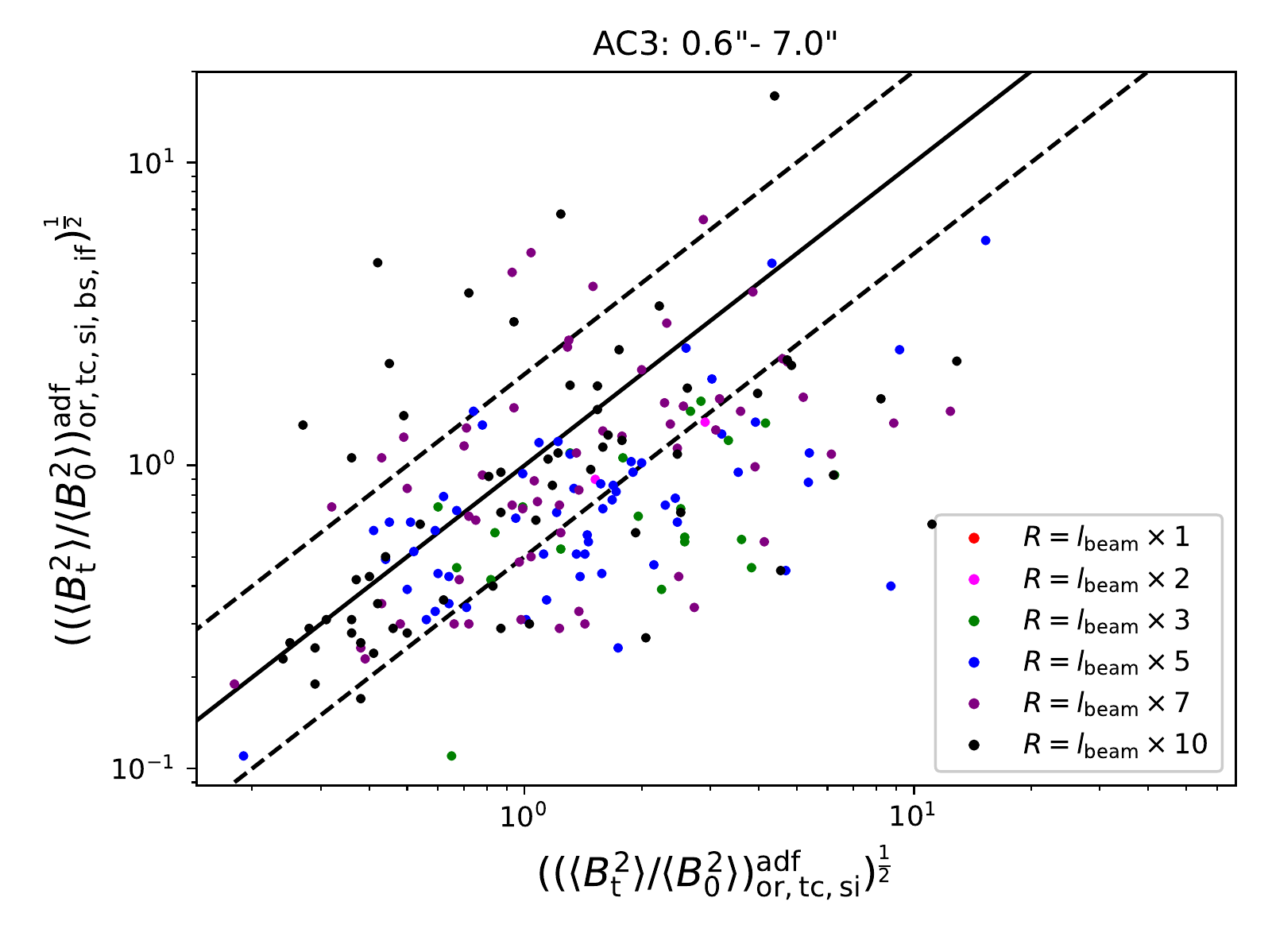}{0.45\textwidth}{e}
 \fig{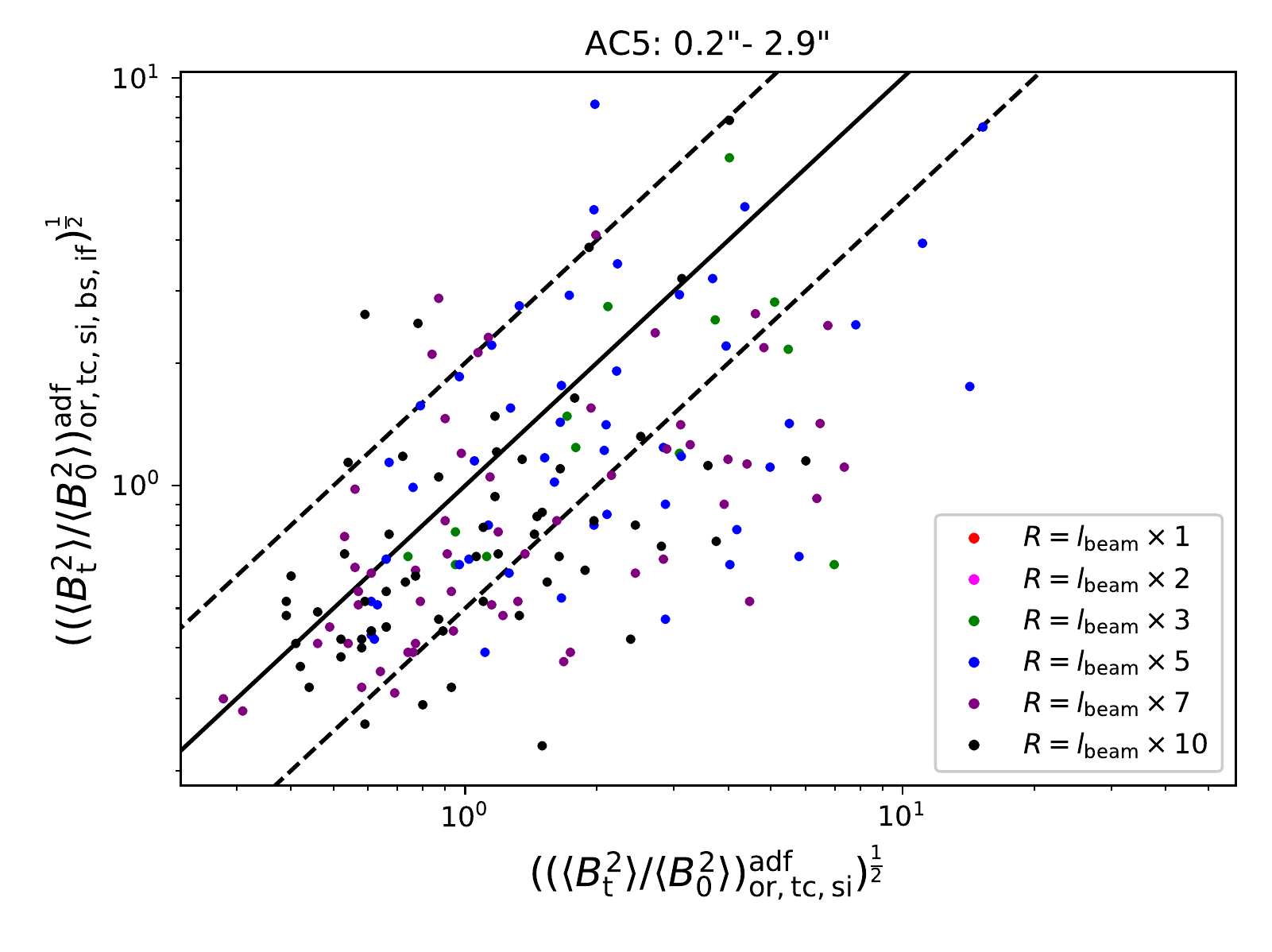}{0.45\textwidth}{f}}
\caption{(a)-(c). Comparison between $(\langle B_{\mathrm{t}}^2 \rangle/ \langle B_0^2\rangle)_{\mathrm{or,tc}}^{\mathrm{adf}}$ and $(\langle B_{\mathrm{t}}^2 \rangle/ \langle B_0^2\rangle)_{\mathrm{or,tc,bs,if}}^{\mathrm{adf}}$. (d)-(f). Comparison between $(\langle B_{\mathrm{t}}^2 \rangle/ \langle B_0^2\rangle)_{\mathrm{or,tc,si}}^{\mathrm{adf}}$ and $(\langle B_{\mathrm{t}}^2 \rangle/ \langle B_0^2\rangle)_{\mathrm{or,tc,si,bs,if}}^{\mathrm{adf}}$. Different colors correspond to different radii in units of multiples of the synthesized beam size. Solid lines correspond to 1:1 relation. Dashed lines correspond to 1:2 and 2:1 relations.  \label{fig:BtB0_BtB0alma}}
\end{figure*}

Figures \ref{fig:BtB0_BtB0alma} (a), (b), and (c) compare the $(\langle B_{\mathrm{t}}^2 \rangle/ \langle B_0^2\rangle)_{\mathrm{or,tc}}^{\mathrm{adf}}$ derived by fitting ADFs of raw (unsmoothed) synthetic polarization maps (see Section \ref{sec:adf}) with the $(\langle B_{\mathrm{t}}^2 \rangle/ \langle B_0^2\rangle)_{\mathrm{or,tc,bs,if}}^{\mathrm{adf}}$ derived by fitting ADFs of synthetic ALMA polarization maps. Generally the $(\langle B_{\mathrm{t}}^2 \rangle/ \langle B_0^2\rangle)_{\mathrm{or,tc}}^{\mathrm{adf}}$ and $(\langle B_{\mathrm{t}}^2 \rangle/ \langle B_0^2\rangle)_{\mathrm{or,tc,bs,if}}^{\mathrm{adf}}$ agree with each other within a factor of two, which indicates that the ADF method does correctly take into account the beam-smoothing effect and the interferometric filtering effect.  

Figures \ref{fig:BtB0_BtB0alma} (d), (e), and (f) compare the $(\langle B_{\mathrm{t}}^2 \rangle/ \langle B_0^2\rangle)_{\mathrm{or,tc,si}}^{\mathrm{adf}}$ derived by fitting ADFs of raw synthetic polarization maps with the $(\langle B_{\mathrm{t}}^2 \rangle/ \langle B_0^2\rangle)_{\mathrm{or,tc,si,bs,if}}^{\mathrm{adf}}$ derived by fitting ADFs of synthetic ALMA polarization maps. Generally the relation between $(\langle B_{\mathrm{t}}^2 \rangle/ \langle B_0^2\rangle)_{\mathrm{or,tc,si}}^{\mathrm{adf}}$ and $(\langle B_{\mathrm{t}}^2 \rangle/ \langle B_0^2\rangle)_{\mathrm{or,tc,si,bs,if}}^{\mathrm{adf}}$ shows large scatters, which indicates that the ADF method may not simultaneously correctly account for the line-of-sight signal integration, beam-smoothing, and interferometric filtering effects. This may be because the beam resolution and the pixel size are insufficient to trace the turbulent correlation scale (see Appendix \ref{app:gridsize}), the turbulent correlation scale is anisotropic \citep{1995ApJ...438..763G}, the turbulent correlation scales at different depth are not uniform, the information of turbulent correlation scale is lost due to observational uncertainty, or because of some other reasons. 

\subsection{Equipartition between the turbulent kinetic energy and the turbulent magnetic energy}\label{sec:equi}

The DCF method assumes the equipartition between the turbulent kinetic energy $E_{\mathrm{K,3d}}^{\mathrm{t}}$ and the turbulent magnetic energy $E_\mathrm{B,3d}^{\mathrm{t}}$.  \citet{2008ApJ...679..537F} proposed that the energy equipartition assumption is fulfilled in sub-Alfv\'{e}nic (i.e., the turbulent kinetic energy $E_{\mathrm{K,3d}}^{\mathrm{t}}$ is smaller than the uniform magnetic field energy $E_\mathrm{B,3d}^{\mathrm{u}}$) cases and found that a super-Alfv\'{e}nic (i.e., $E_{\mathrm{K,3d}}^{\mathrm{t}} > E_\mathrm{B,3d}^{\mathrm{u}}$) model with initial $E_\mathrm{B,3d}^{\mathrm{t}} < E_{\mathrm{K,3d}}^{\mathrm{t}}$ can increase its $E_\mathrm{B,3d}^{\mathrm{t}}$ as time evolves and reach a state close to the energy equipartition ($E_\mathrm{B,3d}^{\mathrm{t}} \sim E_{\mathrm{K,3d}}^{\mathrm{t}}$) after 0.3 $\times$ the crossing timescale. 

\begin{figure}[!htbp]
 \gridline{\fig{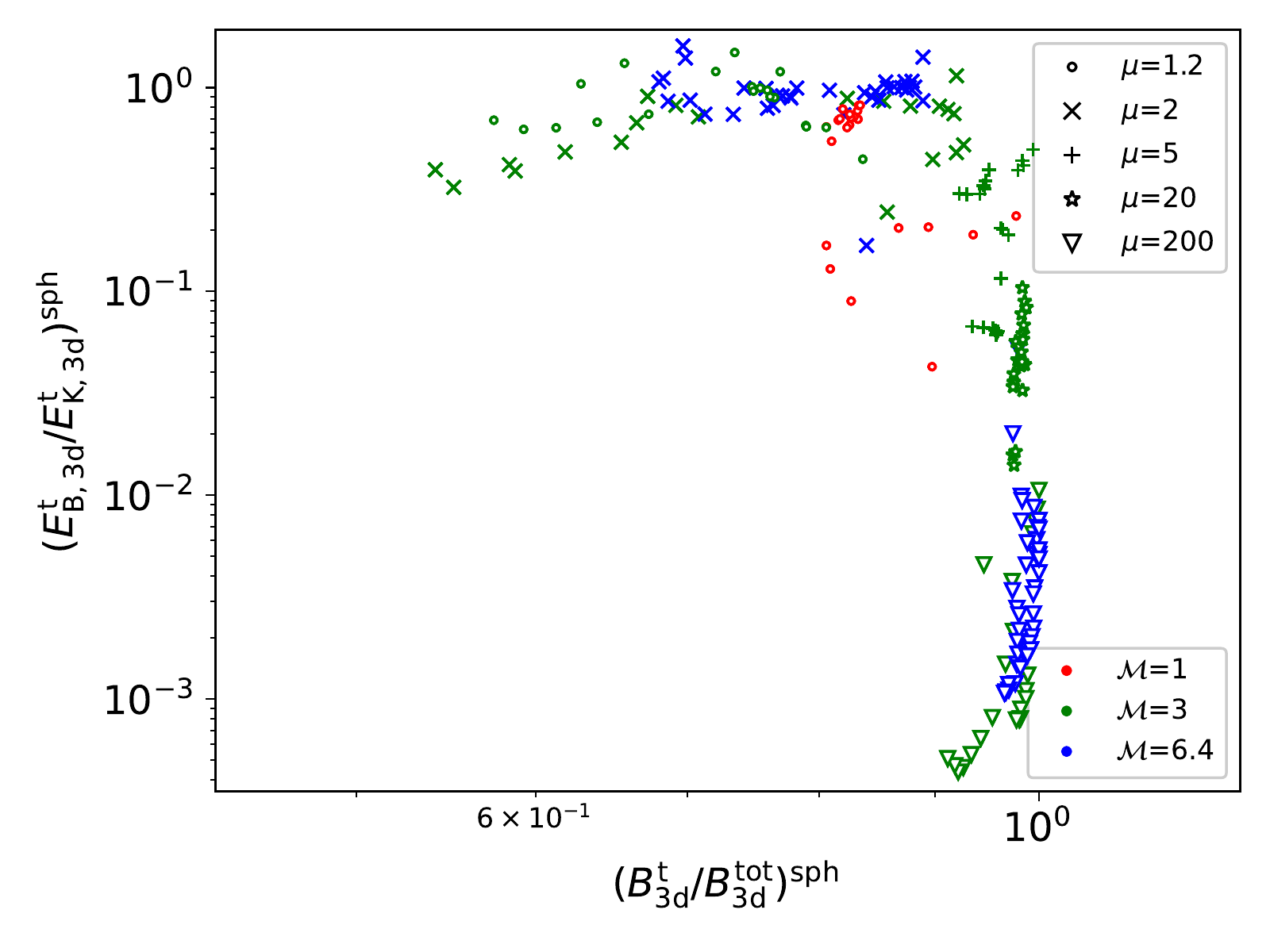}{0.45\textwidth}{a}}
  \gridline{\fig{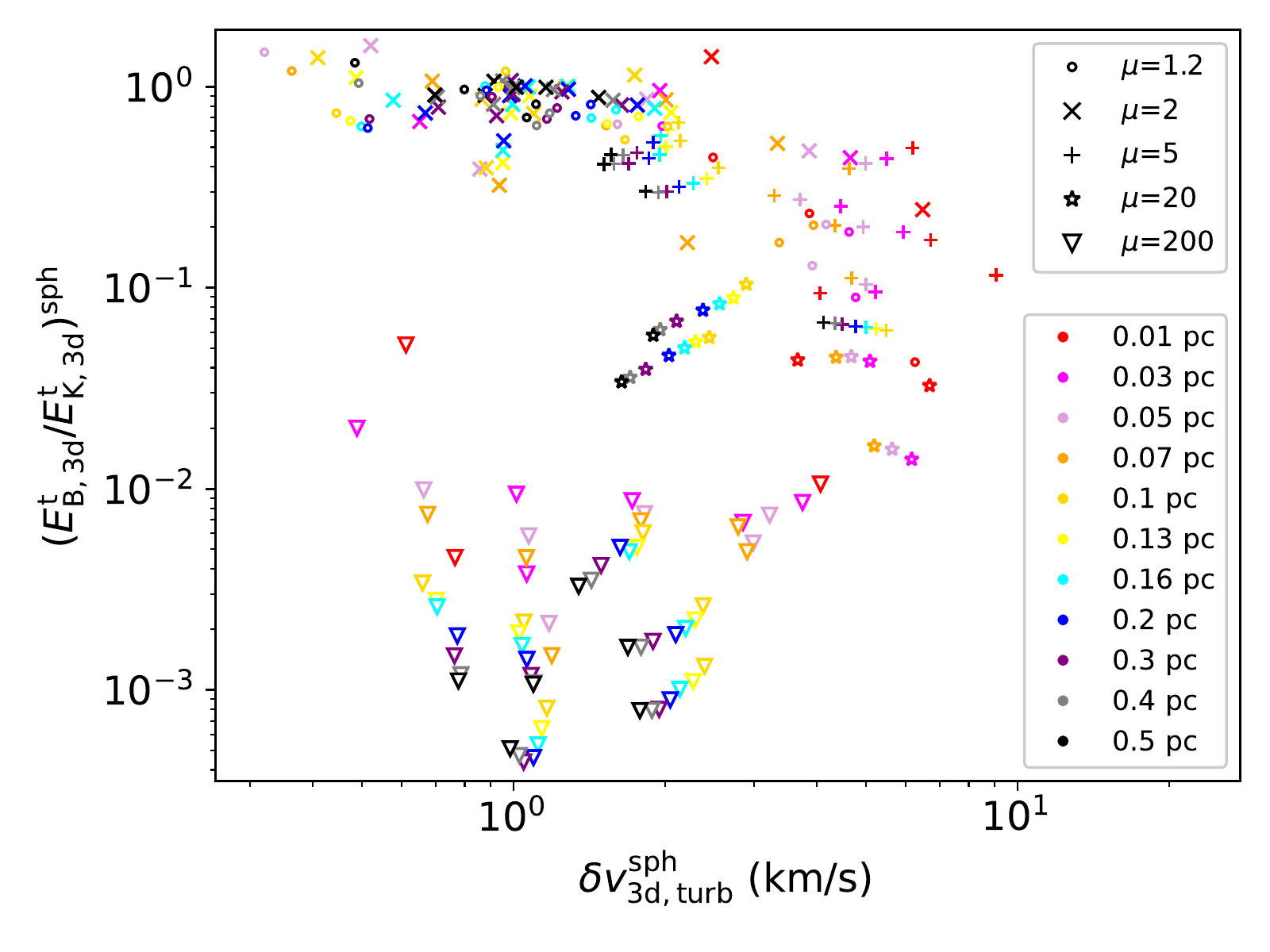}{0.45\textwidth}{b}}
  \gridline{\fig{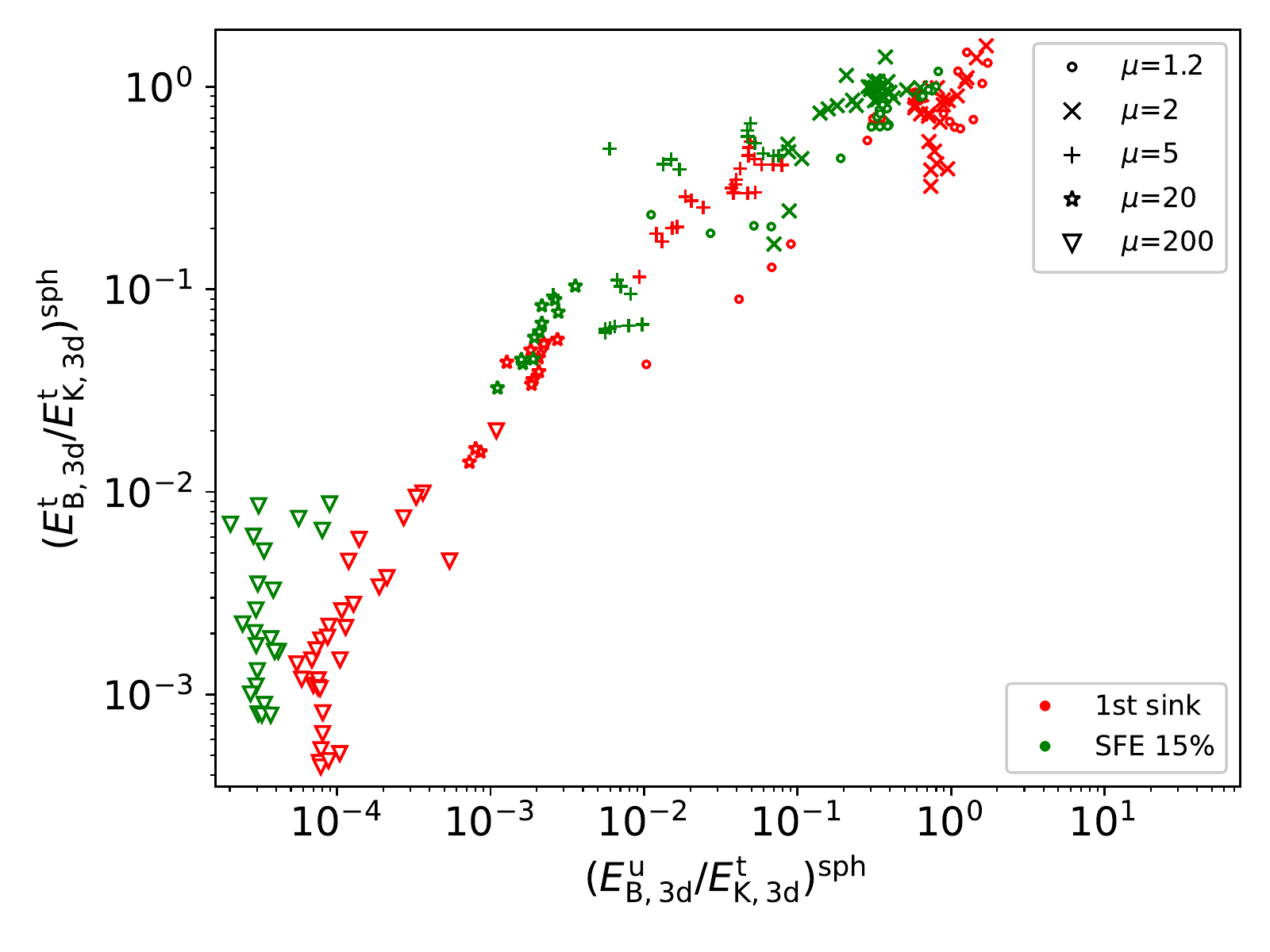}{0.45\textwidth}{c}}
\caption{The relation between the energy ratio $E_\mathrm{B,3d}^{\mathrm{t}}/E_{\mathrm{K,3d}}^{\mathrm{t}}$ and different physical parameters within spheres of different radii for all simulations models. Different symbols represent models with different initial $\mu$ values. In (a) and (b), different colors correspond to models with different initial Mach numbers. In (c), different colors correspond to different time snapshots. \label{fig:EBtEturb}}
\end{figure}

We measure the energies in the 3D volume, $E_\mathrm{B,3d}^{\mathrm{t}}$, $E_\mathrm{B,3d}^{\mathrm{u}}$, $E_{\mathrm{K,3d}}^{\mathrm{t}}$, the 3D turbulent velocity dispersion $\delta v_{\mathrm{3d,turb}}$, and the 3D turbulent-to-total magnetic field strength ratio $B^{\mathrm{t}}_{\mathrm{3d}}/B^{\mathrm{tot}}_{\mathrm{3d}}$ within spheres of different radii for all models. It should be noted that in our estimation of the turbulent magnetic field component, the contribution from large-scale field structure is not excluded, so $E_\mathrm{B,3d}^{\mathrm{t}}$ and $B^{\mathrm{t}}_{\mathrm{3d}}$ could be overestimated and $E_\mathrm{B,3d}^{\mathrm{u}}$ could be underestimated. 

Figure \ref{fig:EBtEturb} (a) shows the relation between the ratios $E_\mathrm{B,3d}^{\mathrm{t}}/E_{\mathrm{K,3d}}^{\mathrm{t}}$ and $B^{\mathrm{t}}_{\mathrm{3d}}/B^{\mathrm{tot}}_{\mathrm{3d}}$. The turbulent magnetic energy is overall less than the turbulent kinetic energy. For large $B^{\mathrm{t}}_{\mathrm{3d}}/B^{\mathrm{tot}}_{\mathrm{3d}}$ values, the ratio $E_\mathrm{B,3d}^{\mathrm{t}}/E_{\mathrm{K,3d}}^{\mathrm{t}}$ shows large scatters over several orders of magnitude. When $B^{\mathrm{t}}_{\mathrm{3d}}/B^{\mathrm{tot}}_{\mathrm{3d}} \lesssim 0.8$ ($B^{\mathrm{t}}_{\mathrm{3d}}/B^{\mathrm{u}}_{\mathrm{3d}} \lesssim 1.3$), the ratio $E_\mathrm{B,3d}^{\mathrm{t}}/E_{\mathrm{K,3d}}^{\mathrm{t}}$ is between $\sim$0.3 and $\sim$1.6 and the average $E_\mathrm{B,3d}^{\mathrm{t}}/E_{\mathrm{K,3d}}^{\mathrm{t}}$ is $\sim$0.9, suggesting that the two energies tend to be in equipartition for small turbulent-to-total field strength ratios. Figure \ref{fig:EBtEturb} (a) also shows that models with stronger initial magnetic field (i.e., smaller initial $\mu$ value) tend to have smaller $B^{\mathrm{t}}_{\mathrm{3d}}/B^{\mathrm{tot}}_{\mathrm{3d}}$ and larger $E_\mathrm{B,3d}^{\mathrm{t}}/E_{\mathrm{K,3d}}^{\mathrm{t}}$ values. However, such relation is not found between the plane-of-sky ratio $B^{\mathrm{t}}_{\mathrm{pos}}/B^{\mathrm{tot}}_{\mathrm{pos}}$ and the initial $\mu$ value (see Figure \ref{fig:disp_BtBtot}). This might be a consequence of the uncertainty from the projection effect when measuring quantities on the plane of sky (see Appendix \ref{app:projection}).  

Figure \ref{fig:EBtEturb} (b) shows the relation between the ratio $E_\mathrm{B,3d}^{\mathrm{t}}/E_{\mathrm{K,3d}}^{\mathrm{t}}$ and the $\delta v_{\mathrm{3d,turb}}$. The same as mentioned above, models with stronger magnetic field tend to be closer to an energy equipartition. There is no strong relation between the energy ratio and the turbulent velocity dispersion in general. However, for strong field models ($\mu=1.2$ or $\mu=2$), data points with smaller turbulent velocity dispersion, mostly at large radii ($>$0.1 pc), seem to be closer to the energy equipartition. For strong field models at radii$>$0.1 pc, the $E_\mathrm{B,3d}^{\mathrm{t}}/E_{\mathrm{K,3d}}^{\mathrm{t}}$ ranges from $\sim$0.4 to $\sim$1.3 and the average $E_\mathrm{B,3d}^{\mathrm{t}}/E_{\mathrm{K,3d}}^{\mathrm{t}}$ is $\sim 0.9$, which is quite close to an energy equipartition.

Figure \ref{fig:EBtEturb} (c) shows that the energy ratio $E_\mathrm{B,3d}^{\mathrm{t}}/E_{\mathrm{K,3d}}^{\mathrm{t}}$ is positively correlated with $E_\mathrm{B,3d}^{\mathrm{u}}/E_{\mathrm{K,3d}}^{\mathrm{t}}$. Models with stronger initial magnetic field tend to have larger  $E_\mathrm{B,3d}^{\mathrm{t}}/E_{\mathrm{K,3d}}^{\mathrm{t}}$ and $E_\mathrm{B,3d}^{\mathrm{u}}/E_{\mathrm{K,3d}}^{\mathrm{t}}$ values. As $E_\mathrm{B,3d}^{\mathrm{u}}/E_{\mathrm{K,3d}}^{\mathrm{t}}$ increases, $E_\mathrm{B,3d}^{\mathrm{t}}/E_{\mathrm{K,3d}}^{\mathrm{t}}$ reaches unity when $E_\mathrm{B,3d}^{\mathrm{u}}/E_{\mathrm{K,3d}}^{\mathrm{t}} \sim 1$, which agrees with the prediction in \citet{2008ApJ...679..537F}. Figure \ref{fig:EBtEturb} (c) also shows an increase of the energy ratio $E_\mathrm{B,3d}^{\mathrm{t}}/E_{\mathrm{K,3d}}^{\mathrm{t}}$ as time evolves. However, there is only a slight increase from the time the first sink forms to the time when SFE=15\%, which cannot bring the state to an energy equipartition for models with small $E_\mathrm{B,3d}^{\mathrm{t}}/E_{\mathrm{K,3d}}^{\mathrm{t}}$ values. 

\subsection{The correction factor $Q_c$}\label{sec:corfac}

The assumptions of the DCF method may not be valid in some cases, therefore the magnetic field strength estimated from the DCF method can deviate from the model magnetic field strength. The correction factor $Q_c$ is defined as the ratio between the model and estimated magnetic field strength and is used to correct for the derived magnetic field strength. Here we derive the correction factors for the magnetic field strength derived from the raw DCF method and the ADF method based on our simulation results. The observational effects (i.e., the beam-smoothing effect of telescopes and the large-scale filtering effect of interferometers) are not considered in the derivation of the correction factors here. 

\begin{figure*}[!htbp]
 \gridline{\fig{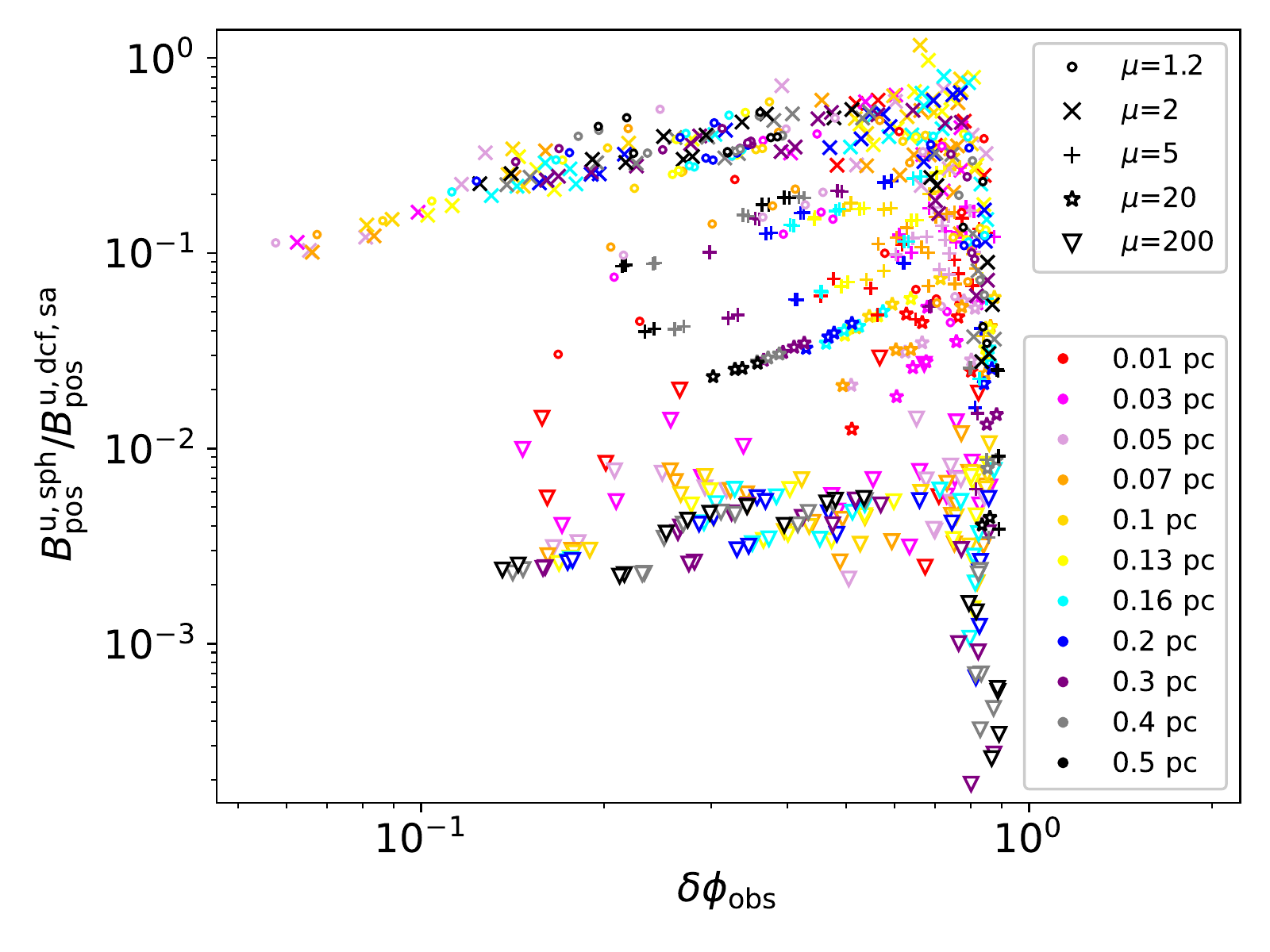}{0.45\textwidth}{a}
 \fig{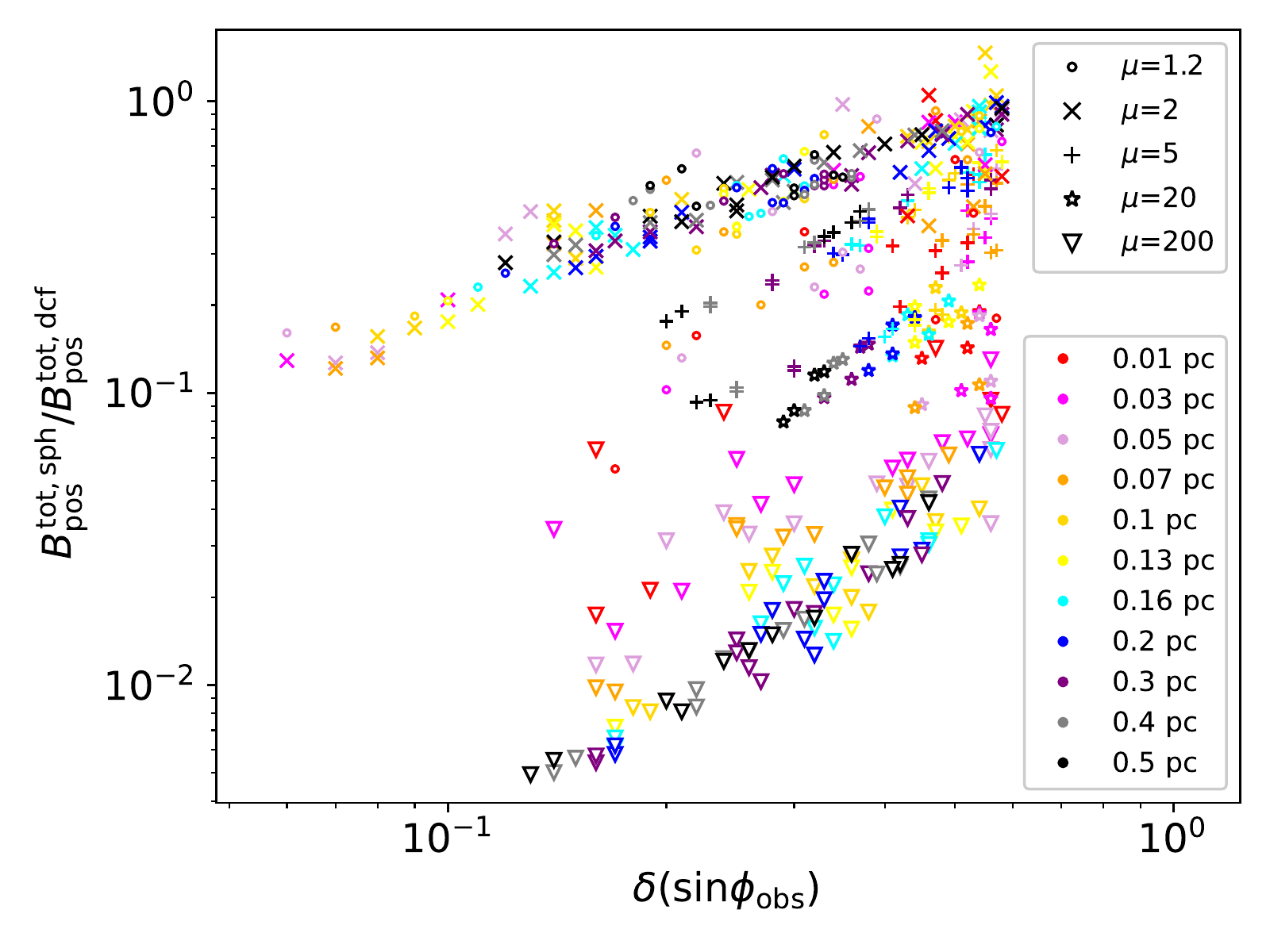}{0.45\textwidth}{b}}
  \gridline{\fig{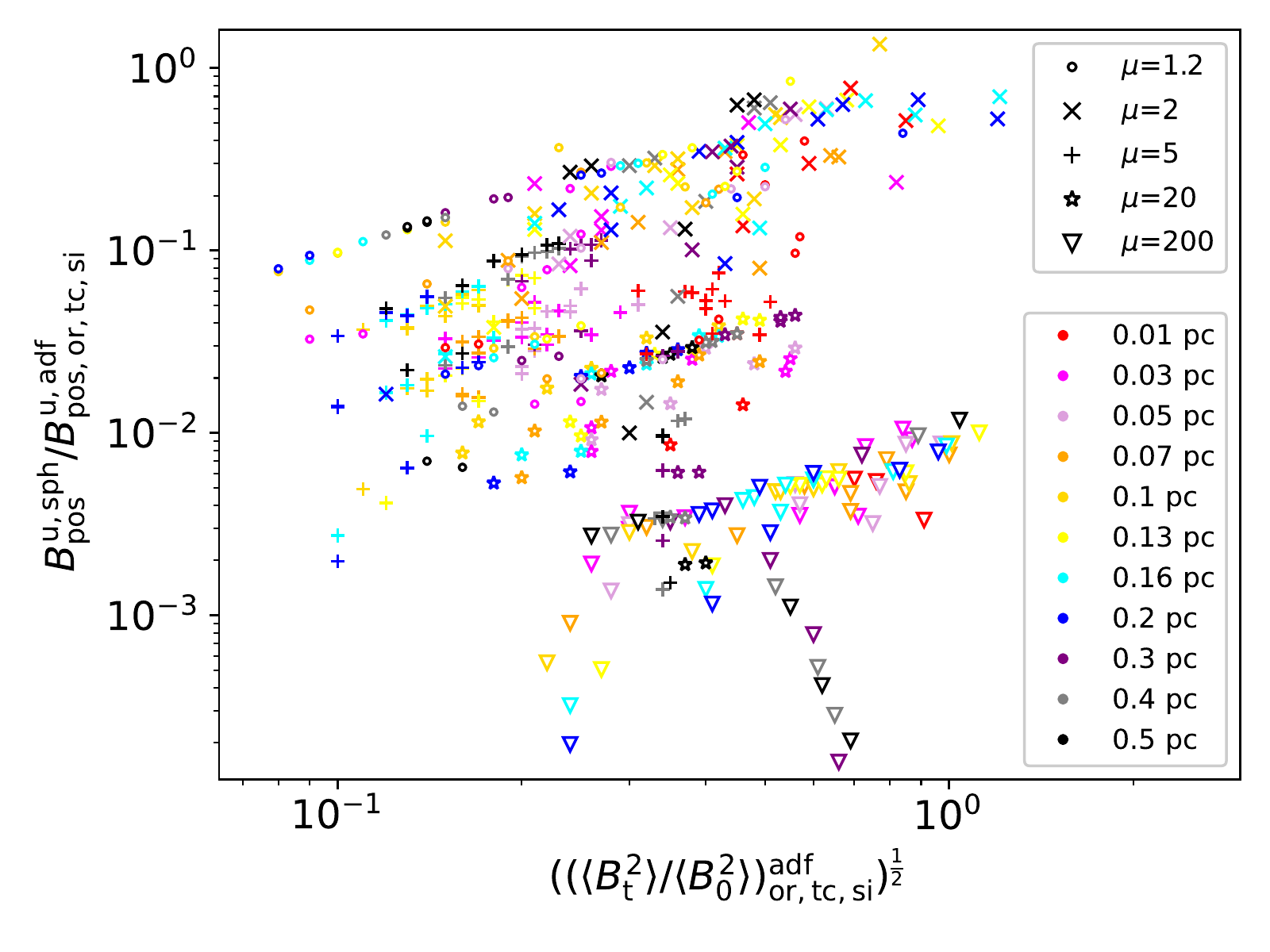}{0.45\textwidth}{c}
  \fig{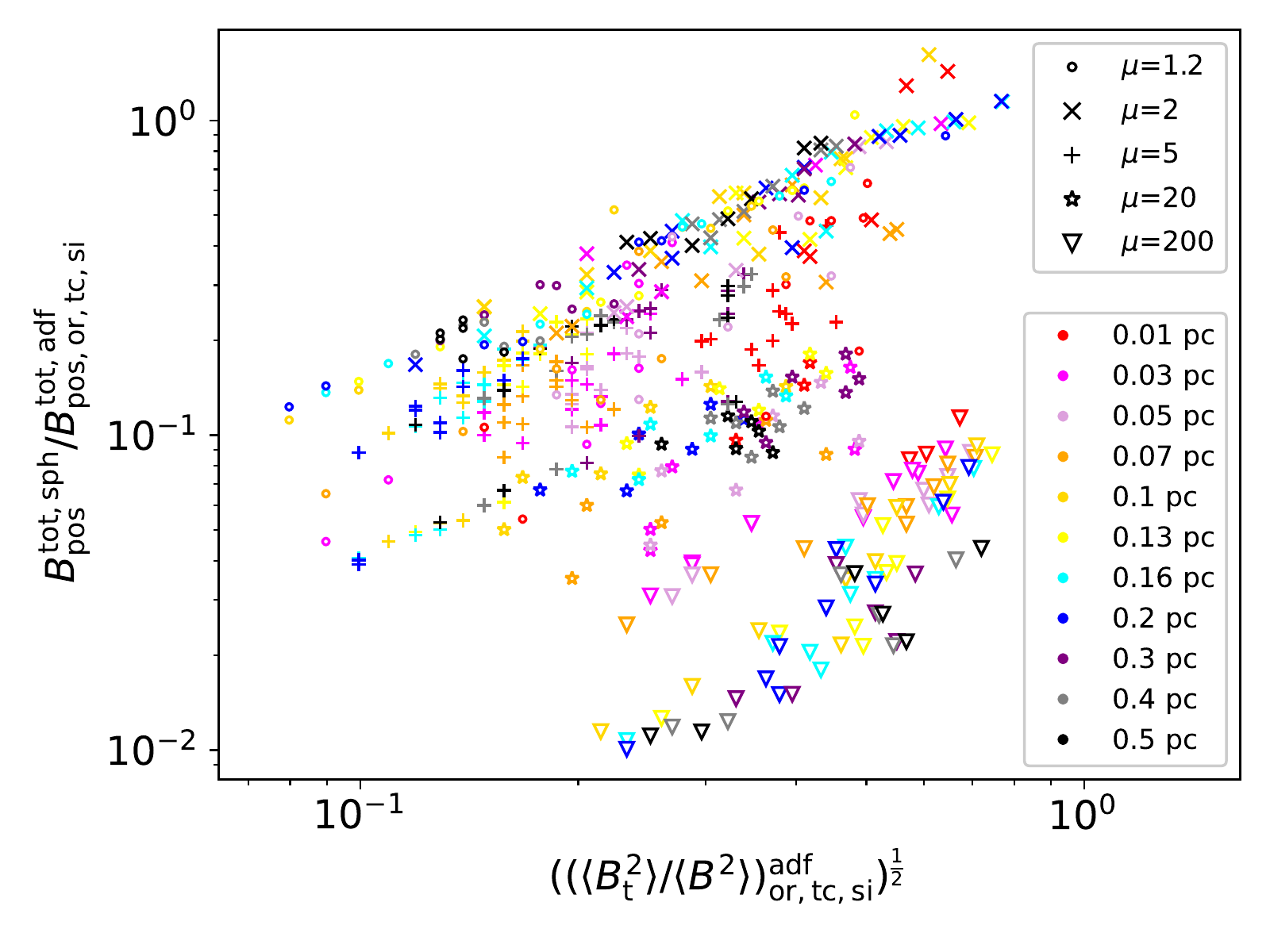}{0.45\textwidth}{d}}
    \gridline{\fig{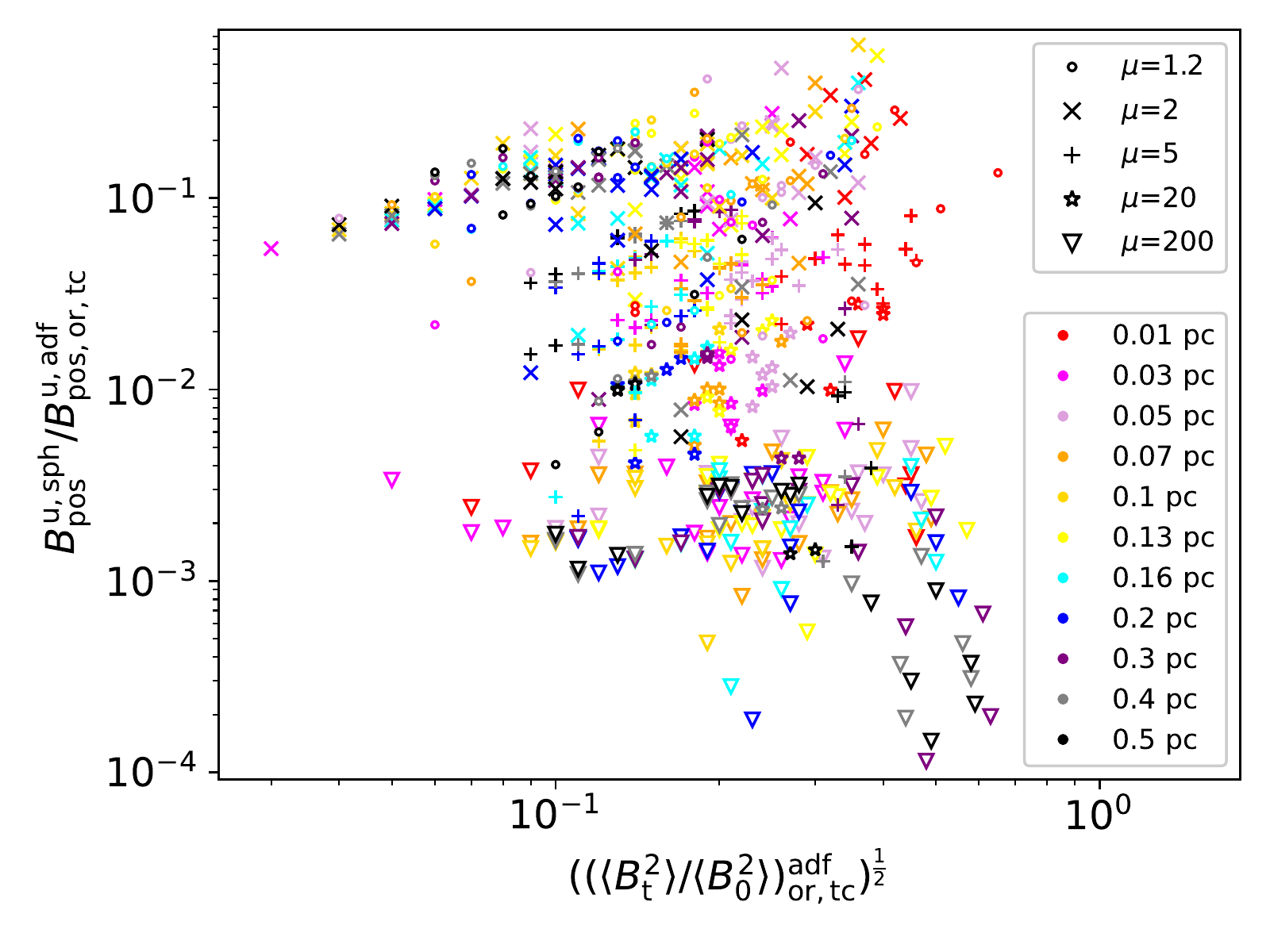}{0.45\textwidth}{e}
  \fig{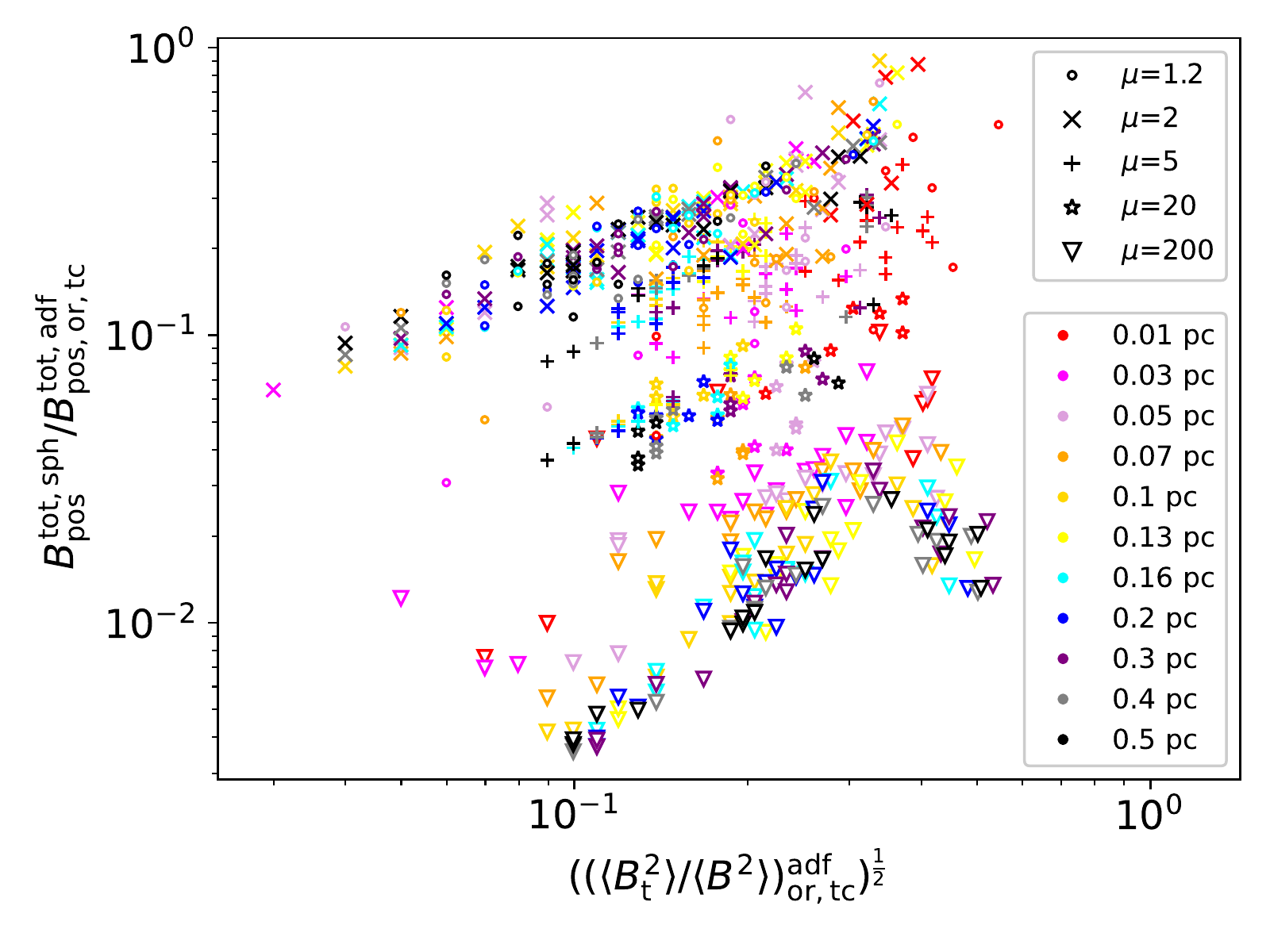}{0.45\textwidth}{f}
}
\caption{(a)-(b). Relation between the correction factor within spheres of different radii and the corresponding angular dispersion in polarization maps for all simulations models. (c)-(f). Relation between the correction factor within spheres of different radii and the corresponding turbulent-to-ordered or -total magnetic field strength ratio derived from the ADF method for all simulations models. Different symbols represent models with different initial $\mu$ values. Different colors correspond to different radii. \label{fig:Q_angdis}}
\end{figure*}

We derive the plane-of-sky mean/total magnetic field strength in the models within spheres or cylinders of different radii with respect to the most massive sink and divide them by the plane-of-sky mean/total magnetic field strength derived from the raw DCF method or the ADF method in the projected circles in the polarization maps to obtain the correction factors. Figure \ref{fig:Q_angdis} relates the correction factors for $B^{\mathrm{u,dcf,sa}}_{\mathrm{pos}}$ (see Equation \ref{eq:eqdcf1}), $B^{\mathrm{tot,dcf}}_{\mathrm{pos}}$ (see Equation \ref{eq:eqdcftot}), $B^{\mathrm{u,adf}}_{\mathrm{pos,or,tc,si}}$ (see Equation \ref{eq:eqbadfu}), $B^{\mathrm{tot,adf}}_{\mathrm{pos,or,tc,si}}$ (see Equation \ref{eq:eqbadftot}), $B^{\mathrm{u,adf}}_{\mathrm{pos,or,tc}}$, and $B^{\mathrm{tot,adf}}_{\mathrm{pos,or,tc}}$, with the measured angular dispersion or the turbulent-to-ordered/-total magnetic field strength ratio derived from the ADF method. The correction factor for $B^{\mathrm{tot,dcf,sa}}_{\mathrm{pos}}$ (see Equation \ref{eq:eqdcftot1}) is very similar to the correction factor for $B^{\mathrm{tot,dcf}}_{\mathrm{pos}}$ and is not shown here.

As shown in Figure \ref{fig:Q_angdis}, overall the magnetic field strength is overestimated by the DCF method. As stated in Section \ref{sec:compangBtB}, the angular dispersion in polarizaton maps do not or only slightly overestimate the turbulent-to-ordered or -total magnetic field strength ratio for radii greater than 0.1 pc, so the overestimation is mostly due to the energy non-equipartition. We find that models with strong initial magnetic field and models with weak initial magnetic field show similar ranges of angular dispersion or derived turbulent-to-ordered/-total magnetic field strength ratio. Thus, for a given angular dispersion or derived turbulent-to-ordered/-total magnetic field strength ratio, the correction factor shows large scatters over different models. So the angular dispersion or derived turbulent-to-ordered/-total magnetic field strength ratio should not be used as indicators for the correction factor. 

The correction factor for models with stronger initial magnetic field is closer to 1, which is because models with stronger magnetic field are closer to the energy equipartition. Since the DCF method can significantly overestimate the magnetic field strength for weak field models, we focus on interpreting the correction factors for strong field models ($\mu=1.2$ or $\mu=2$). For strong field models, it seems that correction factors at radii $<$0.1 pc show larger scatters than those at radii $>$0.1 pc. Overall, the scatter in the correction factor for the uniform magnetic field strength is larger than that for the total magnetic field strength. The correction factor for strong field models seems to be closer to 1 for larger angular dispersions or turbulent-to-ordered/total magnetic field strength ratios. For strong field models, the correction factor for $B^{\mathrm{u,dcf,sa}}_{\mathrm{pos}}$ show small scatters at small angular dispersions (e.g., $\delta \phi_{\mathrm{obs}} \lesssim 25\degr$) and radii $>$0.1 pc (see Figure \ref{fig:Q_angdis} (a)), while the correction factor for $B^{\mathrm{tot,dcf}}_{\mathrm{pos}}$ shows small scatters at radii $>$0.1 pc (see Figure \ref{fig:Q_angdis} (b)). On the other hand, even for strong field models, the correction factors for the uniform magnetic field strength derived from the ADF method taking into account the signal integration effect show large scatters regardless of whether the derived turbulent-to-ordered magnetic field strength ratio is large or small, or the radius is large or small (see Figure \ref{fig:Q_angdis} (c)). For the total magnetic field strength derived from the ADF method taking into account the signal integration effect (see Figure \ref{fig:Q_angdis} (d)), the scatter in the correction factor is smaller than that in Figure \ref{fig:Q_angdis} (c), but still spans over one magnitude. The large scatters in Figures \ref{fig:Q_angdis} (c) and (d) may be because the ADF method does not work well on the effect of the line-of-sight signal integration (See Secton \ref{sec:adf}). So we also derive the correction factors for the uniform/total magnetic field strength estimated from the ADF method without taking into account the signal integration effect (see Figures \ref{fig:Q_angdis} (e) and (f)). The relations in Figures \ref{fig:Q_angdis} (e) and (f) are very similar to those in Figures \ref{fig:Q_angdis} (a) and (b). The critical turbulent-to-ordered magnetic field strength ratio is $\sim$0.1 in Figure \ref{fig:Q_angdis} (e), over which the correction factor for $B^{\mathrm{u,adf}}_{\mathrm{pos,or,tc}}$ for strong field models at radii $>$0.1 pc can show large scatters. We also compare the correction factors within cylinders of different radii in the models with the corresponding angular dispersions or turbulent-to-ordered/-total magnetic field strength ratios and find that the relations are similar to what are shown in Figure \ref{fig:Q_angdis}. 

As mentioned above, the correction factors for strong field models (with inital $\mu=1.2$ or $\mu=2$) are close to 1 and can be confined within a small value range if appropriate radii or angular dispersion selection criteria are applied. The average correction factor for strong field models at radii $>$0.1 pc are derived and listed in Table \ref{tab:Qstrongb}. We also directly divide the 3D magnetic field strength within spheres or cylinders in the model by the plane-of-sky magnetic field strength derived from the raw DCF or ADF method and list the results in Table \ref{tab:Qstrongb}. It should be noted that the correction factors in Table \ref{tab:Qstrongb} are only appropriate when there are other evidences supporting for a strong field case. We notice that the uncertainty for the correction factor of the uniform magnetic field strength derived from the ADF method taking into account the signal integration effect is close to 100\%, so we do not recommend to use this method to derive the uniform magnetic field strength. 

\begin{deluxetable}{ccc}[t!]
\tablecaption{Correction factors for strong magnetic field models (initial $\mu=1.2$ or $\mu=2$) at $R>0.1$ pc. Values in the parenthesis are the relative uncertainty. \label{tab:Qstrongb}}
\tablecolumns{3}
\tablewidth{0pt}
\tablehead{
\colhead{Correction factor} &
\colhead{Spheres} &
\colhead{Cylinders} 
}
\startdata
$B^{\mathrm{tot}}_{\mathrm{pos}}/B^{\mathrm{tot,dcf}}_{\mathrm{pos}}$  &   0.64(37\%) &   0.41(39\%)\\ 
$B^{\mathrm{u}}_{\mathrm{pos}}/B^{\mathrm{u,dcf,sa}}_{\mathrm{pos}}$ \tablenotemark{a} &   0.33(25\%) &   0.22(40\%)\\ 
$B^{\mathrm{tot}}_{\mathrm{pos}}/B^{\mathrm{tot,dcf,sa}}_{\mathrm{pos}}$  &   0.75(44\%) &   0.48(45\%)\\ 
$B^{\mathrm{u}}_{\mathrm{pos}}/B^{\mathrm{u,adf}}_{\mathrm{pos,or,tc,si}}$  &   0.25(83\%) &   0.12(113\%)\\ 
$B^{\mathrm{tot}}_{\mathrm{pos}}/B^{\mathrm{tot,adf}}_{\mathrm{pos,or,tc,si}}$  &   0.49(56\%) &   0.29(57\%)\\ 
$B^{\mathrm{u}}_{\mathrm{pos}}/B^{\mathrm{u,adf}}_{\mathrm{pos,or,tc}}$ \tablenotemark{b} &   0.11(32\%) &   0.08(41\%)\\ 
$B^{\mathrm{tot}}_{\mathrm{pos}}/B^{\mathrm{tot,adf}}_{\mathrm{pos,or,tc}}$  &   0.25(45\%) &   0.16(45\%)\\ \hline
$B^{\mathrm{tot}}_{\mathrm{3d}}/B^{\mathrm{tot,dcf}}_{\mathrm{pos}}$  &   0.84(47\%) &   0.56(54\%)\\ 
$B^{\mathrm{u}}_{\mathrm{3d}}/B^{\mathrm{u,dcf,sa}}_{\mathrm{pos}}$ \tablenotemark{a} &   0.34(24\%) &   0.23(39\%)\\ 
$B^{\mathrm{tot}}_{\mathrm{3d}}/B^{\mathrm{tot,dcf,sa}}_{\mathrm{pos}}$  &   0.99(54\%) &   0.66(60\%)\\ 
$B^{\mathrm{u}}_{\mathrm{3d}}/B^{\mathrm{u,adf}}_{\mathrm{pos,or,tc,si}}$  &   0.41(79\%) &   0.24(82\%)\\ 
$B^{\mathrm{tot}}_{\mathrm{3d}}/B^{\mathrm{tot,adf}}_{\mathrm{pos,or,tc,si}}$  &   0.63(57\%) &   0.38(58\%)\\ 
$B^{\mathrm{u}}_{\mathrm{3d}}/B^{\mathrm{u,adf}}_{\mathrm{pos,or,tc}}$ \tablenotemark{b} &   0.12(28\%) &   0.08(37\%)\\ 
$B^{\mathrm{tot}}_{\mathrm{3d}}/B^{\mathrm{tot,adf}}_{\mathrm{pos,or,tc}}$  &   0.32(55\%) &   0.21(60\%)\\ 
\enddata
\tablenotetext{a}{An additional selection criterion of $\delta \phi_{\mathrm{obs}} \lesssim 25\degr$ ($\sim$0.44) is adopted.}
\tablenotetext{b}{An additional selection criterion of $(\langle B_{\mathrm{t}}^2 \rangle/  \langle B_0^2\rangle)_{\mathrm{or,tc}}^{\mathrm{adf}} \lesssim 0.1$ is adopted.}
\end{deluxetable}

\section{Discussion} \label{section:discussion}
\subsection{Implication of angular dispersions}\label{sec:disang}

The DCF method assumes the turbulent-to-ordered or -total magnetic field strength ratios are traced by the angular dispersion of magnetic field orientations. With a similar amount of turbulent field, the stronger the ordered or total field is, the turbulent-to-ordered or -total strength ratio would be smaller. Thus, in principle, the angular dispersions would also be expected to be smaller for a stronger field. However, we find that strong field models and weak field models do not differ significantly in the value range of angular dispersions, i.e., models with strong field can have large angular dispersions and models with weak field can have small angular dispersions. This could probably be partly explained by the uncertainty from projection effect. Moreover, weak field and strong field models do not produce the same probability distribution functions in density of the formed structures. Since the magnetic field measured is the combination of different structures with different field strength at different scales in the different models, there would be difficulty to associate simply large dispersions to weak fields and small dispersions to strong fields. Thus, one can not distinguish between strong and weak field cases only with the measured angular dispersions. 

The turbulent-to-ordered field strength ratio is expected to be traced by $\delta (\tan \phi)$ or $\delta \phi$ (in small angle approximation). When $\phi \sim 90\degr$, $\tan \phi$ would be very sensitive to $\phi$. $\delta (\tan \phi)$ would be dominated by data points with $\phi \sim 90\degr$ and show large scatters (see Section \ref{sec:compangb}). Thus, $\delta (\tan \phi)$ generally does not correctly estimate the turbulent-to-ordered field strength ratio. The turbulent-to-ordered field strength ratio could be much greater than 1 for weak field cases, while the $\delta \phi$ for a random field is $\sim 52 \degr$ ($\sim$0.91). In that situation, the measured $\delta \phi$ could greatly underestimate the turbulent-to-ordered field strength ratio and overestimate the uniform magnetic field strength. On the other hand, the turbulent-to-total field strength ratio is expected to be traced by $\delta (\sin \phi)$ or $\delta \phi$ (in small angle approximation). Our estimations suggest that the $\delta (\sin \phi)$ or $\delta \phi$ can be well correlated with the turbulent-to-total field strength ratio within a factor of 2 in the simulation models. 
%

Observationally, the angular dispersion is measured from the 2D polarization maps and could be affected be the effect of signal integration along the line of sight. Our results indicate that the $\delta \phi$ or $\delta (\sin \phi)$ in the polarization maps agrees with the $\delta \phi$ or $\delta (\sin \phi)$ in the simulation models within a factor of $\sim$2 when the radius of the structure is greater than $\sim$0.1 pc (approximately at the core scale), but the angular dispersion in the polarization maps could be significantly lower than the angular dispersion in the simulation models when radii$<$0.1 pc. For the effect of signal integration along the line of sight, the ADF method analytically accounts for its effect on the measured angular dispersion, while \citet{2016ApJ...821...21C} proposed a measure of the correction factor for this effect from velocity variations. However, our results suggests that the ADF method is not applicable to account for this effect in the majority of cases. The CY16 method correctly takes into account this effect at radii$>$0.1 pc, while it fails at radii$<$0.1 pc. Since the measured angular dispersion can be significantly underestimated at radii$<$0.1 pc, and neither the ADF method nor the CY16 method works well on the influence of the signal integration along the line of sight below the core scale, we suggest that the DCF methods should be avoided to be applied on scales smaller than the core scale. 
%

The contribution from the ordered field structure can bias the measured angular dispersion in polarization maps toward larger values. If the field structure is highly ordered, the background field structures may by subtracted by fitting with specific field models \citep{2006Sci...313..812G, 2018ApJ...868...51M}. The attempts to universally remove the contribution from the large-scale ordered field structure include the spatial filtering method \citep{2015ApJ...799...74P}, the unsharp masking method \citep{2017ApJ...846..122P}, and the ADF method. The results of our simple Monte Carlo simulations suggest that the ADF method can correctly account for the ordered field contribution (see Appendix \ref{app:ADForder}). Some comparative studies indicate that the magnetic field strengths derived from the unsharp masking method are larger than those derived from the ADF method in the same region \citep{2009ApJ...706.1504H, 2017ApJ...846..122P, 2019ApJ...877...43L}. There are two factors that may explain the systematic difference between the results of the two methods: the turbulent-to-ordered magnetic field strength ratio might be overestimated by the ADF method because of overestimated number of turbulent cells along the line of sight, while the angular dispersion might be underestimated by the unsharp masking method because this method does not consider the beam-smoothing effect and the turbulent correlation effect.

The ADF method also is the only method to analytically take into account the turbulent correlation effect, the beam-smoothing effect, and the interferometric filtering effect. Our results suggest that the ADF method does correctly account for the beam-smoothing effect and the interferometric filtering effect, but do not provide evidence on whether the ADF method correctly accounts for the turbulent correlation effect. 

\subsection{Equipartition between the turbulent kinetic energy and the turbulent magnetic energy}\label{sec:disequi}
The equipartition between the turbulent kinetic energy and the turbulent magnetic energy is a basic assumption of the DCF method. This assumption is expected to be satisfied in sub-Alfv\'{e}nic ($E_{\mathrm{K,3d}}^{\mathrm{t}} < E_\mathrm{B,3d}^{\mathrm{u}}$) cases, where weak turbulence generate Alfv\'{e}nic magnetic perturbations on a strong magnetic field \citep{2008ApJ...679..537F}. For super-Alfv\'{e}nic ($E_{\mathrm{K,3d}}^{\mathrm{t}} > E_\mathrm{B,3d}^{\mathrm{u}}$) models, the perturbation on the magnetic field cannot follow the strong turbulent motions. Thus, the turbulent magnetic energy would be lower than the turbulent kinetic energy and the uniform/total magnetic field strength would be overestimated \citep{2001ApJ...561..800H, 2008ApJ...679..537F}. 

In our calculations, we find that models with stronger magnetic fields tend to be closer to the energy equipartition. However, it could be hard to discern between the strong field cases and the weak field cases from observations. Due to the previously discussed uncertainties, the measured angular dispersion in polarization maps does not give us solid information to qualitatively distinguish between strong and weak field cases. Thus, the measured angular dispersion in polarization maps cannot determine how close the state is to the energy equipartition. The polarization-intensity gradient method proposed by \citet{2012ApJ...747...79K} may help to determine whether the field is strong or weak from polarization maps. We will present the application of this method on our simulation results in a separate work. Comparing the large-to-small magnetic field orientation may also be useful to qualitatively discern the strong field case from the weak field case \citep{2014ApJ...792..116Z, 2014ApJS..213...13H}. Strong field cases and weak field cases may also show observable differences in the magnetic field morphology and density structure. 

Massive star formation regions are usually associated with strong turbulence and large velocity dispersions. Low-mass star-forming regions usually have weaker turbulence and smaller velocity dispersions, so they are expected to be more likely in the energy equipartition if they have similar magnetic field strength to that of massive star forming regions. To investigate whether low-mass star formation region is more likely in the energy equipartition, we perform an additional simulation of a 10 $M_{\odot}$ spherical dense core with an initial radius of 0.067 pc. The inital temperature, $\mu$, and Mach number are 10 K, 1.2, and 1, respectively. We find that the ratio $E_\mathrm{B,3d}^{\mathrm{t}}/E_{\mathrm{K,3d}}^{\mathrm{t}}$ within 0.067 pc is $\sim$1.2 when the first sink forms and $\sim$0.8 when SFE=15\%. For comparison, the ratio $E_\mathrm{B,3d}^{\mathrm{t}}/E_{\mathrm{K,3d}}^{\mathrm{t}}$ within 0.067 pc is $\sim$0.2 for the massive star formation model T10M1MU1 at both time snapshots. Thus, we suggest that the low-mass star-forming region would be closer to the energy equipartition for similar magnetic critical parameters. More simulations of low-mass star-forming regions with different initial parameters would be useful to gain more solid conclusions.


One of the purposes to estimate the magnetic field strength is to compare the relative importance of the magnetic field energy and the turbulent energy. The energy equipartition assumption of the DCF method implicitly requires $E_{\mathrm{K,3d}}^{\mathrm{t}} =  E_\mathrm{B,3d}^{\mathrm{t}} < E_\mathrm{B,3d}^{\mathrm{t}} + E_\mathrm{B,3d}^{\mathrm{u}} = E_\mathrm{B,3d}^{\mathrm{tot}}$, i.e., the turbulent kinetic energy is smaller than the total magnetic energy. If the question of energy equipartition is not properly addressed in the application of the DCF method, the estimated total magnetic field energy would be always larger than the turbulent kinetic energy. Then the relative importance between the magnetic field and the turbulence in the energy balance cannot be properly compared. It also should be noted that the total magnetic energy could be larger than the turbulent kinetic energy in a super-Alfv\'{e}nic case if there is significant magnetic energy in turbulent forms. So a super-Alfv\'{e}nic state does not necessarily mean the turbulence is more important than the magnetic field in the energy balance.

Another main purpose to estimate the magnetic field strength is to compare the relative importance of the magnetic field energy and the gravitational energy. Similarly, if there is significant amount of magnetic field energy in turbulent forms, the total magnetic energy, rather than the uniform magnetic energy, should be used to compare with the gravitational energy in the energy balance (i.e., replace the uniform magnetic field strength by the total magnetic field strength in the derivation of the mass-to-flux-ratio-to-critical-value or the magnetic virial parameter).

Most previous observational DCF studies applied the DCF method on polarization maps to estimate the uniform magnetic field strength, but none of them considered the problem of energy non-equipartition. In practice, if the estimated Alfv\'{e}nic Mach number is greater than 1 (super-Alfv\'{e}nic), the magnetic field strength must have been overestimated due to the effect of energy non-equipartition. On the other hand, the estimated magnetic field strength for strong field cases and weak field cases could be degenerated. Even if the estimated Alfv\'{e}nic Mach number is smaller than 1 (sub-Alfv\'{e}nic), it does not necessarily ensure that there is an equipartition between the turbulent kinetic energy and the turbulent magnetic energy. It is still possible that the magnetic field strength is overestimated by the DCF method and the true situation is super-Alfv\'{e}nic. 

In summary, our simulations suggest that the energy equipartition is only achieved in strong field models (i.e., $\mu$=1.2 or $\mu$=2). In weak field models, the energy equipartition generally could not be reached and the magnetic field strength would be overestimated by the DCF method. On the other hand, low-mass star-forming regions are closer to or more likely in the energy equipartition since they are associated with weaker turbulence than massive star-forming regions. In Section \ref{sec:corfac}, the correction factors accounting for the energy non-equipartition are derived for the strong field models of massive star formation. If the strong field scenario can be confirmed independently from observations, these correction factors can be applied to correct for the magnetic field strength estimated with the DCF method. We note that there is a lack of theoretical efforts to quantify the energy equipartition problem from observations so far. 

\subsection{Comparison with previous simulation works}\label{sec:dissimu}

There have been several studies on testing the reliability of the DCF method with numerical simulations in the literature \citep{2001ApJ...546..980O, 2001ApJ...559.1005P, 2001ApJ...561..800H, 2008ApJ...679..537F, 2020arXiv201015141S}. Here we compare our results with those of the previous numerical works. 

\citet{2001ApJ...546..980O} performed 3D compressible ideal MHD simulations to present the conditions of self-gravitating giant molecular clouds. Their simulations are with a resolution of 256$^3$, a fiducial average temperature of 10 K, and a fiducial average density of 100 cm$^{-2}$. Their models considered 3 different magnitude of magnetic field strength, but they only derived the correction factor for the plane-of-sky uniform magnetic field strength for a strong field model (with initial $B^{\mathrm{u}}_{\mathrm{3d}} = 14 \mu$G) that is sub-Alfv\'{e}nic, magneticallly subcritical, and with a box length of $\sim$8 pc. They found that the correction factor $B^{\mathrm{u}}_{\mathrm{pos}}/B^{\mathrm{u,dcf,sa}}_{\mathrm{pos}}$ is $\sim$0.5 if the angular dispersion in polarization map (without removing the ordered field structure) is smaller than 25$\degr$ ($\sim$0.44). Since the model is sub-Alfv\'{e}nic, it is very likely that the energy equipartition assumption is fullfilled. So the correction factor reflects the ratio between the angular dispersion and the turbulent-to-ordered magnetic field strength ratio. However, their correction factors are appropriate for pc-scale molecular clouds at low densities. At smaller scales of molecular clumps and cores, the gravity become prominent and there could be significant contribution from ordered field structure to the directly measured angular dispersion in polarization maps. Our simulation results indicate that statistically the average ratio between the directly measured angular dispersion in polarization maps and the turbulent-to-ordered magnetic field strength ratio is $\sim$0.25 (see Table \ref{tab:angBtB}) in clump and core scales if $\delta \phi_{\mathrm{obs}} < 25\degr$, which is smaller than the value reported by \citet{2001ApJ...546..980O}.

\citet{2001ApJ...559.1005P} performed 3D compressible MHD simulations of a super-Alfv\'{e}nic and magneticallly supercritical molecular cloud with a size of 6.25 pc and a numerical resolution of 128$^3$. They selected 3 self-gravitating $\sim$1 pc clumps (they used ``cores'' in their terminology) and estimated an average correction factor to be $\sim$0.4 for the plane-of-sky uniform magnetic field strength ($B^{\mathrm{u}}_{\mathrm{pos}}/B^{\mathrm{u,dcf,sa}}_{\mathrm{pos}}$). Since their model is super-Alfv\'{e}nic, it is likely that the energy is not in equipartition and the correction factor derived by their work includes the contribution from energy non-equipartition. Although their estimated $\delta \phi_{\mathrm{obs}}$ are in the range of 0.37-0.73 (21$\degr$-42$\degr$), the correction factors derived by \citet{2001ApJ...559.1005P} show small scatters, which may be because that they only have several estimations. As indicated by our simulation results, $\delta \phi_{\mathrm{obs}}$ could be much smaller than $(B^{\mathrm{t}}_{\mathrm{pos\perp}}/B^{\mathrm{u}}_{\mathrm{pos}})$ and the ratio between the two quantities could have large scatters when $\delta \phi_{\mathrm{obs}} > 25\degr$. The DCF method should be avoided to be applied to derive the uniform magnetic field strength when the angular dispersion in polarization map is large.

\citet{2001ApJ...561..800H} presented a set of 3D ideal MHD scale-free simulations with constant turbulent energy input to simulate the interior of molecular clouds. They considered both super-Alfv\'{e}nic and sub-Alfv\'{e}nic models with resolutions ranging from 128$^3$ to 512$^3$. They used the same equation as Equation \ref{eq:eqdcf} in this paper to estimate the plane-of-sky uniform magnetic field strength. They found that the plane-of-sky uniform magnetic field strength derived from this equation is overestimated by a factor of 2 to 3 for strong field models and show significant scatters for weak fields. They also proposed that the 3D total (rms) magnetic field strength is estimated as $B^{\mathrm{tot,Hei01}}_{\mathrm{3d}} \sim \sqrt{\mu_0 \rho }(\delta v_{\mathrm{los}}/\delta (\tan \phi))(1+3\delta (\tan \phi)^2)^{1/2}$ if the turbulent magnetic energy is isotropic and the mean magnetic field orientation is along the plane of sky. They found that the 3D total magnetic field strength derived from their equation is overestimated by a factor of no more than 3 for strong field models, but the field strength could be overestimated by up to 3 orders of magnitude for weak fields. After correction for energy non-equipartition, they found both the estimated uniform and total magnetic field strength match the model field strength by a factor between 1 and 1.5 for strong field models, but still yield large deviations from the weak field estimates. For comparison, our simulation results show large scatters for estimations including $\delta (\tan \phi)$ regardless of whether the magnetic field is strong or weak, so we do not recommend to use $\delta (\tan \phi)$ as an estimate of the turbulent-to-ordered magnetic field strength ratio. On the other hand, similar to the results in \citet{2001ApJ...561..800H}, our Equations \ref{eq:eqdcftot} and \ref{eq:eqdcftot1} also only slightly overestimate the total magnetic field strength for strong field models and overestimate the total magnetic field strength by up to several orders of magnitude for weak field models (see Figure \ref{fig:Q_angdis}). Unlike the \citet{2001ApJ...561..800H} equation where the total magnetic field strength is significantly overestimated for weak field cases, our Equations \ref{eq:eqdcftot} and \ref{eq:eqdcftot1} only slightly overestimate the total magnetic field strength regardless of whether the magnetic field is strong or weak (see Figure \ref{fig:ang_BtB}) if the energy equipartition assumption is fulfilled. 

\citet{2008ApJ...679..537F} presented 3D ideal high-resolution (512$^3$) MHD simulations with continuous turbulence injection to study the polarization emission in molecular clouds. They considered both  super-Alfv\'{e}nic and sub-Alfv\'{e}nic simulations that are non-self-gravitating and scale-independent. They proposed that the plane-of-sky total magnetic field strength is estimated with $B^{\mathrm{tot,Fal08}}_{\mathrm{pos}} \sim \sqrt{\mu_0 \rho }\delta v_{\mathrm{los}}/\tan (\delta \phi)$ where they assume $(B^{\mathrm{t}}_{\mathrm{pos\perp}}/B^{\mathrm{tot}}_{\mathrm{pos}})^{\mathrm{Fal08}} \sim \delta(\tan \phi) \sim \tan (\delta \phi)$. However, we note that they might have approximated the turbulent-to-ordered magnetic field strength ratio with the turbulent-to-total magnetic field strength ratio because there are $B^{\mathrm{t}}_{\mathrm{pos\perp}}/B^{\mathrm{tot}}_{\mathrm{pos}} \sim \delta(\sin \phi)$ and $B^{\mathrm{t}}_{\mathrm{pos\perp}}/B^{\mathrm{u}}_{\mathrm{pos}} \sim \delta(\tan \phi)$. Due to this approximation, we refrain from discussing their results for the correction factors. We note that $(B^{\mathrm{t}}_{\mathrm{pos\perp}}/B^{\mathrm{tot}}_{\mathrm{pos}})^{\mathrm{Fal08}} \sim \tan (\delta \phi)$ is only valid when there is small angle approximation $\tan (\delta \phi) \sim \delta \phi$. 

Recently, \citet{2020arXiv201015141S} used 3D ideal scale-free sub-Alfv\'{e}nic MHD simulations with no gravity to investigate the none-Alfv\'{e}nic modes in ISM. In contrast to the assumption of the raw DCF method that there is an energy equipartition between the turbulent kinetic energy and turbulent magnetic energy, they assumed the turbulent kinetic energy equals the cross-term of magnetic energy. They proposed that the uniform plane-of-sky magnetic field strength could be estimated with $B^{\mathrm{u,Ska20}}_{\mathrm{pos}} \sim \sqrt{\mu_0 \rho /2}\delta v_{\mathrm{los}}/\sqrt{\delta \phi}$ in small angle approximation. They claimed that their new equation can yield accurate estimate of the true uniform field strength within a mean relative deviation of 17\%. However, for different ratios between the angular dispersion and the turbulent-to-ordered field strength ratio and different ratios between the turbulent kinetic energy and the cross-term of magnetic energy, the estimated correction factor may degenerate with each other. As they did not compare the two relations separately and they only have 5 estimations of the correction factors for models with an Alfv\'{e}nic Mach number of 0.7, it is unknown whether their correction factor $\sim$1 is a coincidence since there might not be equipartition between turbulent kinetic energy and cross-term magnetic energy in sub-Alfv\'{e}nic models. 



\section{Summary} \label{section:summary}

In this paper we present a set of 3D high-resolution ideal MHD simulations and radiative transfer simulations of self-gravitating clustered massive star-forming regions to study the accuracy of the DCF method, which is the most widely used method to estimate the plane-of-sky magnetic field strength with statistics of polarization position angles from polarization maps. There are several assumptions in the DCF method: there is a mean and prominent magnetic field component (i.e., small angle approximation); there is an energy equipartition between the turbulent kinetic energy and the turbulent magnetic energy; the turbulence is isotropic; and the perturbation on the magnetic field is traced by angular dispersions in polarization maps. The estimated angular dispersion in polarization maps is subject to uncertainties from the line-of-sight signal integration, the turbulent correlation effect, the contribution from ordered field structure, the beam-smoothing effect of telescopes, and additional large-scale filtering effect of inteferometric observations. We tested the assumptions of the DCF method and the factors affecting the measured angular dispersion step by step with our simulation results. The main conclusions are as follows:
\begin{enumerate}
\item The magnetic critical parameter (mass-to-flux ratio over its critical value) of a simulation is determined by its initial value. There is no strong relation between the magnetic critical parameter and the spatial scale in our simulations. On the other hand, simulations with different initial Mach numbers show similar levels of turbulent velocity dispersion, which indicates the initial environmental turbulence decays very quickly and the turbulence is most likely generated from star formation activities when the self-gravity is dominant. 
\item If the energy equipartition between the turbulent magnetic energy and the turbulent kinetic energy is fulfilled and the directly measured angular dispersion in the polarization map is smaller than 25$\degr$, the average correction factor between the model and estimated plane-of-sky uniform magnetic field strength would be $\sim$0.25 at clump and core scales, which is smaller than the correction factor of $\sim$0.5 found in previous numerical studies at larger scales. When the energy equipartition is met, the average correction factor between the model and estimated plane-of-sky total magnetic field strength would be $\sim$0.81 or $\sim$0.57, respectively, depending on whether the angular dispersion is in the forms of $\delta \phi$ or $\delta (\sin \phi)$. 
\item Our simulations indicate that models of stronger magnetic fields are closer to an energy equipartition. Since low-mass star-forming regions are associated with weaker turbulence, they are closer to or more likely in the energy equipartition compared to massive star-forming regions with similar magnetic critical parameters. When the turbulent magnetic energy is smaller than the turbulent kinetic energy, the magnetic field strength would be overestimated by the DCF method. In our simulations, only strong field models are close to the energy equipartition. We estimate and catalogue the correction factors for the uniform and total magnetic field strength accounting for the energy non-equipartition for strong field massive star formation models with initial magnetic critical parameters of 1.2 or 2. These correction factors are applicable to observational data to determine the correct magnetic field strength at clump and core scales in strong field cases. 
\item  The ADF method \citep{2008ApJ...679..537F, 2009ApJ...696..567H, 2009ApJ...706.1504H, 2016ApJ...820...38H} analytically takes into account the factors that affect the estimated angular dispersion. We test the accuracy of the ADF method with our simulation results and find that the ADF method does correctly account for the ordered field structure, the beam-smoothing effect, and the interferometric filtering effect. However, the ADF method might not be applicable to account for the line-of-sight signal integration effect from observational data in most cases. 
\item We test the \citet{2016ApJ...821...21C} method and find that this method correctly accounts for the effect of line-of-sight signal integration at $>$0.1 pc, but fails at $<$0.1 pc.
\item The line-of-sight signal integration only underestimate the angular dispersion by a factor of no more than 2 at radii$>$0.1 pc, while the angular dispersion may be significantly underestimated at scales below $\sim$0.1 pc (the core scale). Since the signal integration effect may not be correctly accounted for at $<$0.1 pc, the DCF method should be avoided to be applied on scales below the core scale.
\item The DCF method is valid to derive the uniform magnetic field strength only when the directly measured angular dispersion in the polarization map is small (i.e., $\delta \phi_{\mathrm{obs}} < 25\degr$). There is no limitation on the angular dispersion for the DCF method to derive the total magnetic field strength. Overall, the ratio between the model and estimated uniform magnetic field strength show larger scatters than the ratio for the total magnetic field strength. To properly compare the relative importance between the magnetic field and other forces (e.g., turbulence and gravity), the total magnetic field strength should be used instead of the uniform magnetic field strength if there is significant turbulent magnetic field.

\end{enumerate}

\acknowledgments 
We are indebt to the anonymous referee whose constructive comments improved the presentation and clarity of the paper. J.L. thanks Prof. Martin Houde for helpful discussions on the principle of the angular dispersion function method, and Dr. Robert Brauer for the help on the installation of POLARIS. J.L. also thanks Dr. Xing Lu, Mr. Yu Cheng, and Dr. Shanghuo Li for helpful discussions on the components of velocity dispersions. We thank Dr. Stefan Reissl for helpful discussions on the POLARIS code. K.Q. and J.L. are supported by National Key R\&D Program of China No. 2017YFA0402600. K.Q. and J.L. acknowledge the support from National Natural Science Foundation of China (NSFC) through grants U1731237, 11590781, and 11629302. J.L. acknowledges the support from the program of China Scholarship Council (No. 201806190134), the Smithsonian Astrophysical Observatory pre-doctoral fellowship, and the support by the EAO Fellowship Program under the umbrella of the East Asia Core Observatories Association. This study was enabled by the MagneticYSOs project, an European Research Council (ERC) funded collaboration, under the European Union Horizon 2020 research and innovation programme (grant agreement N. 679937, PI: Maury). B.C. acknowledges support by the "Programme National de Physique Stellaire" (PNPS) of CNRS/INSU co-funded by CEA and CNES”. The 3D RAMSES simulations were performed using HPC resources from GENCI-CINES (Grant A0080407247). This research made use of  APLpy, an open-source plotting package for Python \citep{2012ascl.soft08017R}, Astropy, a community-developed core Python package for Astronomy \citep{2013A&A...558A..33A}, and Matplotlib, a Python 2D plotting library for Python \citep{2007CSE.....9...90H}.
\software{APLpy \citep{2012ascl.soft08017R}, Astropy \citep{2013A&A...558A..33A}, Matplotlib \citep{2007CSE.....9...90H}.}


\appendix
\section{Examples of the synthetic images}\label{app:imexam}

\begin{figure*}[!htbp]
 \gridline{\fig{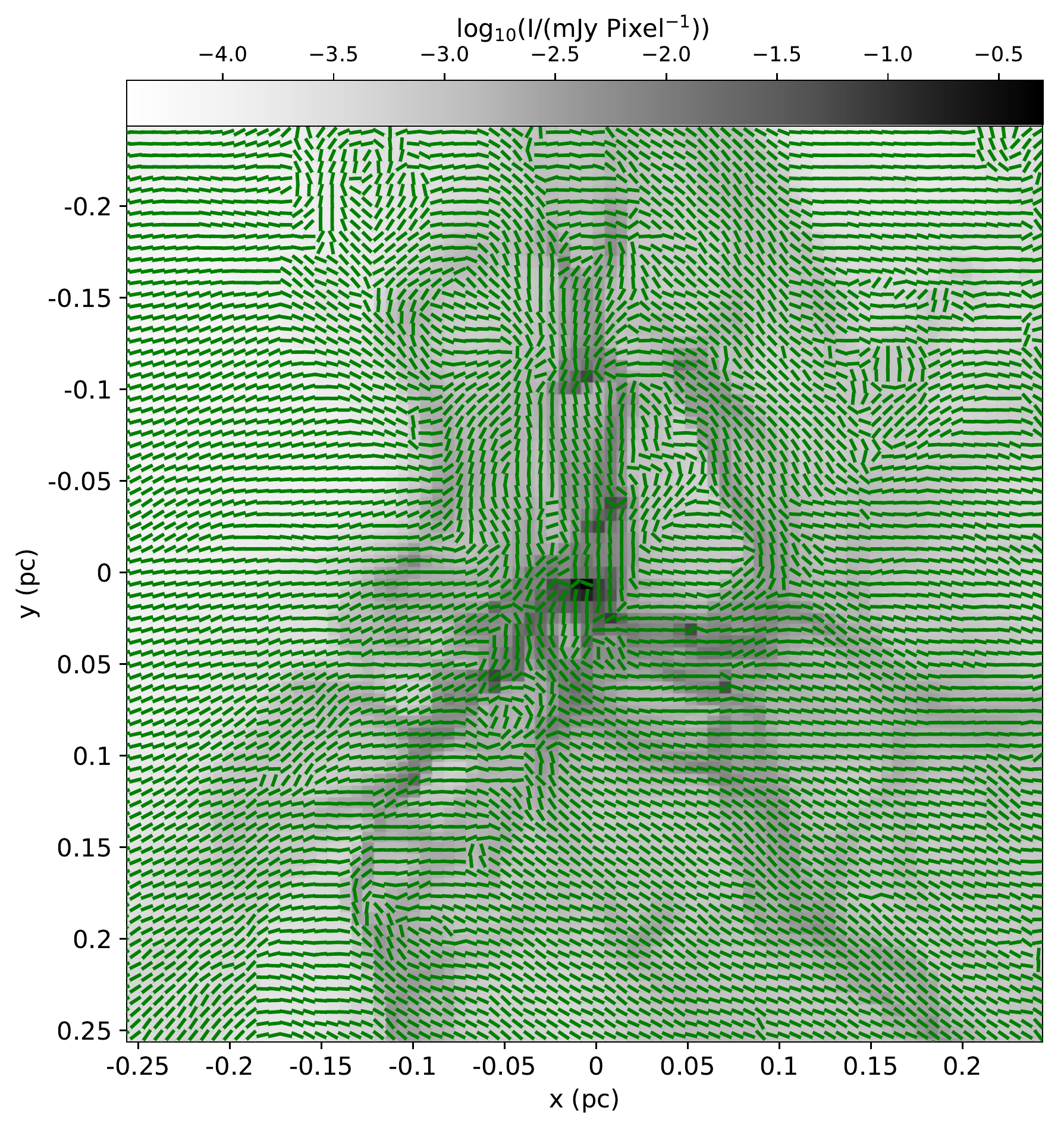}{0.5\textwidth}{(a)}
      \fig{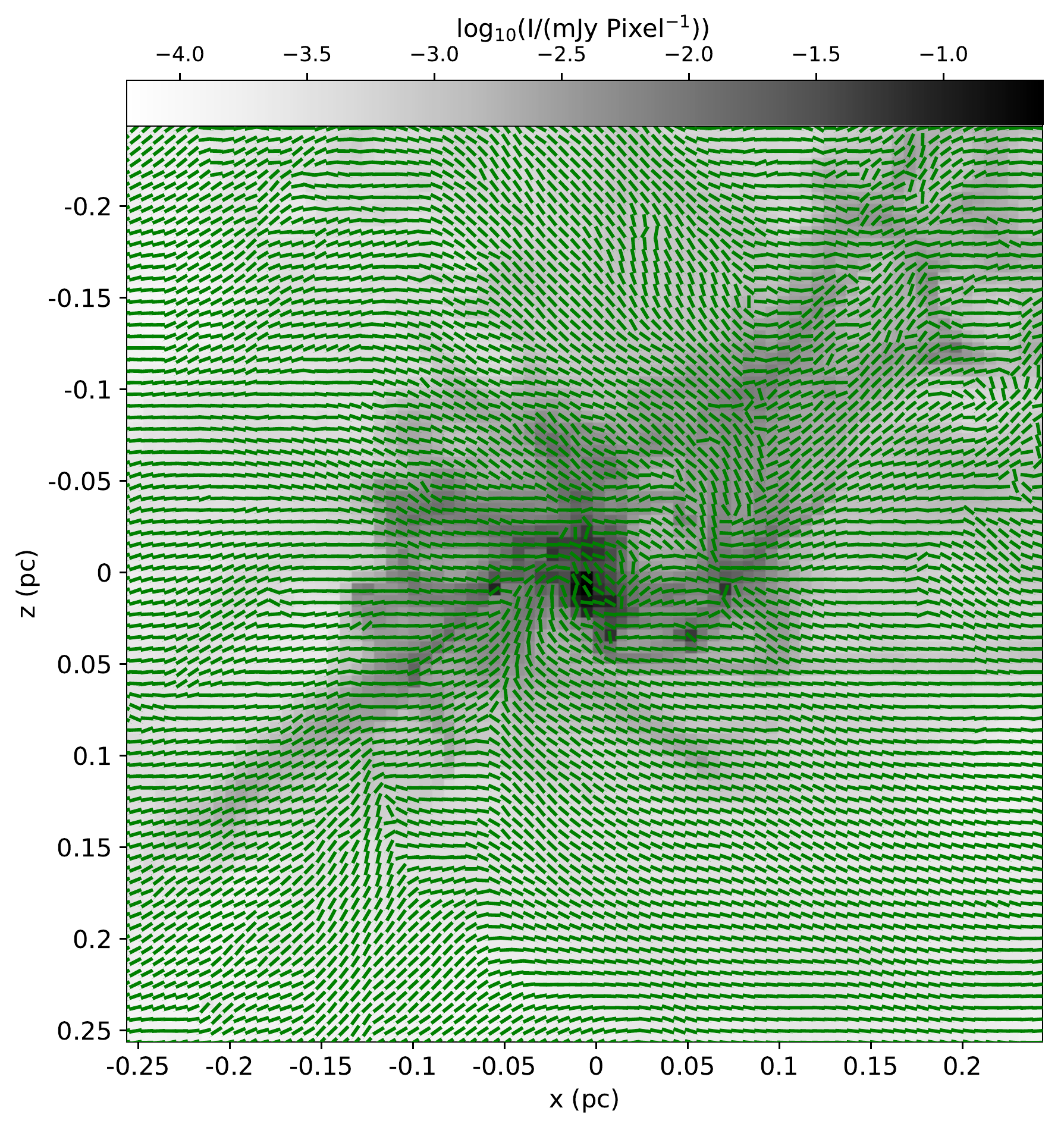}{0.5\textwidth}{(b)}}
\gridline{\fig{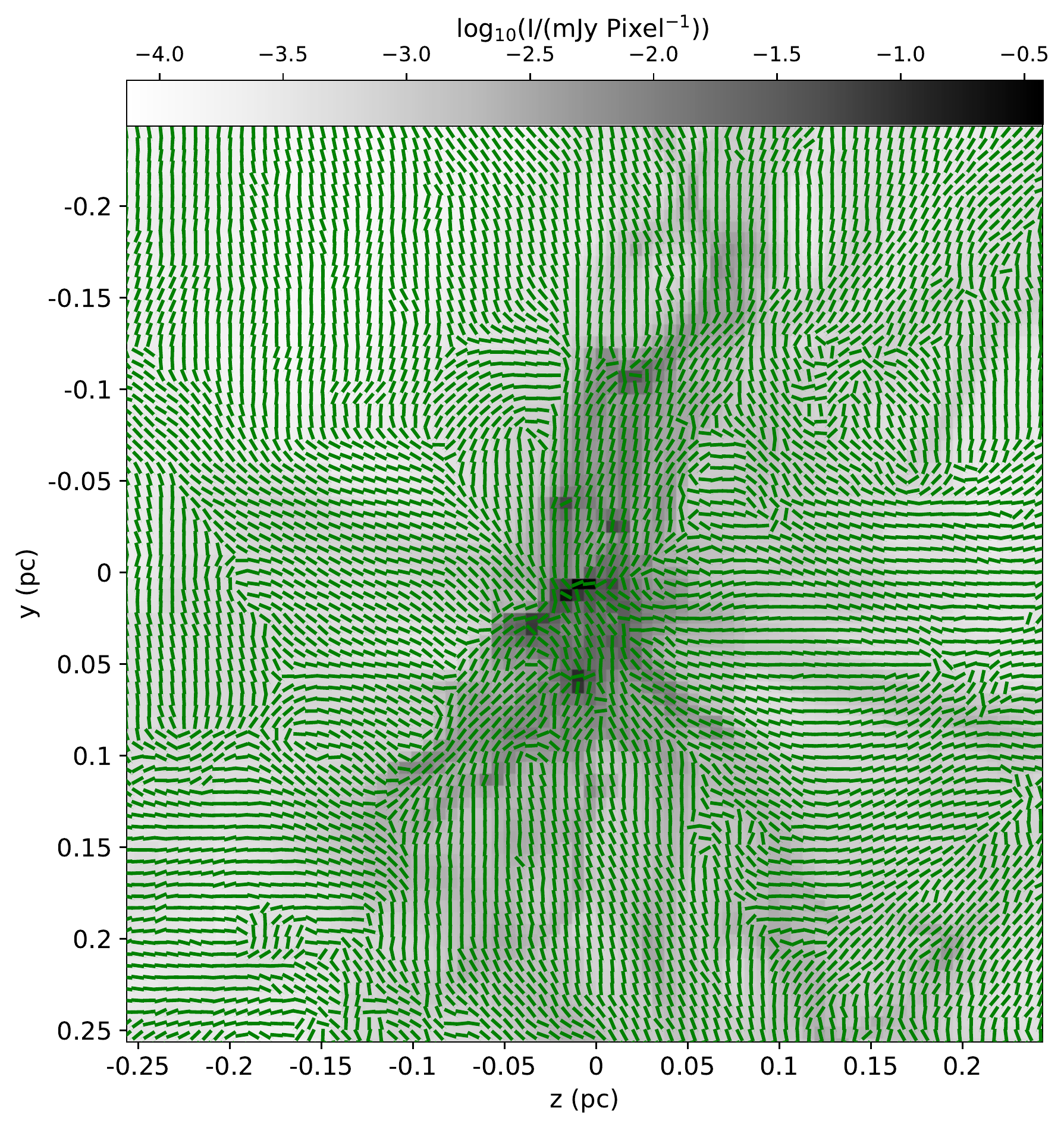}{0.5\textwidth}{(c)}}

\caption{Synthetic observations of model T10M6MU200 at SFE=15\% and at 3 orthogonal projections. Gray scales indicate Stokes $I$ emissions. Line segments represent the orientation of the magnetic field (rotating the orientation of the observed linear polarization by 90$^{\circ}$). Line segments length are arbitrary.   \label{fig:mhdmapmu200}}
\end{figure*}

\begin{figure*}[!htbp]
 \gridline{\fig{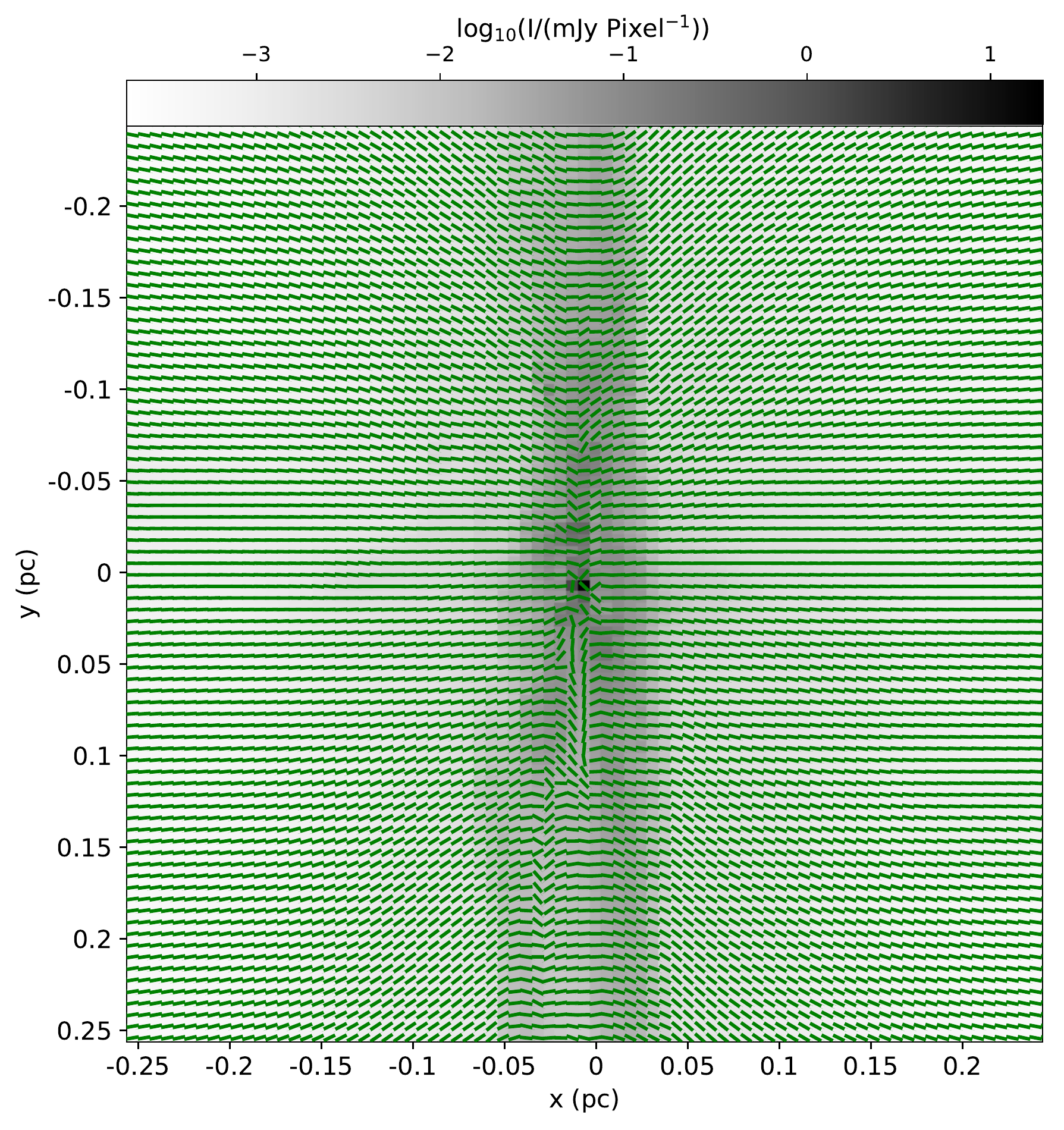}{0.5\textwidth}{(a)}
      \fig{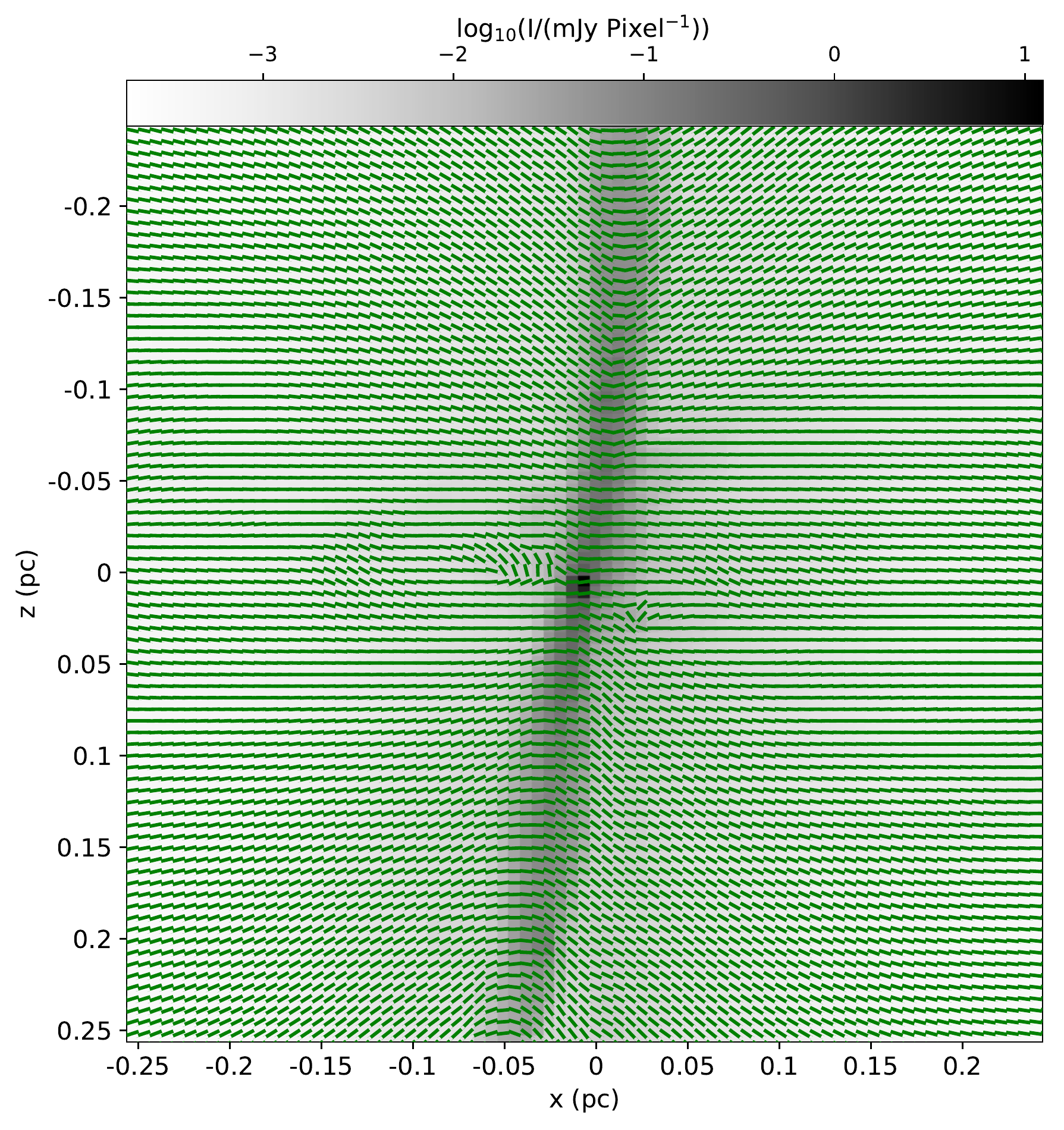}{0.5\textwidth}{(b)}}
\gridline{\fig{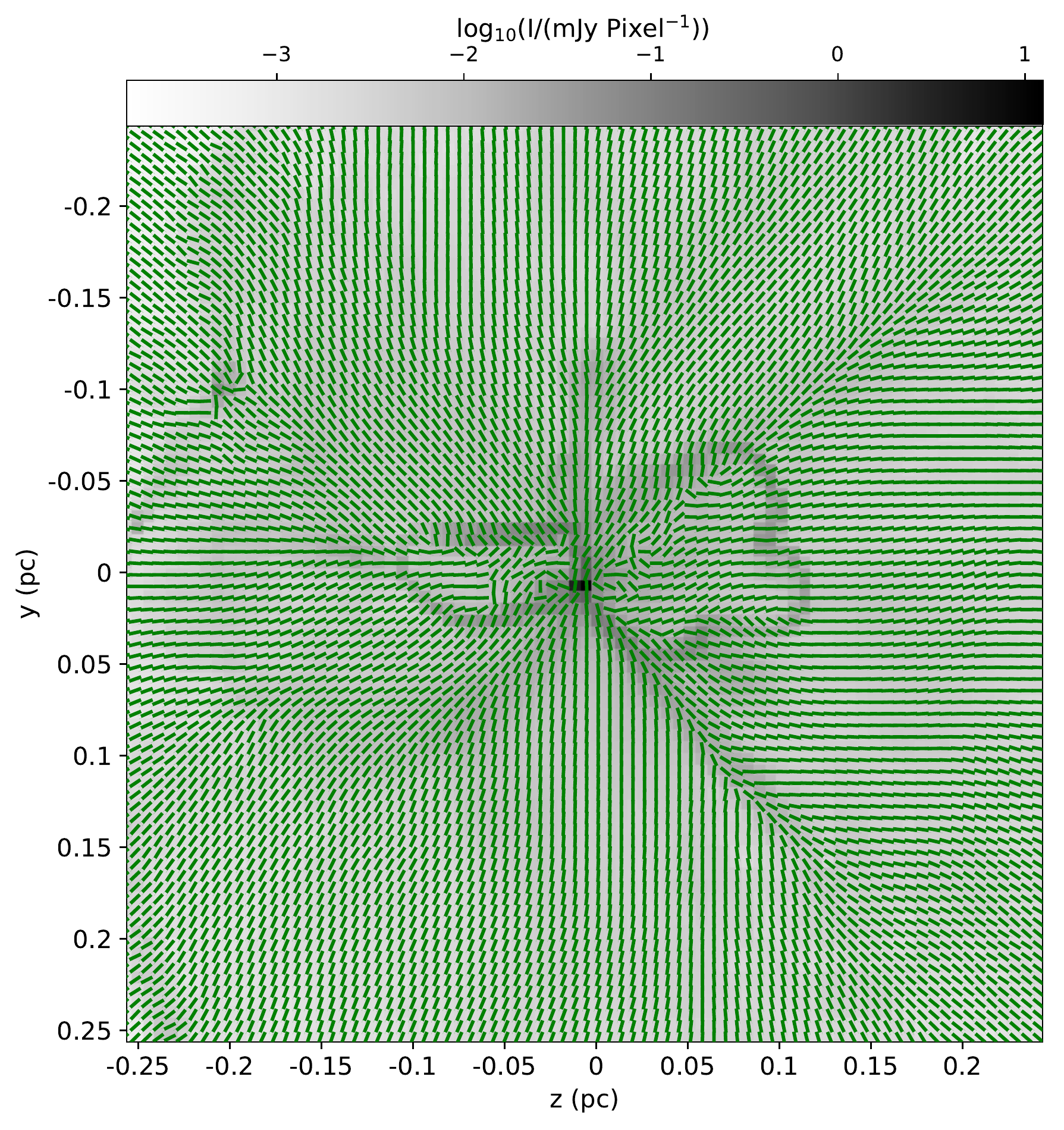}{0.5\textwidth}{(c)}}

\caption{Synthetic observations of model T10M1MU1 at SFE=15\% and at 3 orthogonal projections. Gray scales indicate Stokes $I$ emissions. Line segments represent the orientation of magnetic field (rotating the orientation of the observed linear polarization by 90$^{\circ}$). Line segments length are arbitrary. \label{fig:mhdmapmu1}}
\end{figure*}

\begin{figure*}[!htbp]
 \gridline{\fig{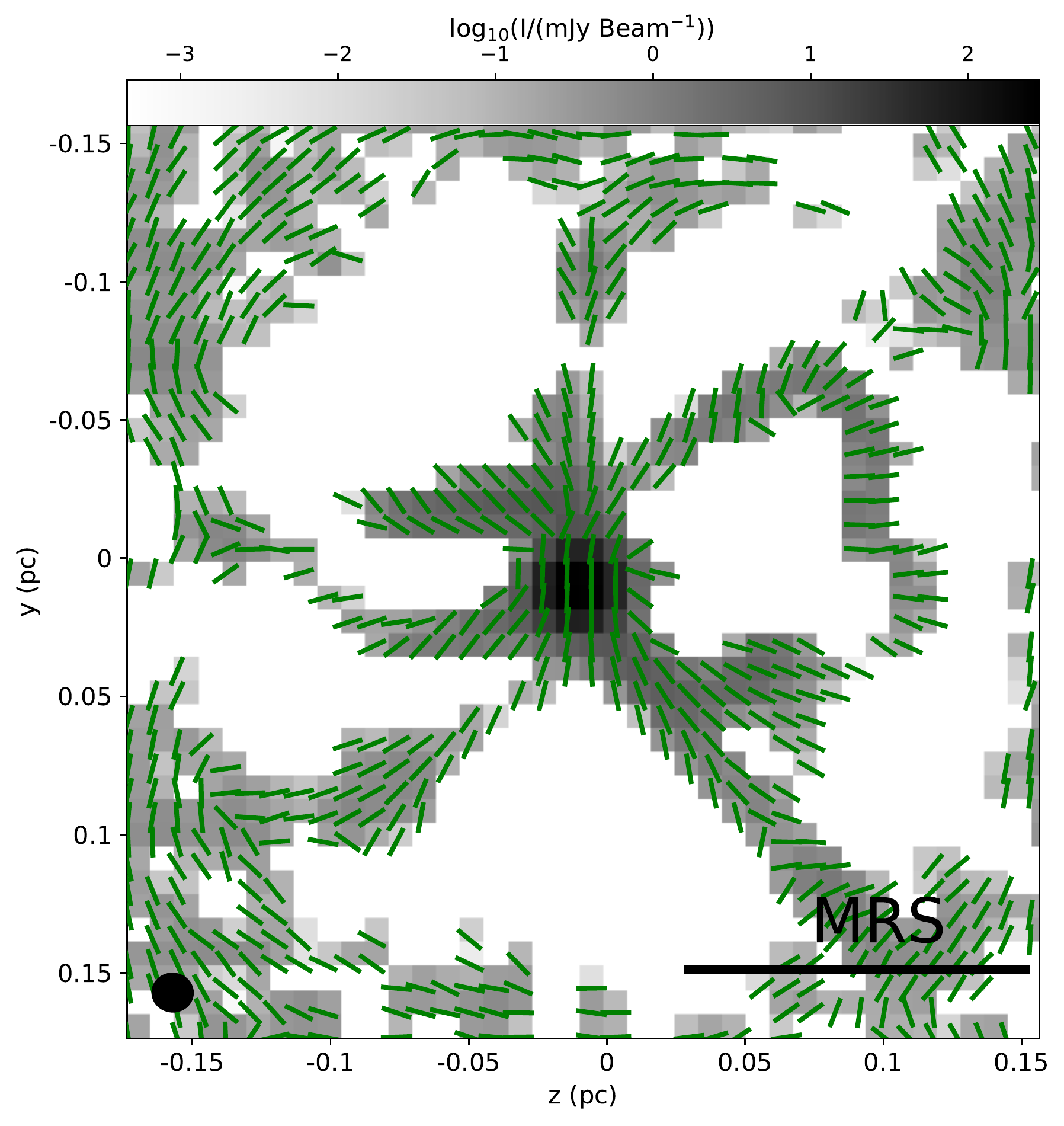}{0.5\textwidth}{(a)}
      \fig{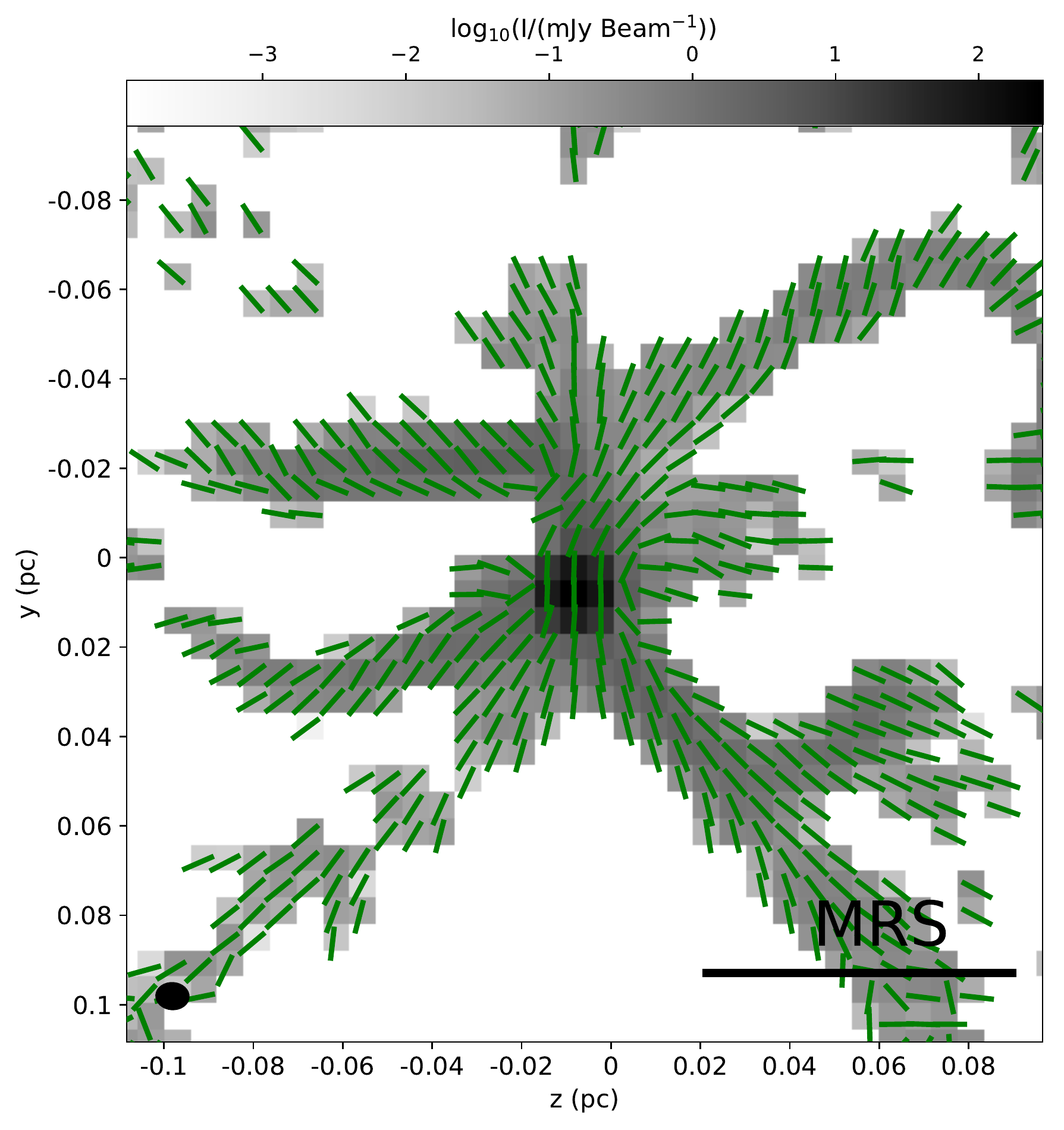}{0.5\textwidth}{(b)}}
\gridline{\fig{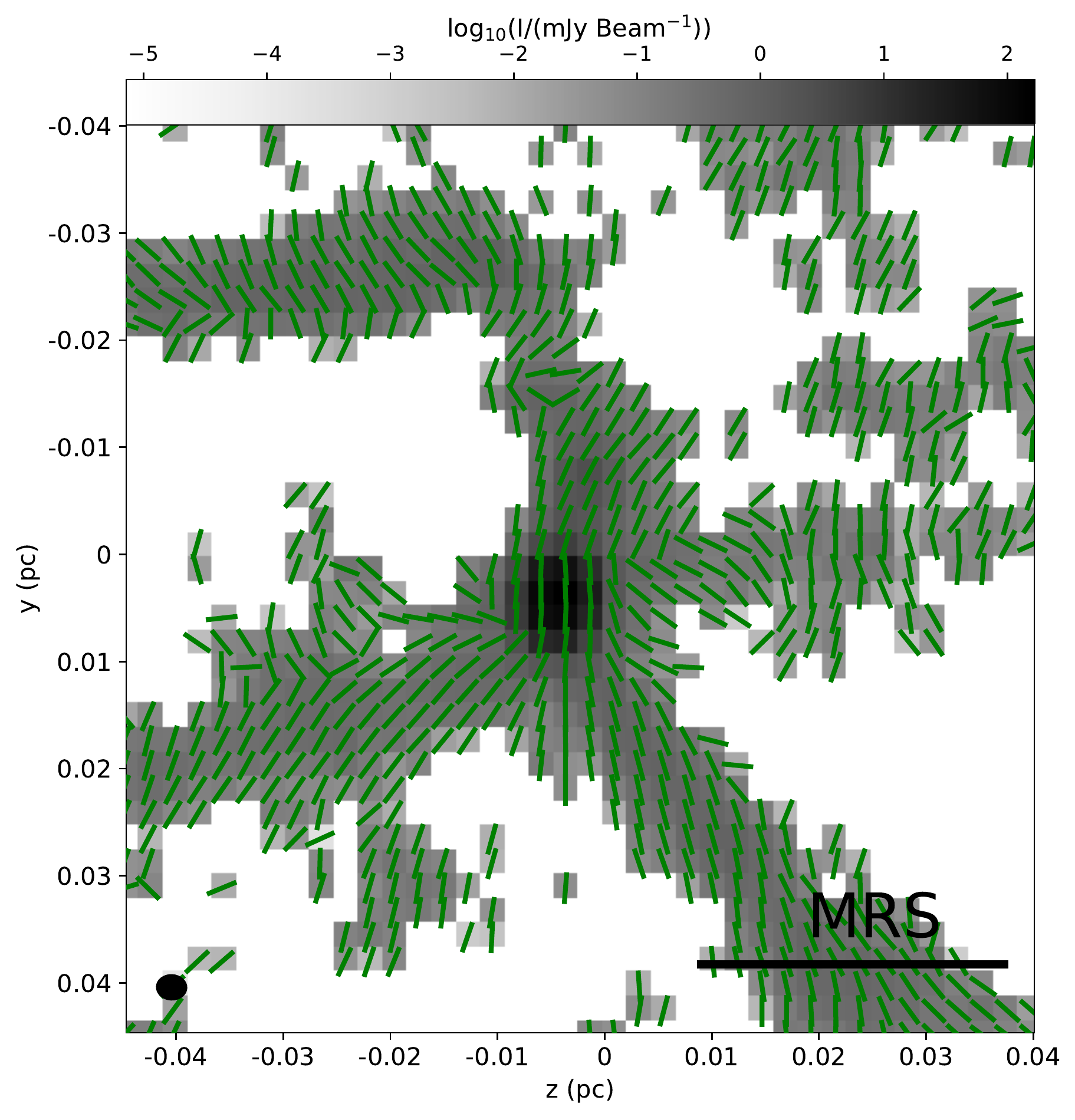}{0.5\textwidth}{(c)}}

\caption{Synthetic ALMA observations of model T10M1MU1 at SFE=15\% and at zy plane. Gray scales indicate Stokes $I$ emissions. Regions with negative $I$ values are not shown. Line segments represent the orientation of magnetic field (rotating the orientation of the observed linear polarization by 90$^{\circ}$). Line segments length are arbitrary. (a) AC1. (b) AC3. (c) AC5. The size of the synthesized beam is shown in the lower left corner of each panel. The size of the maximum recoverable scale is indicated in the lower right corner of each panel. \label{fig:mhdmapmu1zyalma}}
\end{figure*}

Figure \ref{fig:mhdmapmu200} shows the synthetic images of model T10M6MU200 at SFE=15\% as an example of the synthetic observations of a model with weak magnetic field and strong turbulence. Figure \ref{fig:mhdmapmu1} shows the synthetic images of model T10M1MU1 at SFE=15\% as an example of the synthetic observations of a model with strong magnetic field and weak turbulence. Figure \ref{fig:mhdmapmu1zyalma} shows the synthetic ALMA images of model T10M1MU1 at SFE=15\% and at the zy plane as an example of the simulated ALMA observations. All images are centered at the projected position of the most massive sink. 

\section{Effect of grid size}\label{app:gridsize}

The DCF method assumes there is magnetic field perturbations around the mean field component. If the perturbation only dominants at scales smaller than a turbulent correlation scale $l_\delta$, increasing the pixel size at scales smaller than $l_\delta$ could increase the angular dispersion from field perturbation and increasing the pixel size at scales larger than $l_\delta$ does not change the turbulent contribution to the angular dispersion \citep{2001ApJ...561..800H, 2009ApJ...696..567H}. In the presence of field curvature, the pixel size may also affect the directly measured angular dispersion if the complex small-scale field structure is smoothed in large pixel size. Here we investigate how the grid (2D pixel or 3D cell) size can affect the directly measured angular dispersions from 2D polarization maps and 3D simulation models. 

The 2D polarization maps and 3D simulation parameter value grids used in this paper are gridded with sizes of 0.13$\arcsec$ ($\sim$0.0013 pc or $\sim$260 AU). Due to the limitation of computer disk space and time, it is hard for us to perform the analysis with grid sizes smaller than 0.13$\arcsec$. In a similar way to that in Section \ref{sec:compangb}, we measure the angular dispersions from binned polarization maps and magnetic field parameter grids with bin sizes of 5, 13, and 21 and compare them with the angular dispersion from unbinned polarization maps and field parameter grids. 

\begin{figure}[!htbp]
 \gridline{\fig{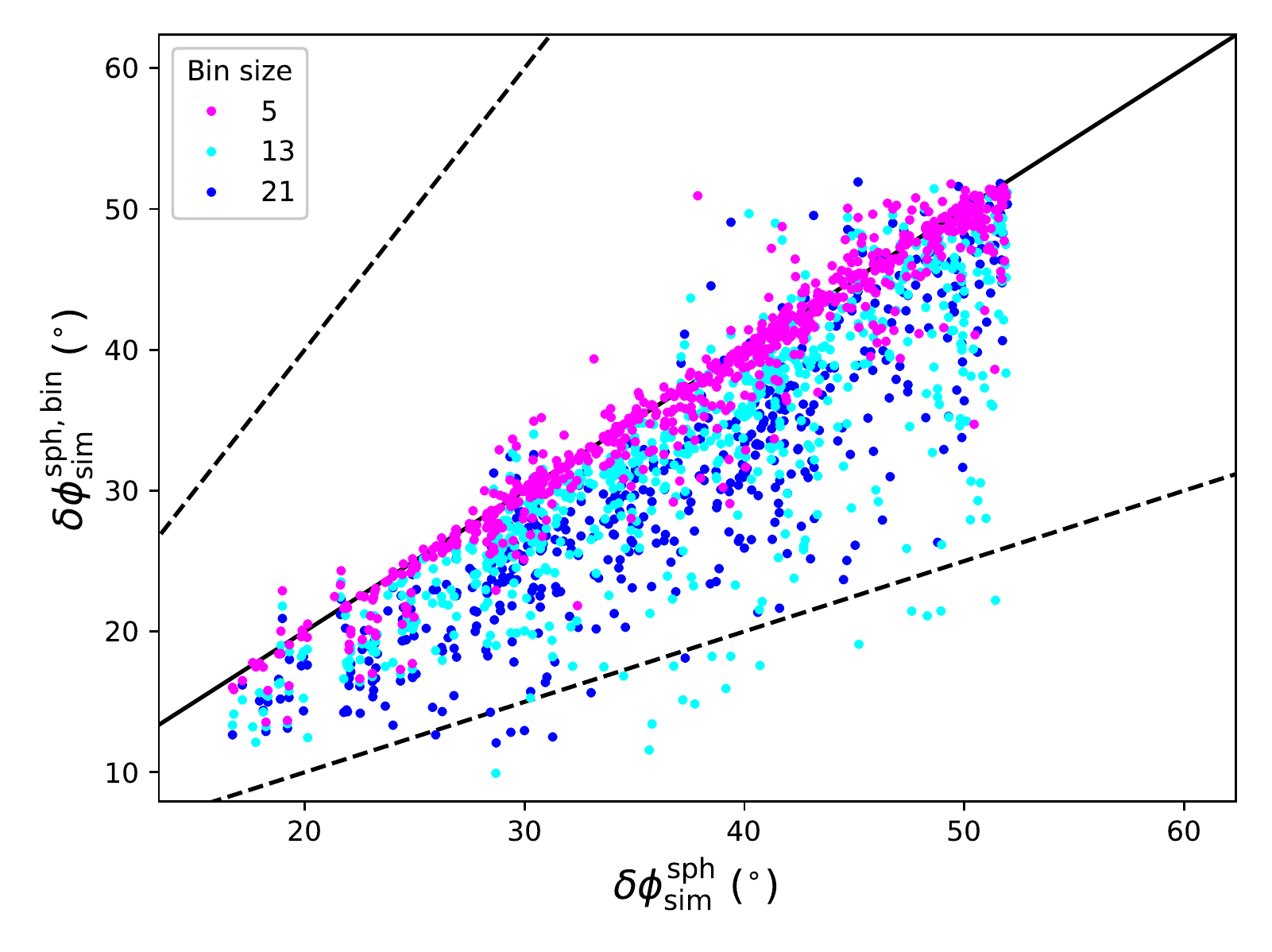}{0.45\textwidth}{(a)}}
 \gridline{\fig{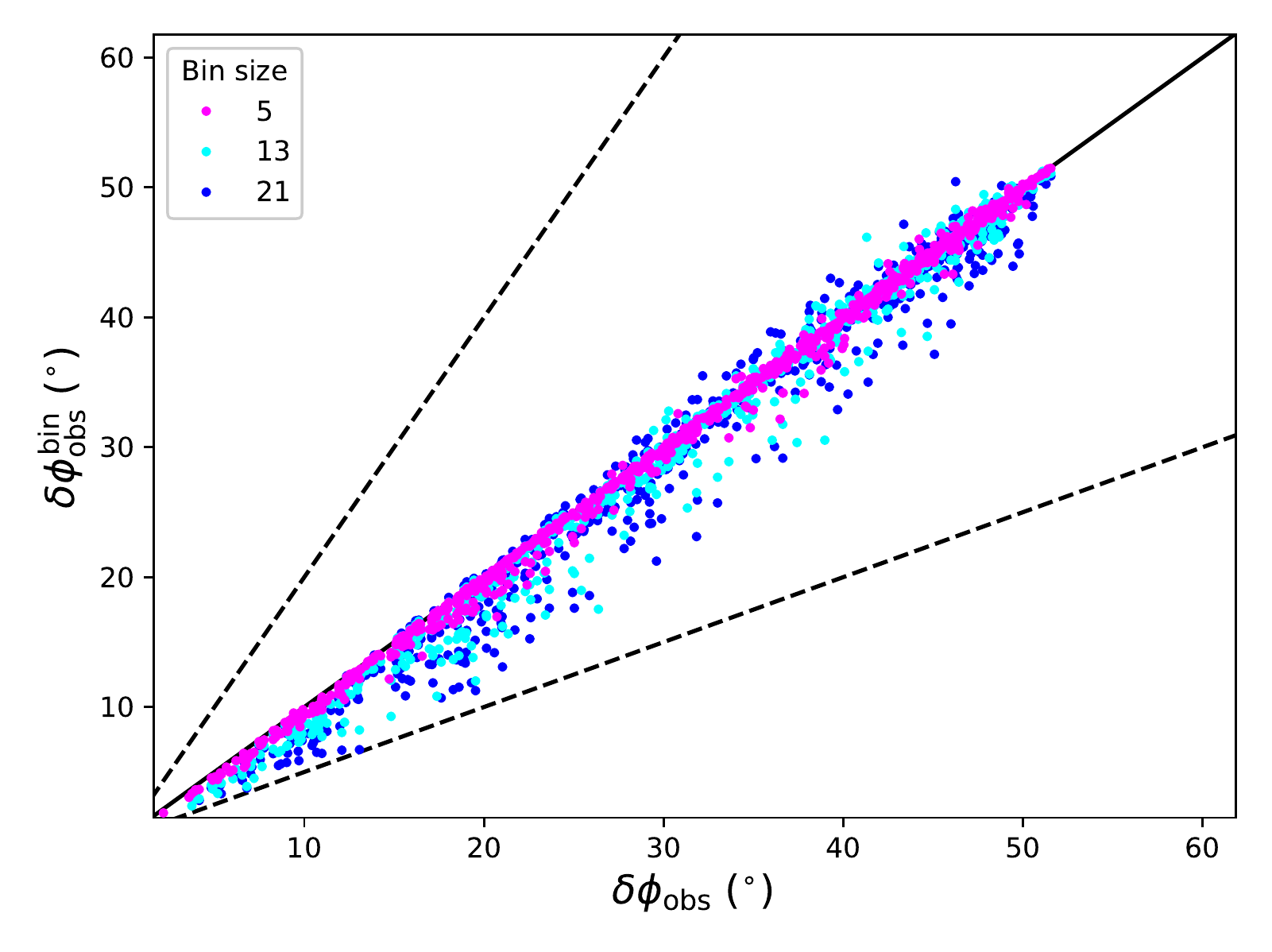}{0.45\textwidth}{(b)}}
\caption{Comparison between the binned angular dispersions and the unbinned angular dispersions in (a) simulations grids and (b) polarization maps. Different colors indicate different bin size. The solid lines indicates the 1:1 relation and the dashed lines indicate the 1:2 and 2:1 relations. \label{fig:gridsize}}
\end{figure}

Figure \ref{fig:gridsize}(a) compares the angular dispersions within spheres of different radii measured from magnetic field parameter grids of different cell size for all simulation models. We do not estimate the angular dispersion if the number of cell is smaller than 50 within a sphere. For most of the estimations, the ratio between the binned angular dispersions and unbinned angular dispersions is 0.5-1, which could be an effect of smoothing below $l_\delta$ (i.e., $l_\delta > 0.0013$ pc) and/or because the ordered field structure is smoothed. For a small number of estimations, the binned angular dispersions are larger than the unbinned angular dispersions, which could be because of combined effects of $l_\delta \lesssim 0.0013$ pc and random measurement scatters.

Figure \ref{fig:gridsize}(b) compares the angular dispersions within circles of different radii measured from polarization maps of different pixel size for all simulation models. We do not estimate the angular dispersion if the number of pixel is smaller than 50 within a circle. The binned angular dispersons generally agree with the unbinned angular dispersions with small scatters. There is only a slight tendency that the binned angular dispersons would underestimate the unbinned angular dispersons. The scatter of Figure \ref{fig:gridsize}(b) is much smaller than Figure \ref{fig:gridsize}(a), which might be because the line-of-sight length is much larger than the turbulent correlation length and the polarization position angles in polarization maps are smoothed along the line of sight. 

Calculation of the ADFs are very time-consuming if there are large numbers of pixels included in the analysis, so we only test a few cases. We find that in the cases we test, the derived $(\langle B_{\mathrm{t}}^2 \rangle/  \langle B_0^2\rangle)_{\mathrm{or,tc}}^{\mathrm{adf}}$ for different pixel sizes agree with each other with small scatters. However, the derived $(\langle B_{\mathrm{t}}^2 \rangle/  \langle B_0^2\rangle)_{\mathrm{or,tc,si}}^{\mathrm{adf}}$ for different pixel sizes can disagree with each other by more than one orders of magnitude, which might be because larger pixel sizes are insufficient to trace the turbulent correlation scale. Thus, we conclude that the ADF method may not correctly account for the signal integration along the line of sight if an inappropriate pixel size is used. 

\section{Accuracy of the ADF method on accounting for the ordered field structure}\label{app:ADForder}

\begin{figure}[!htbp]
 \gridline{\fig{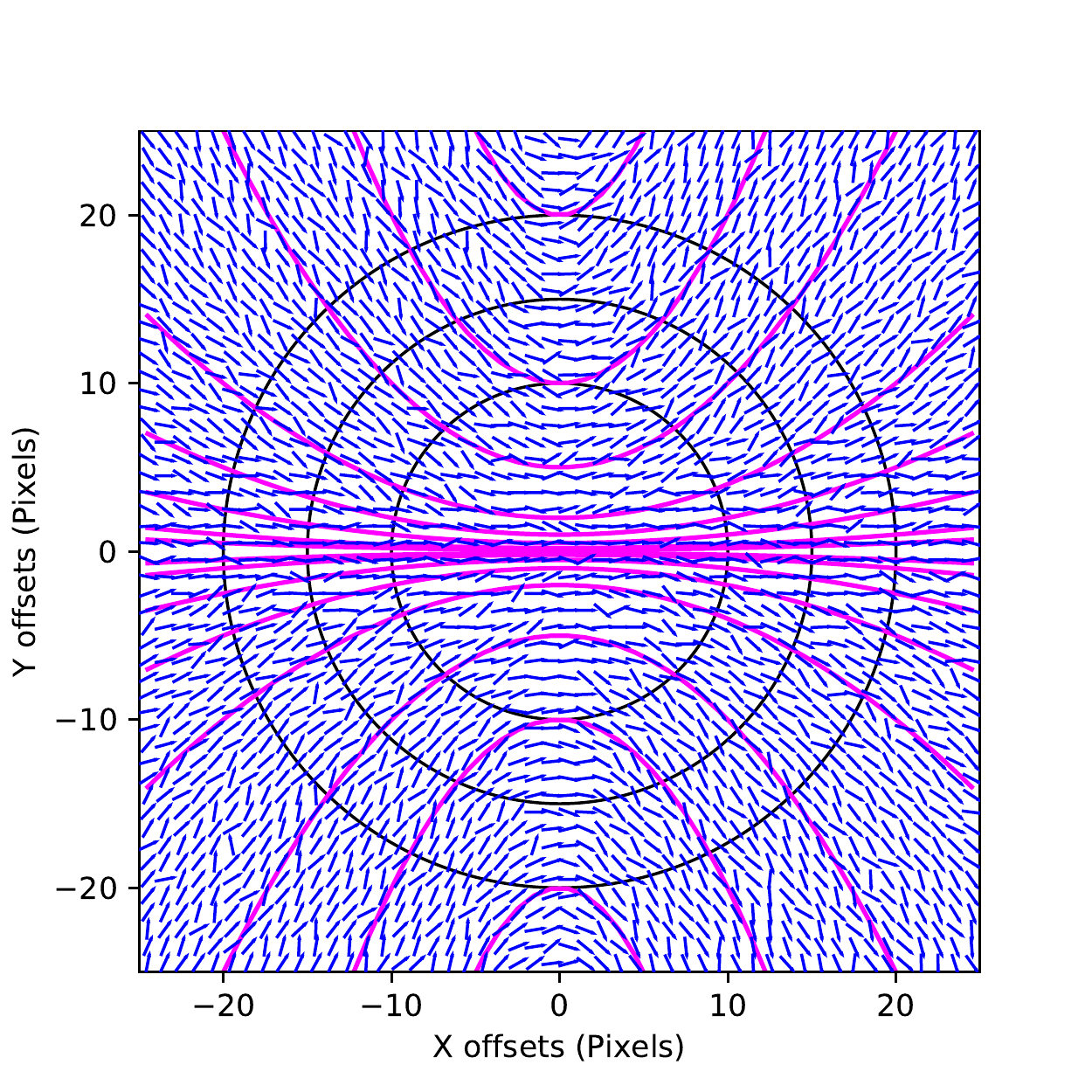}{0.45\textwidth}{}}
\caption{The modeled B segments with an angular dispersion of $\delta \phi = 0.1$ are shown in blue. The interval of B segments is 1 pixel. Magenta curves denote the underlying parabola field models with $C = 0.01 $. Line segments are of unit length. The circles mark the regions within which we derive the ADFs.  \label{fig:mcmap}}
\end{figure}

Here we independently investigate whether the ADF method can accurately account for the contribution from the large-scale field structure with simple Monte Carlo simulations of randomly generated Gaussian angular dispersions on modeled ordered fields. 
Similar to the approach in the Appendix of \citet{2019ApJ...877...43L}, we firstly generate a set of underlying parabola models with the form 
\begin{equation}
y = g + gCx^2.
\end{equation}
We only investigate three different curvatures: $C = 1$, $C = 0.1$, and $C = 0.01$. Then we apply different magnitudes of Gaussian angular dispersions in the forms of $\delta \phi$, $\delta (\sin \phi)$, or $\delta (\tan \phi)$ on the underlying field models. The angular dispersions for randomly distributed $\phi$ and $\sin \phi$ are $\sim$0.91 ($\sim$52$^{\circ}$) and $\sim$0.58, respectively. An example of the field model with $C=0.01$ and $\delta \phi = 0.1$ ($\sim$6$^{\circ}$) is shown in Figure \ref{fig:mcmap}. 

The ADF only accounting for the contribution from the large-scale ordered field is given by \citep{2009ApJ...696..567H, 2009ApJ...706.1504H, 2016ApJ...820...38H}:
\begin{equation} \label{eq:adforder}
1 - \langle \cos \lbrack \Delta \Phi (l)\rbrack \rangle \simeq a_2' l^2 +  (\frac{\langle B_{\mathrm{t}}^2 \rangle}{\langle B^2 \rangle})_{\mathrm{or}}^{\mathrm{adf}} ,
\end{equation}
where $(\langle B_{\mathrm{t}}^2 \rangle/  \langle B^2\rangle)_{\mathrm{or}}^{\mathrm{adf}} = (\langle B_{\mathrm{t}}^2 \rangle/ \langle B_0^2\rangle)_{\mathrm{or}}^{\mathrm{adf}} / ((\langle B_{\mathrm{t}}^2 \rangle/ \langle B_0^2\rangle)_{\mathrm{or}}^{\mathrm{adf}}+1) $ is the turbulent-to-total energy ratio and $(\langle B_{\mathrm{t}}^2 \rangle/  \langle B_0^2\rangle)_{\mathrm{or}}^{\mathrm{adf}}$ is the turbulent-to-ordered energy ratio. \textbf{Since the average of $\cos \lbrack \Delta \Phi (l)\rbrack$ for a randomly distributed $\Delta \Phi (l)$ is $\sim$0.64 and $a_2'$ is defined as positive values, the maximum values of the derived turbulent-to-ordered strength ratio $((\langle B_{\mathrm{t}}^2 \rangle/  \langle B_0^2\rangle)_{\mathrm{or}}^{\mathrm{adf}})^{1/2}$ and turbulent-to-total strength ratio $((\langle B_{\mathrm{t}}^2 \rangle/  \langle B^2\rangle)_{\mathrm{or}}^{\mathrm{adf}})^{1/2}$ are $\sim$0.76 and $\sim$0.6, respectively.} We derive the ADFs within radii of 10, 15, and 20 pixels with respect to the center of each model and fit each ADF with Equation \ref{eq:adforder} in a similar way to that in Section \ref{sec:adf}.  

\begin{figure}[!htbp]
 \gridline{\fig{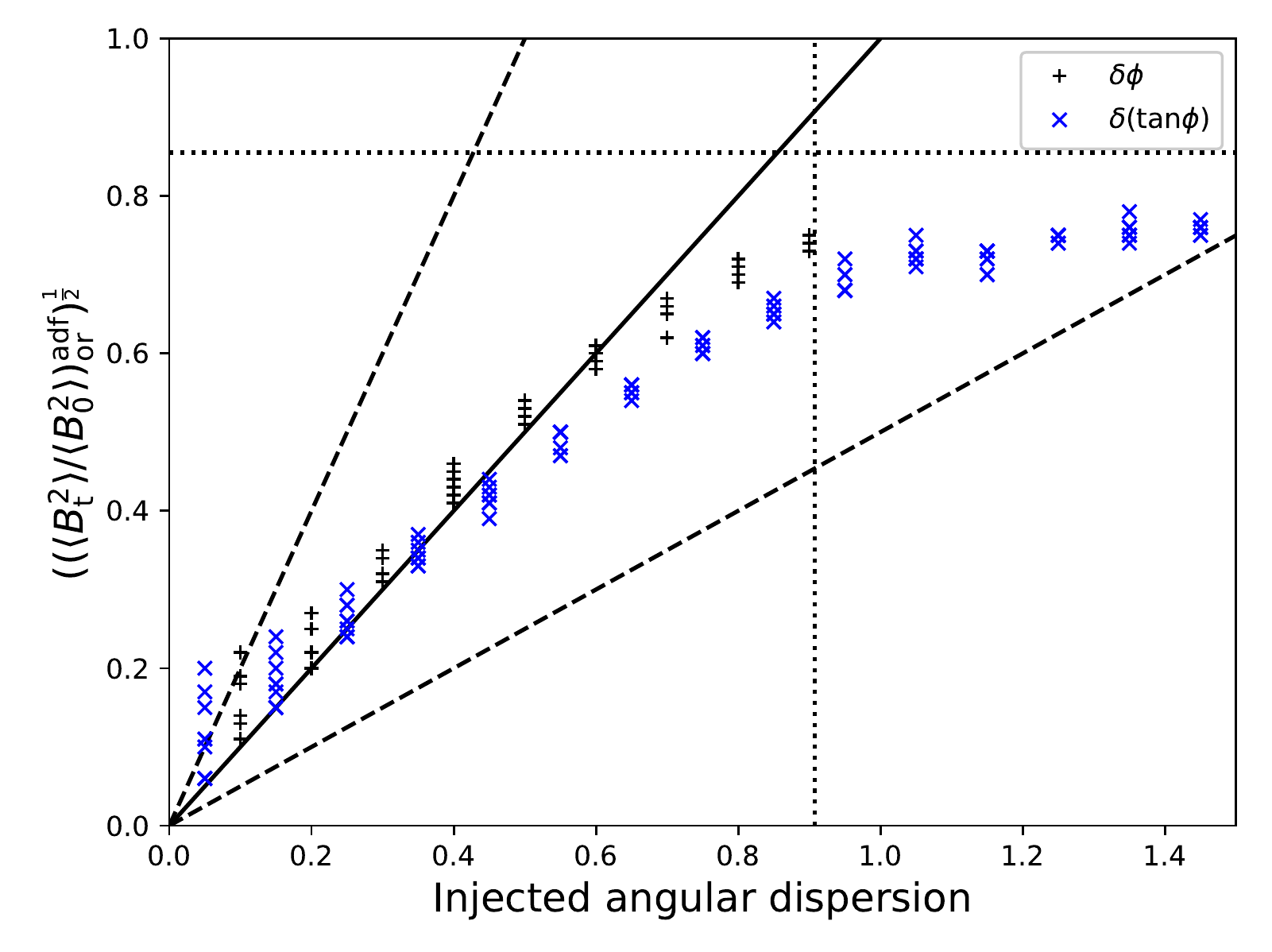}{0.45\textwidth}{(a)}}
\gridline{\fig{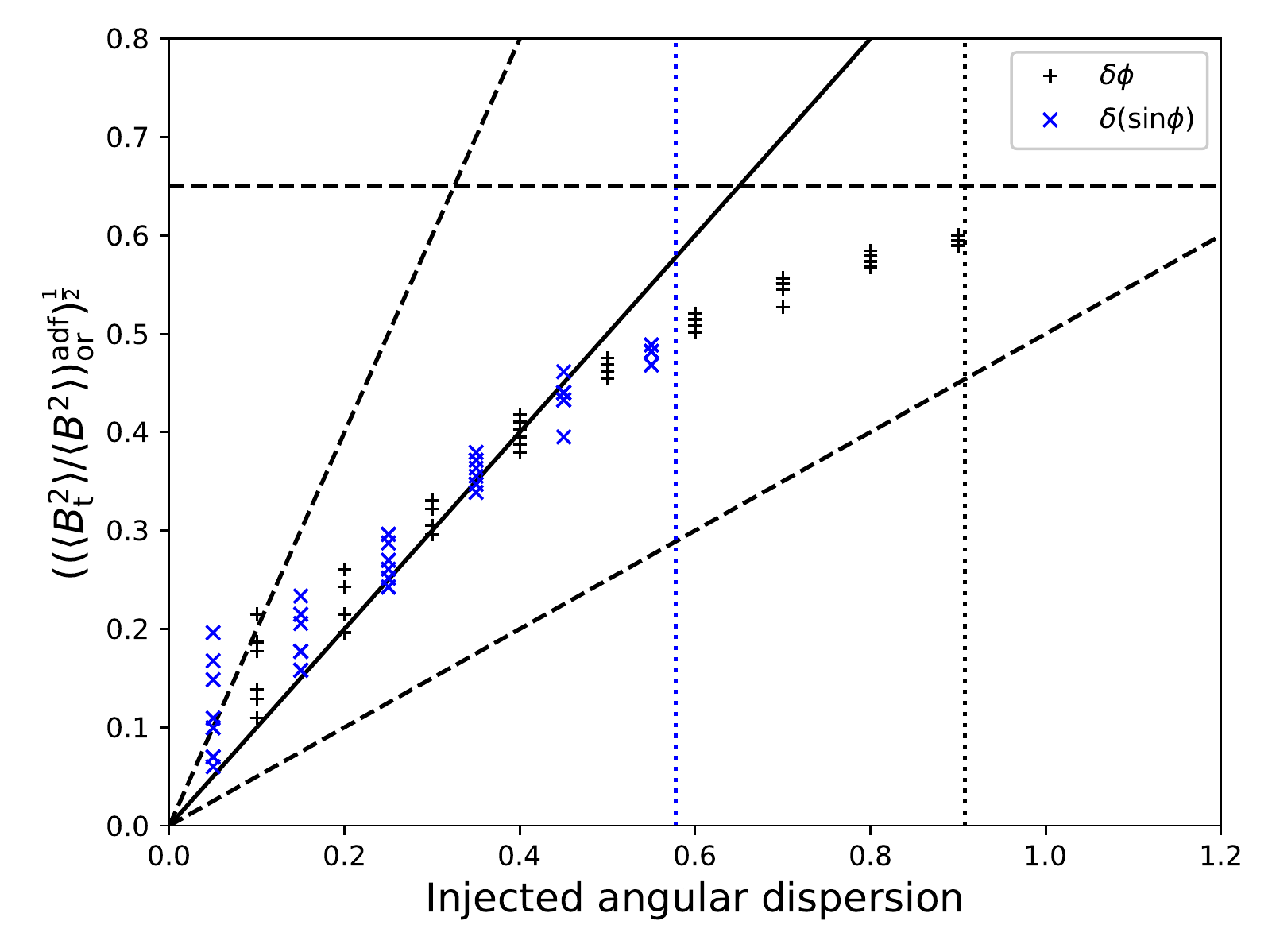}{0.45\textwidth}{(b)}}
\caption{(a). Estimated turbulent-to-ordered field strength ratio versus injected angular dispersion in forms of $\delta \phi$ (black) and $\delta (\tan \phi)$ (blue). The horizonal dashed line corresponds to $((\langle B_{\mathrm{t}}^2 \rangle/  \langle B_0^2\rangle)_{\mathrm{or}}^{\mathrm{adf}})^{1/2} = 0.86$. The vertical dashed line corresponds to $\delta \phi = 0.91$. (b). Estimated turbulent-to-total field strength ratio versus injected angular dispersion in forms of $\delta \phi$ (black) and $\delta (\sin \phi)$ (blue). The horizonal dotted line corresponds to $((\langle B_{\mathrm{t}}^2 \rangle/  \langle B^2\rangle)_{\mathrm{or}}^{\mathrm{adf}})^{1/2} = 0.65$. The vertical dotted lines correspond to $\delta \phi = 0.91$ and $\delta (\sin \phi) = 0.58$. In both (a) and (b), the solid lines indicates the 1:1 relation and the dashed lines indicate the 1:2 and 2:1 relations. \label{fig:mc_adfresult}}
\end{figure}

Figures \ref{fig:mc_adfresult}(a) shows the derived turbulent-to-ordered strength ratio $((\langle B_{\mathrm{t}}^2 \rangle/  \langle B_0^2\rangle)_{\mathrm{or}}^{\mathrm{adf}})^{1/2}$ compared to the injected angular dispersion in forms of $\delta \phi$ and $\delta (\tan \phi)$. As shown by Figures \ref{fig:mc_adfresult}(a), the derived $((\langle B_{\mathrm{t}}^2 \rangle/  \langle B_0^2\rangle)_{\mathrm{or}}^{\mathrm{adf}})^{1/2}$ may overestimate the injected angular dispersion by a factor of more than 2 when the $((\langle B_{\mathrm{t}}^2 \rangle/  \langle B_0^2\rangle)_{\mathrm{or}}^{\mathrm{adf}})^{1/2}$ is smaller than $\sim$0.2. Other than the overestimation at small angles, the derived $((\langle B_{\mathrm{t}}^2 \rangle/  \langle B_0^2\rangle)_{\mathrm{or}}^{\mathrm{adf}})^{1/2}$ generally agrees with the injected $\delta \phi$ within a factor of 2. However, $\delta (\tan \phi)$ could be very large when $\phi$ approaches -90$^{\circ}$ or 90$^{\circ}$. Thus the derived $((\langle B_{\mathrm{t}}^2 \rangle/  \langle B_0^2\rangle)_{\mathrm{or}}^{\mathrm{adf}})^{1/2}$ would become insensitive to $\delta (\tan \phi)$ and greatly underestimate $\delta (\tan \phi)$ as $\delta (\tan \phi)$ increases.

Figures \ref{fig:mc_adfresult}(b) shows the derived turbulent-to-total strength ratio $((\langle B_{\mathrm{t}}^2 \rangle/  \langle B^2\rangle)_{\mathrm{or}}^{\mathrm{adf}})^{1/2}$ compared to the injected angular dispersion in forms of $\delta \phi$ and $\delta (\sin \phi)$. Similarly, the derived $((\langle B_{\mathrm{t}}^2 \rangle/  \langle B^2\rangle)_{\mathrm{or}}^{\mathrm{adf}})^{1/2}$ may overestimate the injected angular dispersion by a factor of more than 2 when the $((\langle B_{\mathrm{t}}^2 \rangle/  \langle B^2\rangle)_{\mathrm{or}}^{\mathrm{adf}})^{1/2}$ is smaller than $\sim$0.2. Other than the overestimation at small angles,  the derived $((\langle B_{\mathrm{t}}^2 \rangle/  \langle B^2\rangle)_{\mathrm{or}}^{\mathrm{adf}})^{1/2}$ generally agrees with the injected angular dispersion within a factor of 2.

\begin{figure}[!htbp]
 \gridline{\fig{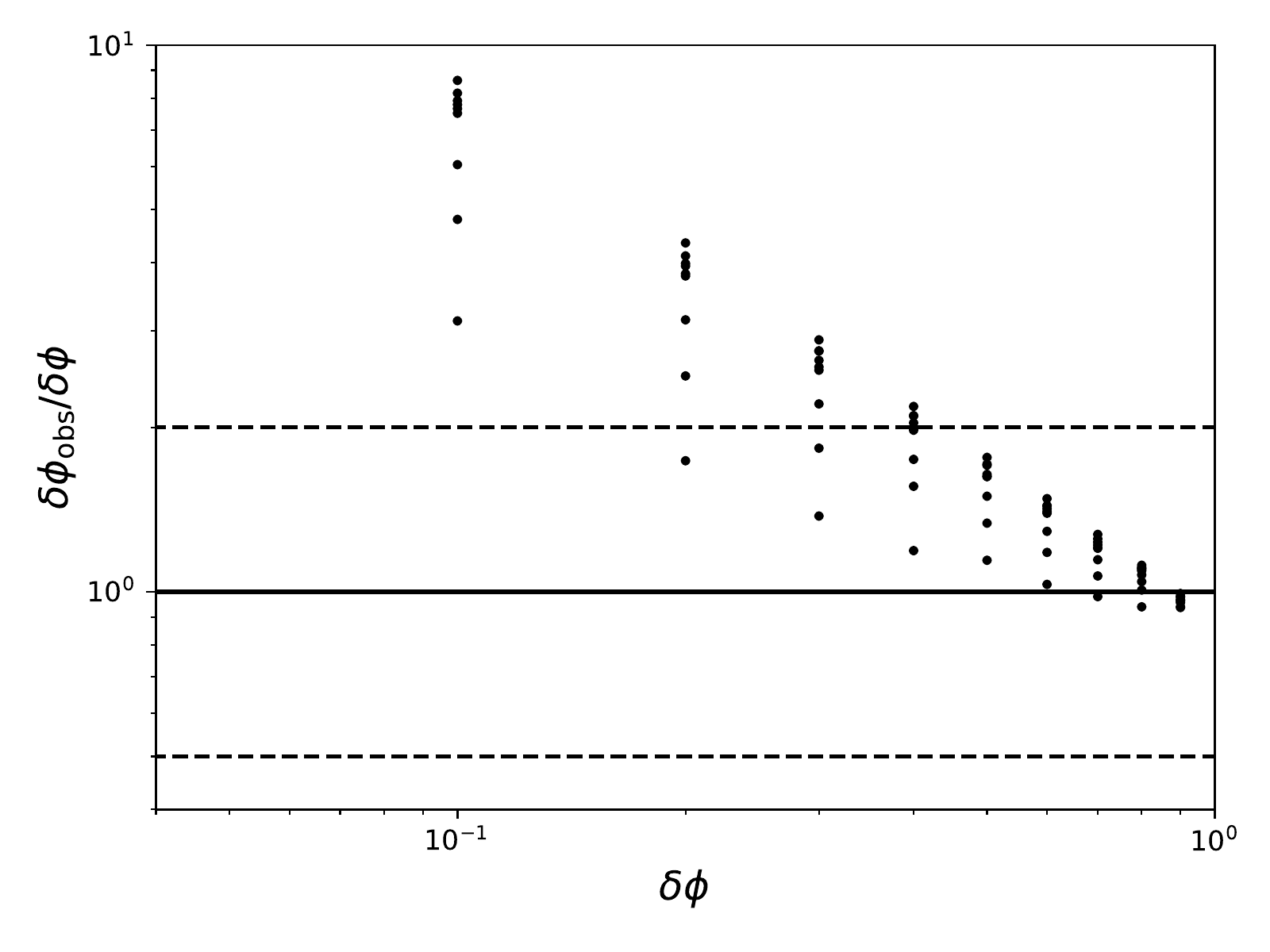}{0.45\textwidth}{(a)}}
\gridline{\fig{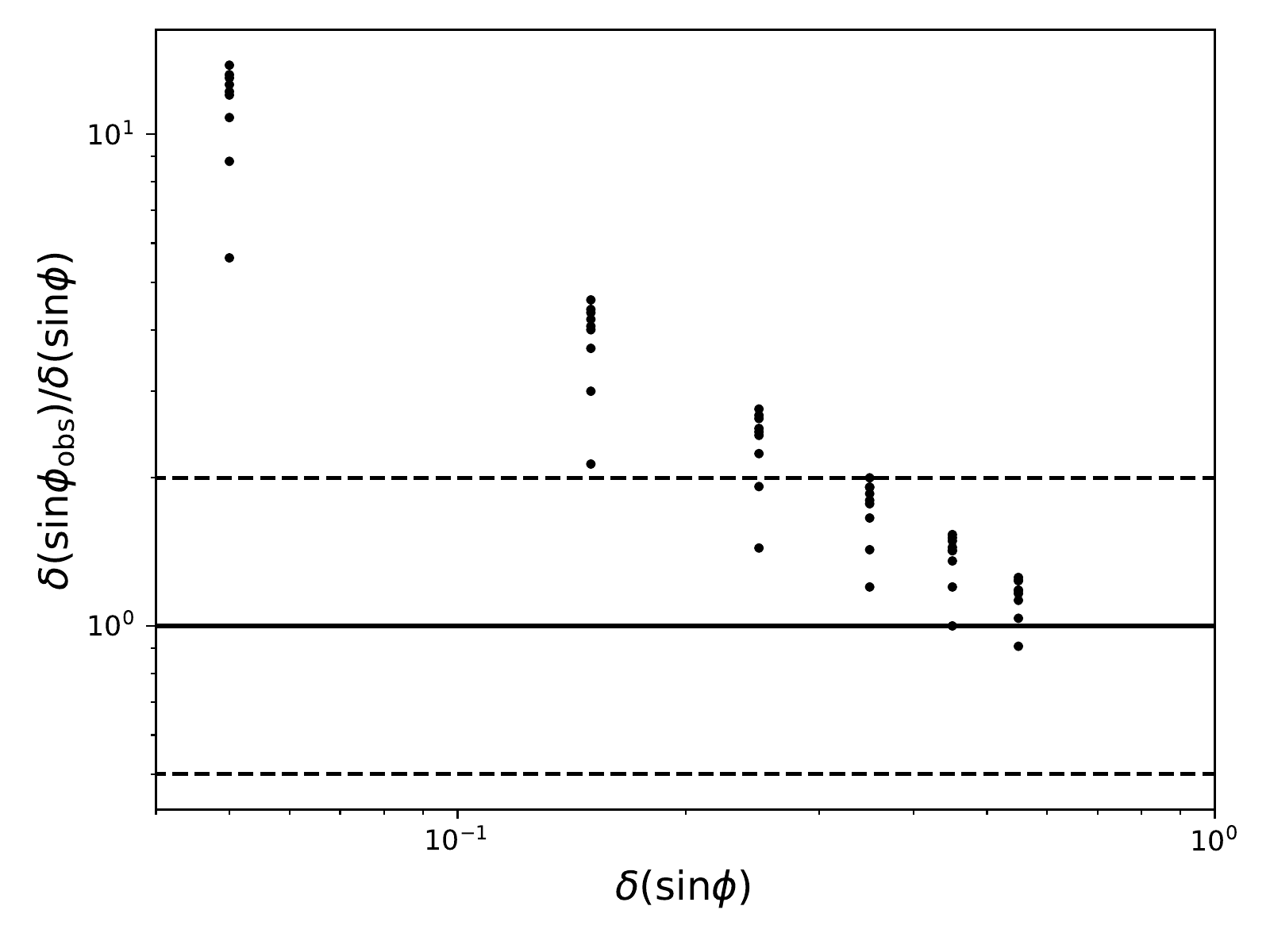}{0.45\textwidth}{(b)}}
\gridline{\fig{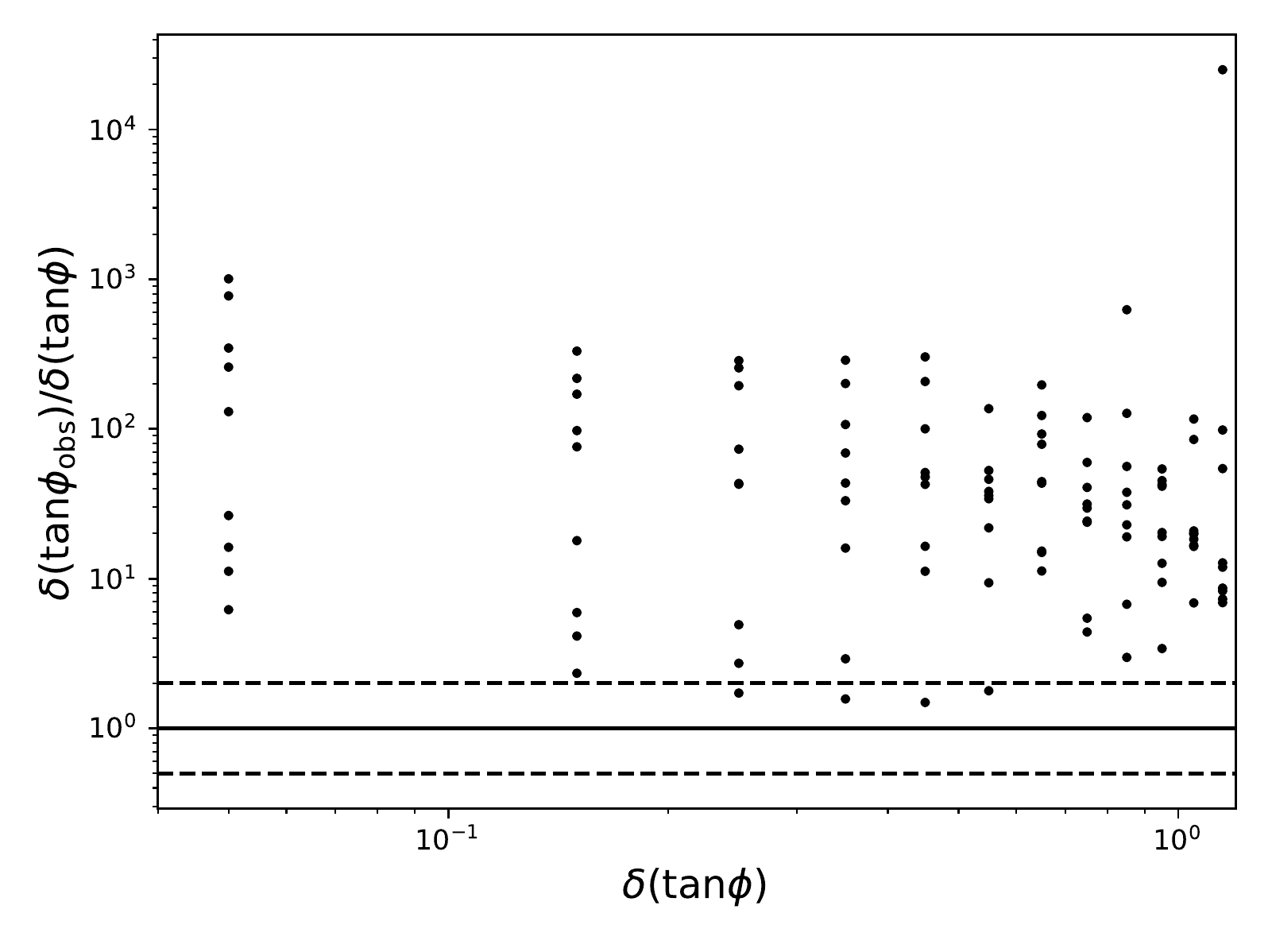}{0.45\textwidth}{(c)}}
\caption{Estimated angular dispersions compared to injected angular dispersions. (a). $\delta \phi_{\mathrm{obs}}$ versus $\delta \phi$. (b). $\delta (\sin\phi_{\mathrm{obs}})$ versus $\delta (\sin\phi)$. (c). $\delta (\tan\phi_{\mathrm{obs}})$ versus $\delta (\tan\phi)$.   \label{fig:mc_ang_ang}}
\end{figure}

We also measured the raw angular dispersions, $\delta \phi_{\mathrm{obs}}$, $\delta (\sin \phi_{\mathrm{obs}})$, and $\delta (\tan \phi_{\mathrm{obs}})$, in a similar way to that in Section \ref{sec:angobs_angsim}. Figure \ref{fig:mc_ang_ang} compares the estimated angular dispersions with the injected angular dispersions. Generally, the estimated angular dispersions are greater than the injected angular dispersions because of the contribution from large-scale field structure. As shown in Figures \ref{fig:mc_ang_ang} (a) and (b), the direct estimation can significantly overestimate the angular dispersion for small injected angular dispersions. Figure \ref{fig:mc_ang_ang} (c) indicates that the estimated $\delta (\tan \phi_{\mathrm{obs}})$ generally would greatly overestimate the injected angular dispersion not only because of the contribution from large-scale field structure but also because $\delta (\tan \phi)$ is very sensitive to $\phi$ when $\phi \sim \pm 90 \degr$. 

\section{Projection effect for magnetic fields and velocity fields}\label{app:projection}

Ultimately we want to obtain the 3D magnetic field strength, while the DCF method only gives the information about the plane-of-sky magnetic field component. The line-of-sight magnetic field strength can be determined by Zeeman observations. However, the tracers of Zeeman splitting are more sensitive in the low-density materials \citep{2012ARA&A..50...29C} and we lack the methods to determine the line-of-sight field component in high-density regions. 

Alternatively, \citet{2004ApJ...600..279C} proposed that there is a statistical relation between the 3D uniform magnetic field strength $B^{\mathrm{u}}_{\mathrm{3d}}$ and the plane-of-sky uniform magnetic field strength $B^{\mathrm{u}}_{\mathrm{pos}}$:
\begin{equation}\label{eq:eqstatpos3du}
B^{\mathrm{u}}_{\mathrm{3d}} = \frac{4}{\pi}\overline{B^{\mathrm{u}}_{\mathrm{pos}}}.
\end{equation}
Similarly, we propose that there is also a statistical relation between the 3D total magnetic field strength $B^{\mathrm{tot}}_{\mathrm{3d}}$ and the plane-of-sky total magnetic field strength $B^{\mathrm{tot}}_{\mathrm{pos}}$:
\begin{equation}\label{eq:eqstatpos3dtot}
B^{\mathrm{tot}}_{\mathrm{3d}} = \sqrt{\frac{3}{2}}\overline{B^{\mathrm{tot}}_{\mathrm{pos}}}
\end{equation}
since $(B^{\mathrm{tot}}_{\mathrm{3d}})^2 = (B^{\mathrm{tot}}_{\mathrm{los}})^2 + (B^{\mathrm{tot}}_{\mathrm{pos\perp}})^2 + (B^{\mathrm{tot}}_{\mathrm{pos\parallel}})^2$.

\begin{figure}[!htbp]
 \gridline{\fig{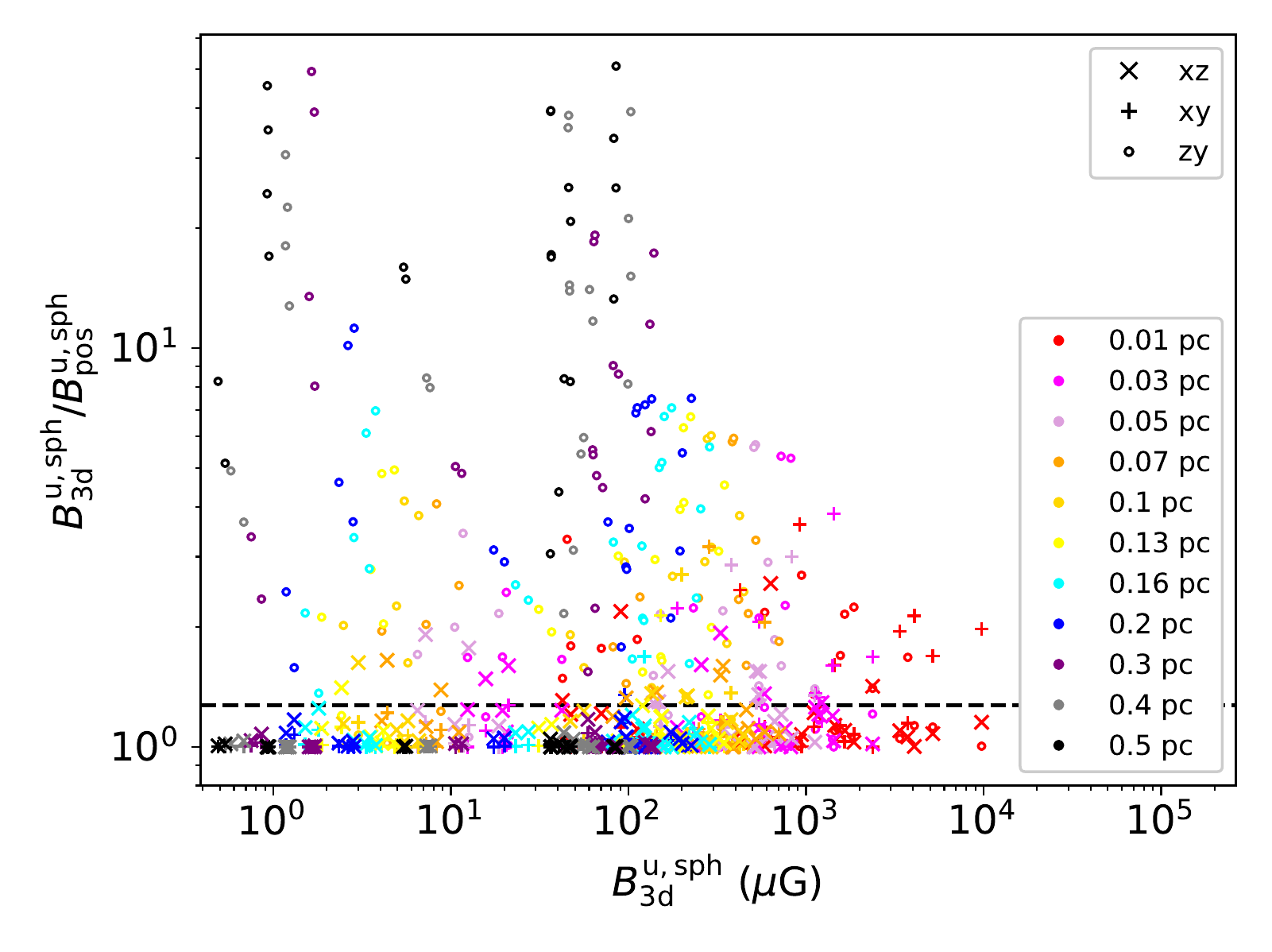}{0.45\textwidth}{a}}
 \gridline{\fig{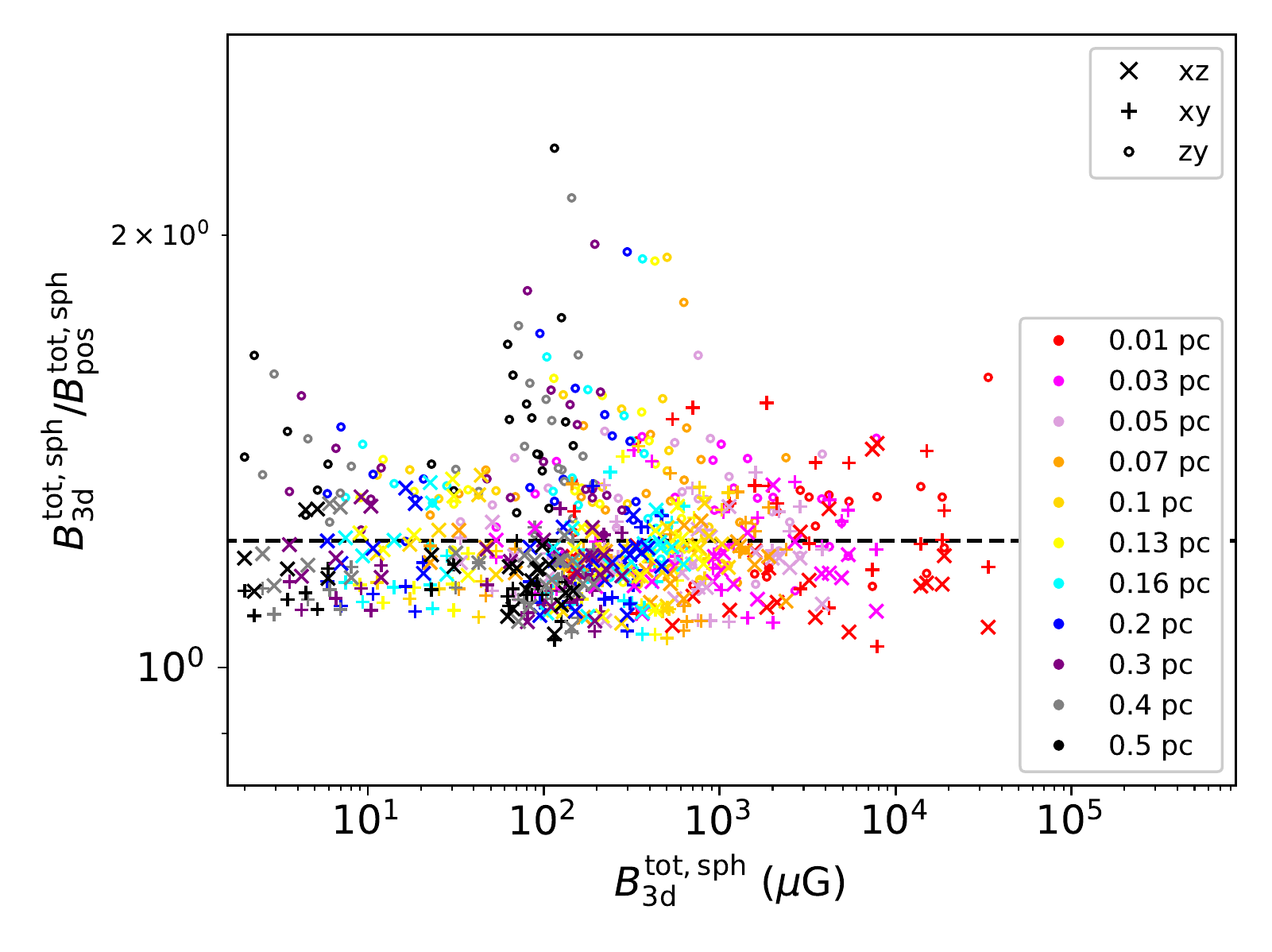}{0.45\textwidth}{b}}
\caption{(a). The ratio between the 3D uniform magnetic field strength and the plane-of-sky uniform magnetic field strength within spheres of different radii for all models. (b). The ratio between the 3D total magnetic field strength and the plane-of-sky total magnetic field strength within spheres of different radii for all models. Different symbols correspond to different projections. Different colors represents models with radii. \label{fig:Bproj}}
\end{figure}

Figure \ref{fig:Bproj}(a) shows the ratio between the 3D uniform magnetic field strength and the plane-of-sky uniform magnetic field strength within spheres of different radii in all simulation models. It turns out that the ratio between the 3D and plane-of-sky uniform field strength are below $4/\pi$ and only slightly greater than 1 if the line-of-sight (i.e., y or z) is perpendicular to the initial magnetic field orientation (i.e., x). On the other hand, if the line-of-sight is along the initial magnetic field orientation, the ratio between the 3D and plane-of-sky uniform field strength could be greater than 10 and the ratio seems to be higher for larger radius. Thus, we conclude that this statistical relation for the uniform magnetic field strength could be very unreliable if the mean magnetic field orientation is along the line of sight. 

Figure \ref{fig:Bproj}(b) shows the ratio between the 3D total magnetic field strength and the plane-of-sky total magnetic field strength within spheres of different radii in all simulation models. For most estimations, the ratio between the 3D and plane-of-sky total field strength is around $\sqrt{3/2}$. The strength ratio for line-of-sights perpendicular to the initial magnetic field orientation tends to be slight lower than $\sqrt{3/2}$. The strength ratio for line-of-sight along the initial magnetic field orientation tends to be higher than $\sqrt{3/2}$, with a upper value of $\sim$2.3 (smaller than 2 $\times \sqrt{3/2} \approx 2.4$). Thus, we conclude that the statistical relation for the total magnetic field strength can give reliable estimates of the 3D total magnetic field strength that is accurate within a factor of 2. 

\begin{figure}[!htbp]
 \gridline{\fig{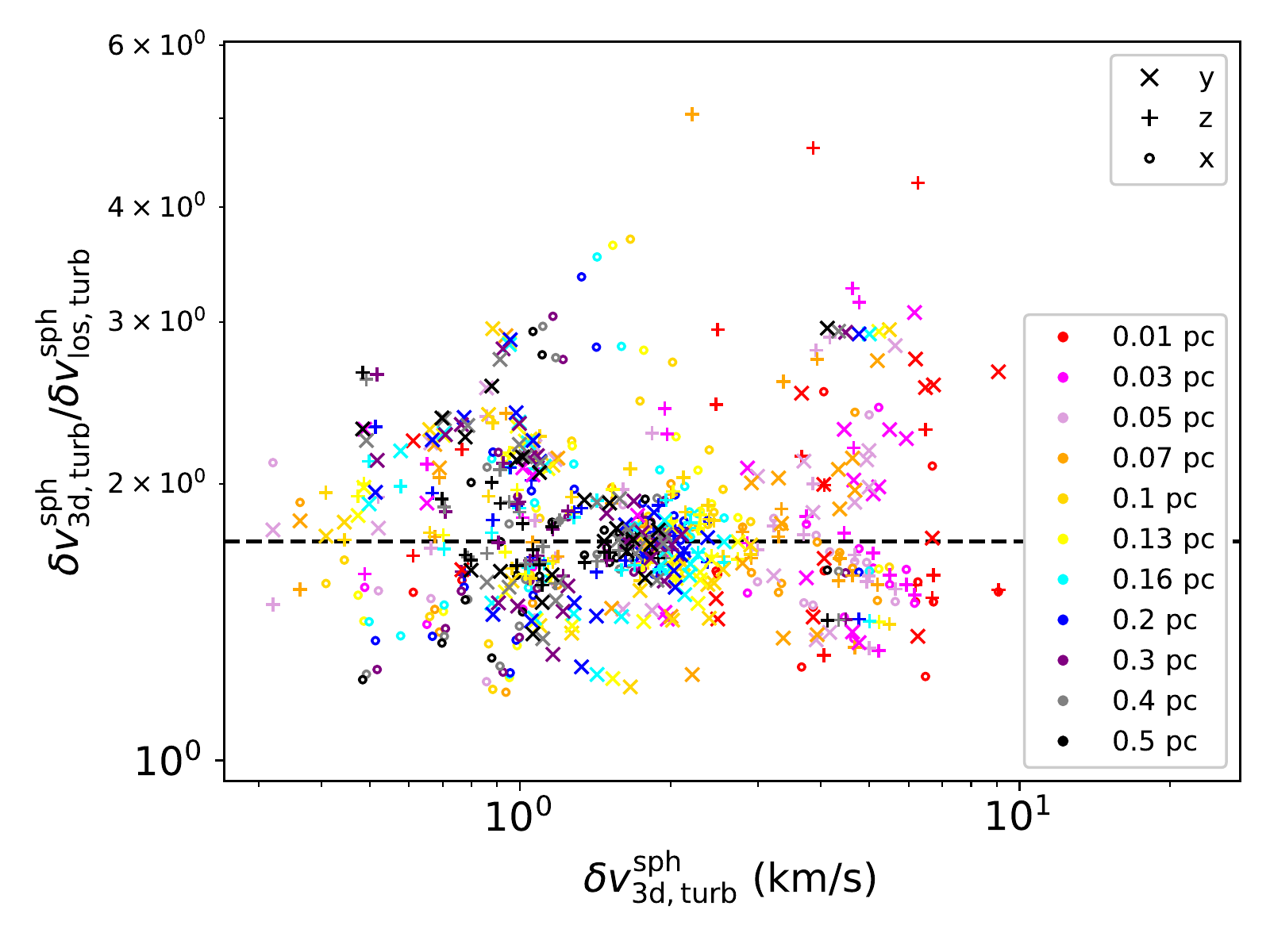}{0.45\textwidth}{a}}
  \gridline{\fig{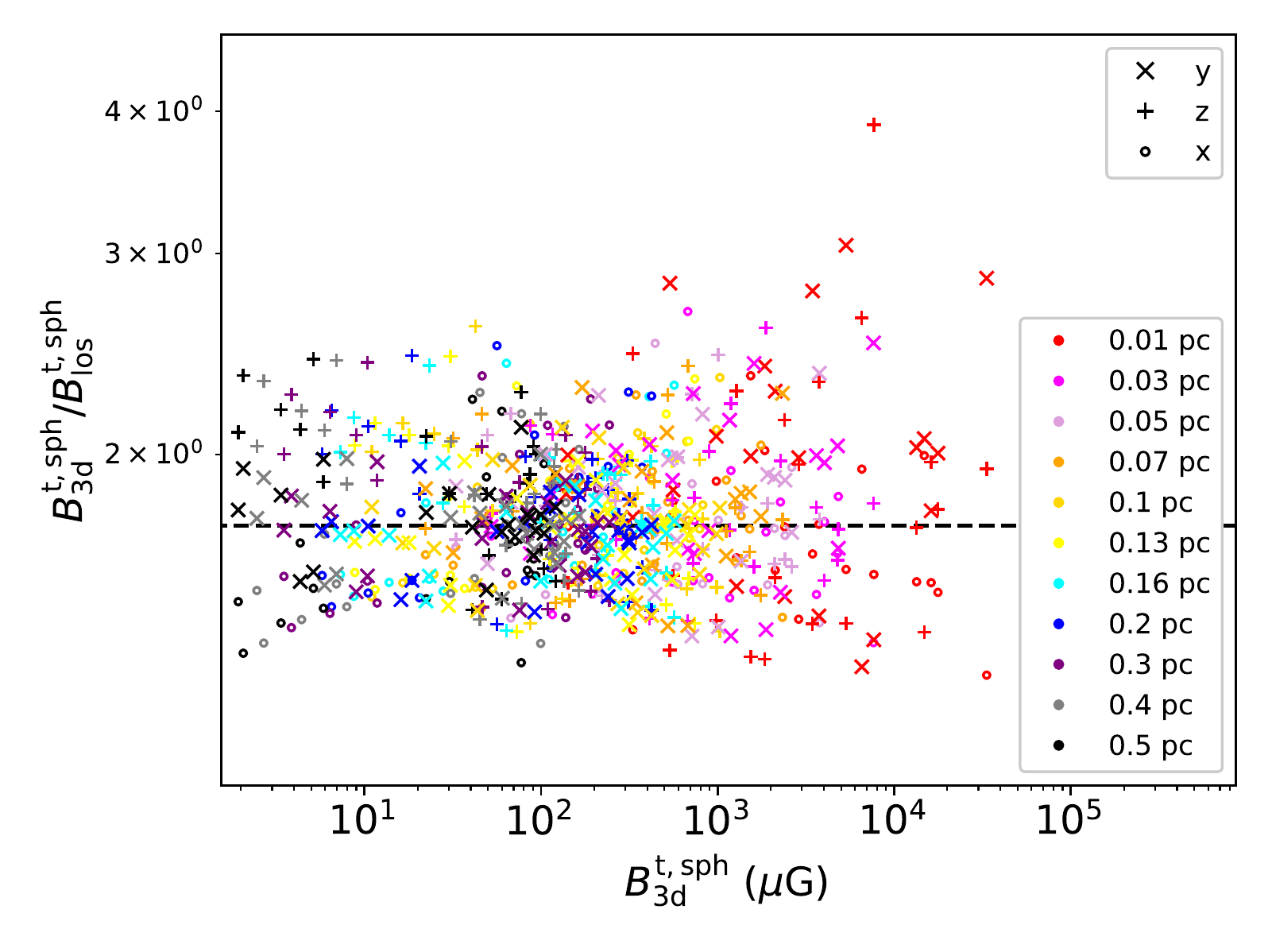}{0.45\textwidth}{b}}
\caption{(a). The ratio between the 3D turbulent velocity dispersion and the line-of-sight turbulent velocity dispersion within spheres of different radii for all models. (b). The ratio between the 3D turbulent magnetic field strength and the line-of-sight turbulent magnetic field strength within spheres of different radii for all models. Different symbols correspond to different line-of-sights. Different colors represents models with different initial $\mu$ values. \label{fig:dvproj}}
\end{figure}

One of the assumptions of the DCF method is isotropic turbulence, which requires 
\begin{equation}\label{eq:eqstatpos3ddv}
\delta v_{\mathrm{3d,turb}} = \sqrt{3} \delta v_{\mathrm{los,turb}}.
\end{equation}
The turbulence may be highly anisotropic if the magnetic field is strong \citep{1995ApJ...438..763G}. As stated in Section \ref{sec:paras}, the rotation can have significant contribution to the total velocity dispersion. So we remove the contribution from the rotation in the velocity dispersion. The 3D turbulent velocity dispersion is estimated as $\delta v_{\mathrm{3d,turb}} = (\delta v_{\mathrm{3d}}^2 - v_{\mathrm{t,rot}}^2)^{0.5}$ and the line-of-sight turbulent velocity dispersion is estimated as $\delta v_{\mathrm{los,turb}} = (\delta v_{\mathrm{los}}^2 - v_{\mathrm{t,rot}}^2/3)^{0.5}$ (see Section \ref{sec:paras} for the derivation of $v_{\mathrm{t,rot}}$). We show the ratio between the 3D turbulent velocity dispersion and the line-of-sight turbulent velocity dispersion within spheres of different radii in all simulation models in Figure \ref{fig:dvproj}(a). For most estimations, the 3D to line-of-sight ratio of turbulent velocity dispersion is between $\sim$1 and $\sim$3.5. Only for a few estimations, the line-of-sight turbulent velocity dispersion is much smaller than the 3D turbulent velocity dispersion. In our simulations, the models with initial $\mu=1.2$ or $\mu=2$ are tran-Alfv\'{e}nic and other models are super-Alfv\'{e}nic. However, there is no significant difference between the 3D to line-of-sight velocity dispersion ratios for models with different initial $\mu$ parameters. Nor do we see difference in ratios of different projections. Thus, we conclude that the assumption of isotropic turbulence generally holds in our simulations and the relation $\delta v_{\mathrm{3d,turb}} = \sqrt{3} \delta v_{\mathrm{los,turb}}$ is accurate within a factor of $\sim$2.

The turbulent magnetic field would also be expected to be isotropic if the turbulence is isotropic. We show the ratio between the 3D turbulent magnetic field strength and the line-of-sight turbulent magnetic field strength within spheres of different radii in all simulation models in Figure \ref{fig:dvproj}(b). The relation is quite similar to what is found for the velocity dispersion but shows slightly less scatters. 

\end{CJK*}
\end{document}